\documentclass[twoside,12pt]{article}
\usepackage{epsfig}

\newcommand{\be}{\begin{equation}}
	\newcommand{\ee}{\end{equation}}
\newcommand{\bea}{\begin{eqnarray}}
	\newcommand{\eea}{\end{eqnarray}}

\usepackage{subfigure}
\usepackage{graphicx}
\usepackage[toc,page]{appendix}
\usepackage{epstopdf}
\usepackage{amsmath,amssymb,amsfonts}
\usepackage{array}
\usepackage{url}
\usepackage{hyperref}
\usepackage{lineno}
\usepackage{xspace}

\begin{document}
\title{ \vspace{1cm} Search for the QCD Critical Point in High Energy Nuclear Collisions}
\author{A.\ Pandav$^{a,b}$, D.\ Mallick,$^{a,b}$ B.\ Mohanty,$^{a,b}$ \\
	\\
	$^a$School of Physical Sciences, National Institute of Science Education and Research, \\ Jatni 752050, India \\
        $^b$Homi Bhabha National Institute, Training School Complex, \\ Anushaktinagar, Mumbai 400094, India}

\maketitle

\begin{abstract}
QCD critical point is a landmark region in the QCD phase diagram outlined by temperature as a function of baryon chemical potential. To the right of this second-order phase transition point, one expects first order quark-hadron phase transition boundary, towards the left a crossover region, top of it lies the quark gluon plasma phase and below it the hadronic phase. Hence locating the QCD critical point through relativistic heavy-ion collision experiments is an active area of research. Cumulants of conserved quantities in strong interaction, such as net-baryon, net-charge, and net-strangeness, are suggested to be sensitive to the physics of QCD critical point and are therefore useful observables in the study of the phase transition between quark-gluon plasma and hadronic matter. We review the experimental status of the search for the QCD critical point via the measurements of cumulants of net-particle distributions in heavy-ion collisions. We discuss various experimental challenges and associated corrections in such fluctuation measurements. We also comment on the physics implications of the measurements by comparing them with theoretical calculations. This is followed by a discussion on future experiments and measurements related to high baryonic density QCD matter.
\end{abstract}


\section{Introduction: The phase diagram of the QCD and heavy-ion collisions}
\begin{figure}[!htb]
	\centering 
	\includegraphics[scale=0.65]{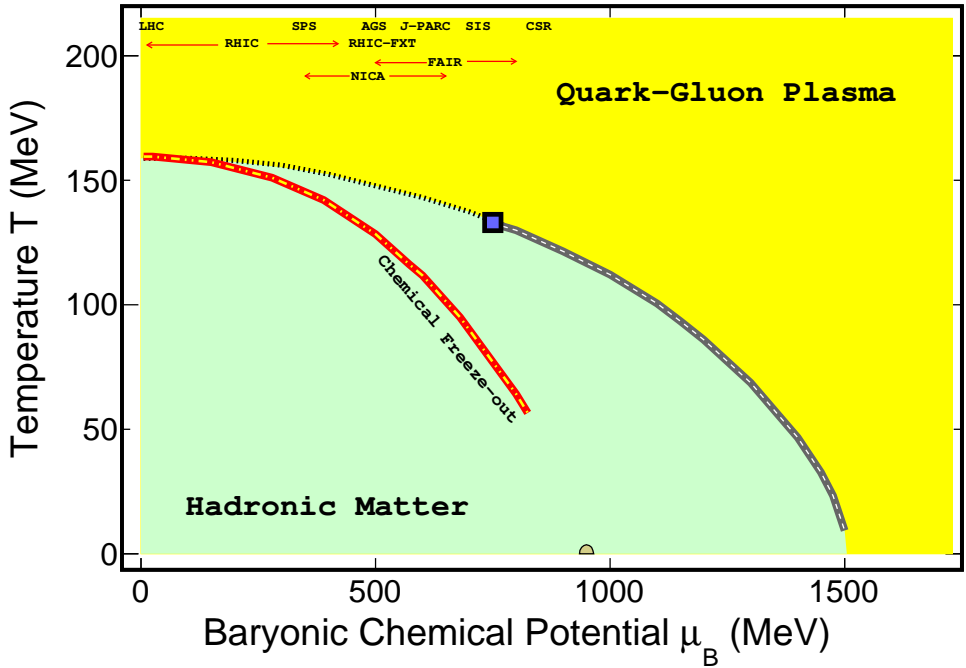}
	\caption{A conjectured QCD phase diagram is shown. The phase boundary separating the hadronic phase and quark-gluon phase at large $\mu_{B}$ is conjectured to be a first-order phase transition line (shown in solid black line). The first-order phase transition line ends at the QCD critical point represented by a solid square marker. The exact location of this point is under active investigation both experimentally and theoretically. At smaller $\mu_{B}$ there is a crossover indicated by a dashed line. The
		red-yellow dotted line corresponds to the chemical freeze-out (where inelastic collisions among the constituents of the system cease) inferred from particle yields in heavy-ion collisions using a thermal model. The ground state of nuclear matter ($T$ $\approx$ 0 and  $\mu_{B}$  $\approx$ 925 MeV)~\cite{Fukushima:2010bq} is shown at the bottom of the figure. The regions of the phase diagram accessed by past (AGS and SPS), ongoing (LHC, RHIC, SPS, and RHIC operating in fixed target mode), and future (FAIR, NICA, CSR, and J-PARC) experimental facilities are also indicated on the top of the figure. }
	\label{phase_dia} 
\end{figure}
Quantum Chromodynamics (QCD) is the underlying theory of strong interaction where the fundamental constituents are quarks and gluons. Under high temperature and pressure, the nuclear matter undergoes a transition from a confined state of quarks and gluons called the hadronic phase~\cite{Fukushima:2010bq,BraunMunzinger:2009zz,McLerran:2007qj}, to a state where quarks and gluons are deconfined, known as the Quark-Gluon Plasma (QGP)~\cite{Shuryak:1980tp}. Experimental study of heavy-ion collisions has revealed the strongly coupled nature of this state of matter~\cite{Romatschke:2007mq,Kovtun:2004de,Csernai:2006zz}. The investigation of the phase diagram of QCD is one of the primary goals of the heavy-ion collision experiments and has been an active subject of research for many years. Unlike the phase diagram of water which has gathered a competent understanding having electromagnetism as the nature of underlying interactions, the QCD phase diagram remains largely conjectured.
QCD phase diagram can be represented by a graph showing the variation of temperature ($T$) vs. chemical potential ($\mu$) associated with conserved charges like baryon number ($B$), electric charge ($Q$), and strangeness number ($S$). In heavy-ion collisions, the value of $\mu_Q$ and $\mu_S$ are fixed by the ratio of electric charge to baryon number and net-strangeness content of the colliding nuclei. Thus, the experimentally accessible QCD phase diagram effectively reduces to a two-dimensional graph of $T$ vs. $\mu_{B}$. A conjectured QCD phase diagram is shown in Fig.~\ref{phase_dia}.

QCD calculations on lattice at vanishing baryonic chemical potential and large temperature have established that the nature of quark-hadron phase transition is a smooth crossover~\cite{Aoki:2006we}. In lattice-QCD calculations, the chiral susceptibility is the order parameter distinguishing the phases. It is defined as, $\chi(N_s,N_\tau) = (\partial^2/\partial m^2_{ud}) (T/V)\log Z$ where $Z$ is the partition function, $m_{ud}$ is the mass of the light $u$,$d$ quarks, $N_s$ is the spatial extension, $N_\tau$ is euclidean time extension, and V the system volume in lattice-QCD. The smooth crossover was proved by showing that the temperature dependence of the peak and the width of the chiral susceptibility are independent of the system volume. For a typical first-order phase transition, the height of the susceptibility peak should have been proportional to the volume, and the width would vary inversely with volume. For a second-order transition, a singular behavior should have been observed with the volume of the system ($V^{\alpha}$, $\alpha$ is a critical exponent).
The point of sharpest change in the temperature dependence of the chiral susceptibility, the strange quark number susceptibility, and the renormalized Polyakov-loop (an approximate order parameter for quark deconfinement in a hot gluonic medium~\cite{Fukushima:2017csk} defined as $L(T) \sim \lim_{r\rightarrow \infty} \exp{(-V(r)/T)}$,
where $V(r)$ is the potential between a static quark-antiquark pair separated by a distance $r$) are used to estimate the QCD transition temperature in lattice calculations. The crossover transition temperature at $\mu_{B}$ = 0 is reported to be 156.5 $\pm$ 1.5 MeV~\cite{HotQCD:2018pds}.  The chiral phase transition temperature at $\mu_{B}$ = 0, which is expected to put an upper bound on the temperature for the possible critical point at nonzero baryon chemical potential, has been recently reported to be
 $136^{+3}_{-6}$ MeV~\cite{HotQCD:2019xnw} for vanishing two light quark masses and with the strange quark mass fixed to its physical value.  The temperature reduces to $98^{+3}_{-6}$ MeV~\cite{Dini:2021hug} for  3-flavor QCD.


Various QCD-based model calculations at large $\mu_{B}$ predict the phase transition to be of first order~\cite{Ejiri:2008xt,Bowman:2008kc}. The point on the $T$ vs. $\mu_{B}$ graph, where the first-order phase transition changes to crossover, is called the QCD critical point (CP)~\cite{Stephanov:1998dy,Stephanov:1999zu,Stephanov:2008qz}. The existence of the QCD critical point guarantees the occurrence of the first-order phase transition, and the vice-versa also holds true. There are several theoretical challenges in determining the location of the QCD critical point. The first principle lattice calculations at finite $\mu_{B}$ faces the issue of sign-problem~\cite{Goy:2016egl}. Even though the location of QCD critical point remains highly uncertain, recent estimates from the lattice-QCD calculations suggest, critical point, if it exists, should be in the region $\mu_{B}$ $\geq$ 300 MeV~\cite{Bazavov:2017tot}. 

\begin{figure}[!htb]
	\centering 
	\includegraphics[scale=0.65]{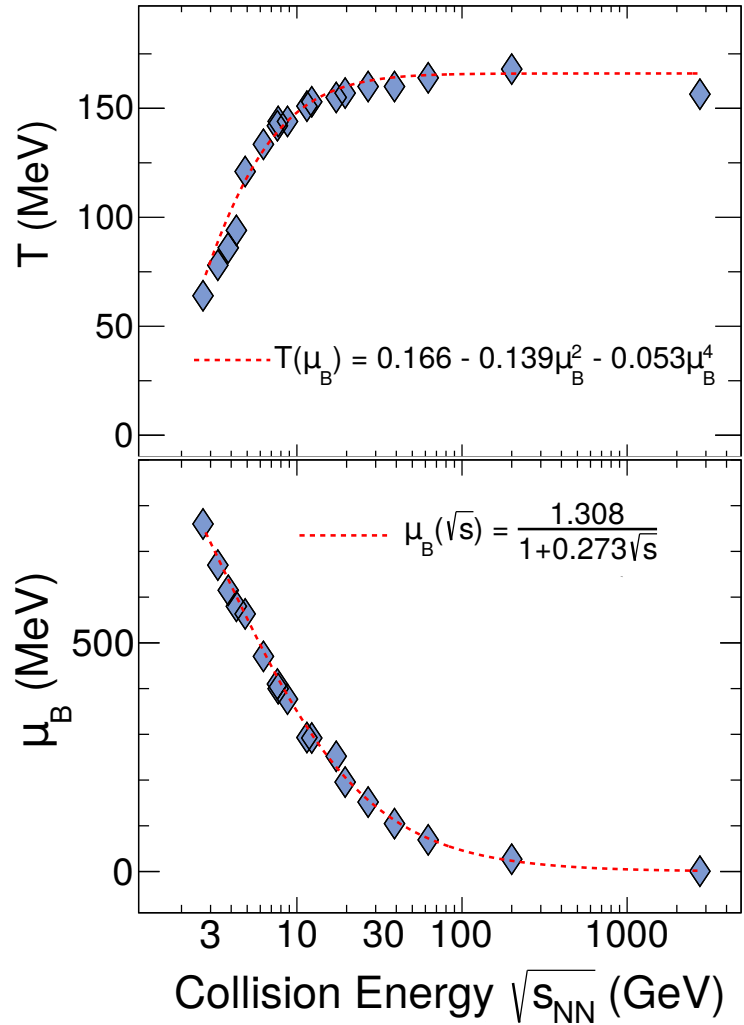}
	\caption{Chemical freeze-out parameters T and $\mu_{B}$ obtained by thermal model fits to experimental data on particle yields as a function of collision energies~\cite{Andronic:2005yp}. Red line represents parametrisations of T and $\mu_{B}$ as a function of $\sqrt{s_{NN}}$ taken from Ref.~\cite{Cleymans:2005xv}  }
	\label{ener_TmuB} 
\end{figure}
Experimentally, changing the energy and species of colliding nuclei can probe various regions of the QCD phase diagram. The measurements on particle yields and their ratios in the heavy-ion collision experiments can be compared to statistical thermal models to extract $T$ and $\mu_{B}$ at the chemical freeze-out~\cite{Cleymans:2005xv, Andronic:2005yp,BraunMunzinger:2007zz}. Figure~\ref{ener_TmuB} shows the extracted freeze-out parameters as a function of collision energy. The measurements suggest that $\mu_{B}$ decreases with increasing center-of-mass energy ($\sqrt{s_{NN}}$) while $T$ exhibits the opposite dependence for the lower collision energies ($\sqrt{s_{NN}} \lesssim $ 20 GeV) then saturates towards higher $\sqrt{s_{NN}}$. Thus by varying the collision energy of heavy-ions the $T$ and $\mu_{B}$ axes of the phase diagram can be scanned. Varying the impact parameter of the collisions, the species of colliding ions, and the rapidity acceptance could possibly further broaden the experimentally accessible region of the phase diagram.

In the scenario where freeze-out happens close to the phase transition, event-by-event fluctuations of conserved charge multiplicity distributions like the net-baryon, net-charge, and net-strangeness are suggested as sensitive observables to explore the QCD phase structure and search for the QCD critical point~\cite{Stephanov:1999zu,Asakawa:2009aj}. The cumulants of event-by-event conserved charge multiplicity distributions are related to the correlation length ($\xi$) of the hot and dense medium formed in heavy-ion collision experiments. Upon approaching the critical point, the correlation length should diverge for a static and infinite system. This is also true for any response functions like the susceptibilities, compressibilities or heat capacities. However, the matter formed in heavy-ion collisions is of femto-scopic in size, short-lived, and dynamically evolving with time, which limits the growth of the correlation length.
Model calculations suggest that the finite size and finite time effects attained in high energy heavy-ion collisions limits the value of the $\xi$ $\approx$ 2-3 fm~\cite{Berdnikov:1999ph}. This makes it extremely challenging to observe signals of criticality in experiments. Fortunately, model-based calculations suggest that the higher the order of the cumulants, the more is the sensitivity to the correlation length~\cite{Stephanov:2011pb}. Further, it is expected that non-Gaussian features in net-particle multiplicity distributions will increase if the system freezes-out closer to the critical point.  Hence, higher-order cumulants are favored as they could provide enhanced and observable critical signals.
The study of cumulants also provides measurements that can be used to establish thermalization in heavy-ion collisions, to find the crossover temperature between normal nuclear matter and a deconfined phase called the quark gluon plasma, and set a scale for the phase diagram of QCD~\cite{Gupta:2011wh}.
Cumulants up to the $4^{th}$ order ($C_{n}, n$ $\leq$ 4) of event-by-event distributions of net-charge, net-kaon and net-proton were measured by the STAR Collaboration in the phase I of Beam Energy Scan (BES) program at RHIC~\cite{starNC,starNK,starNP10,starNP,starNP21,starNP21_long}. 

This review will focus on the most recent measurements on net-particle cumulants~\cite{starNC,starNK,starNP21,starNP21_long} and their implication in the hunt for the QCD critical point. Net-proton and net-kaon are used as proxies for net-baryons and net-strangeness, respectively, as charge-neutral particles are generally not identified in the experiments on an event-by-event basis. Theoretical calculations have shown that the net-proton number fluctuations could reliably reflect the baryon number susceptibility in the vicinity of critical point~\cite{Hatta:2003wn}. The prescription to link net-proton number fluctuations to that of net-baryon are discussed in Ref.~\cite{Kitazawa:2012at}.

\section{Theoretical status of QCD critical point}
\begin{figure}[!htb]
	\centering 
	\includegraphics[scale=0.65]{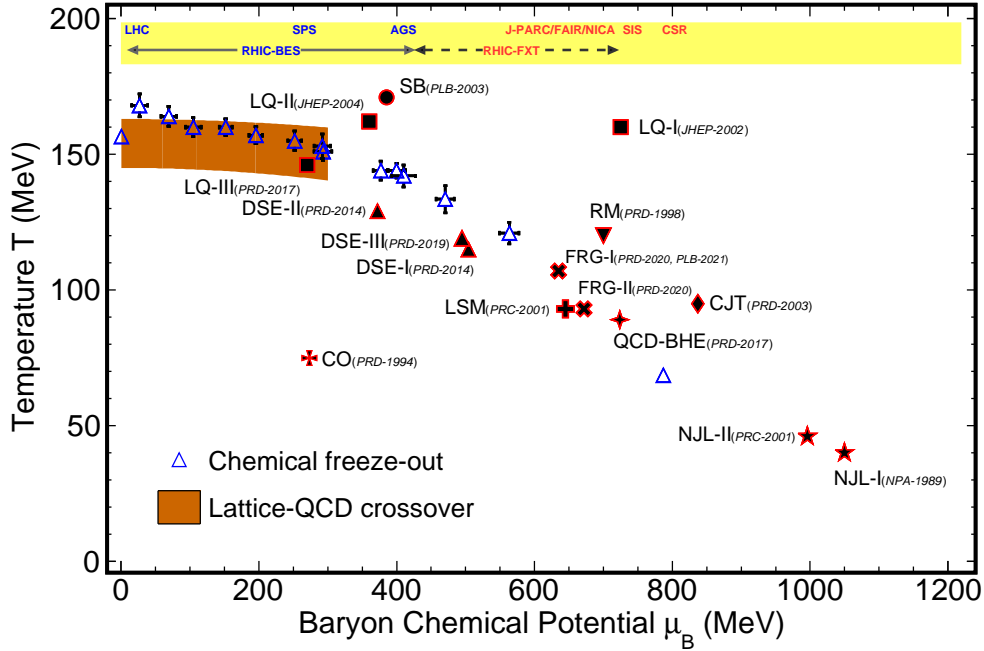}
	\caption{Theoretical predictions on location of critical points from lattice-QCD calculations and QCD-based models. The labels are abbreviated for the models and publications (LQ: Lattice-QCD, DSE: Dyson-Schwinger equations, SB: Statistical Bootstrap Model, RM: Random Matrix Model, FRG: Functional Renormalization Group, LSM: Linear Sigma Model, CJT: Cornwall-Jackiw-Tomboulis effective potential, CO: Composite Operator formalism, and QCD-BHE: QCD from Black Hole Engineering). The band in light orange color represents the crossover from lattice-QCD calculations. The open blue markers are freeze-out locations obtained from the statistical model fits of yields measured in SPS, AGS, RHIC, and LHC experiments. Predicted locations of CP from different calculations is scattered widely in the QCD phase diagram~\cite{Fodor:2001pe,Fodor:2004nz,Datta:2016ukp,Fischer:2014ata,Xin:2014ela,Asakawa:1989bq,Scavenius:2000qd,Hatta:2002sj,Antoniou:2002xq,Fu:2019hdw, Isserstedt:2019pgx,Halasz:1998qr,Barducci:1993bh,Gao:2020qsj,Gao:2020fbl,Critelli:2017oub}. Also shown on the top are the regions in $\mu_{B}$ covered by various facilities and experimental programs.}
	\label{CP_theory} 
\end{figure}

First principle calculations for strongly coupled systems are, in general, very challenging. Theoretically, one would need to calculate the partition function of the QCD and obtain the singularities corresponding to the first-order phase transition line to locate the ($T$, $\mu_{B}$) of the CP. This demands solving infinite-dimensional integrals, which are beyond our current capabilities, and hence one resorts to the numerical methods, $i.e.$ the lattice-QCD (LQCD) simulations. At zero $\mu_{B}$, LQCD allows calculation of the equation of state of QCD as a function of $T$, indicating the nature of phase transition to be a smooth crossover. However, at finite $\mu_{B}$, which is the most suitable regime for the search of QCD critical point, LQCD suffers from the sign problem. The sign problem can be understood as follows: Typically, for any lattice computation one needs to evaluate the expectation value of an observable $X$, $\langle X(m_v) \rangle = {{ \int \scriptstyle{D} U \exp (-S_G) X(m_v)~ {\rm Det}~M(m_s)}
	\over { \int \scriptstyle{D} U \exp (-S_G)~ {\rm Det}~M(m_s)} }~~,$ where $M$ is the Dirac matrix in co-ordinate, colour, spin, flavour space for sea quarks of mass $m_s$, $S_G$ is the gluonic action, and the observable $X$ may contain fermion propagators of mass $m_v$. The Det M for non zero $\mu$ is not positive definite, hence numerical methods of evaluation of the expectation values are difficult; this is commonly referred to as the {\it sign problem}. There are some suggested ways to overcome this issue. (i) Reweighting the partition function in the vicinity of transition temperature and $\mu$ = 0~\cite{Fodor:2004nz}, (ii) Taylor expansion of thermodynamic observables in $\mu$/T about $\mu$ = 0~\cite{Gavai:2003mf}, and (iii) Choosing the chemical potential to be imaginary will make the fermionic determinant positive~\cite{Philipsen:2009yg}.  Due to these difficulties in lattice calculations in the finite or large $\mu_{B}$ region, one requires QCD-based theories to make predictions on the presence and location of the critical point. Various theoretical calculations (LQCD and QCD-based models) on the location of critical point are shown in Fig.~\ref{CP_theory}. The location of QCD critical point from theory calculations is scattered over region of $\mu_{B}$ = 200--1100 MeV and $T$ = 40--180 MeV~\cite{Fodor:2001pe,Fodor:2004nz,Datta:2016ukp,Fischer:2014ata,Xin:2014ela,Asakawa:1989bq,Scavenius:2000qd,Hatta:2002sj,Antoniou:2002xq,Fu:2019hdw, Isserstedt:2019pgx,Halasz:1998qr,Barducci:1993bh,Gao:2020qsj,Gao:2020fbl,Critelli:2017oub}. While the accuracy of predictions for CP from the first principle lattice-QCD calculations worsens towards very large $\mu_{B}$, various model calculations wildly vary in their predictions. Therefore, experimental search of critical point is crucial to help establish it in the phase diagram.

\begin{figure}[!htb]
	\centering 
	\includegraphics[scale=0.99]{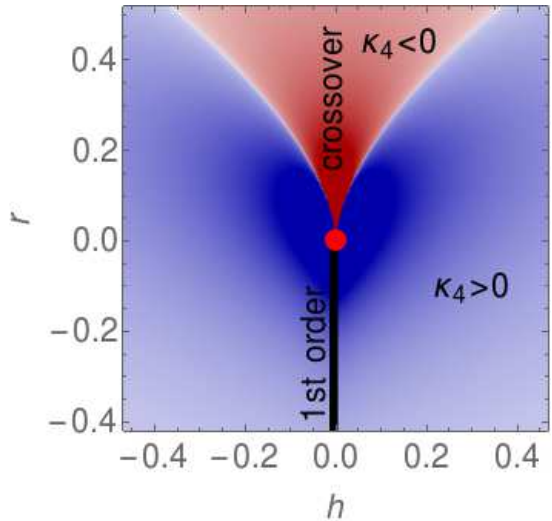}
	\caption{Quartic cumulant $\kappa_{4}$ of the Ising model magnetization near the critical point~\cite{Stephanov:2011pb}. The $y$ axis is the reduced temperature, or the difference of temperature in the Ising model from the critical temperature and the $x$ axis reflects the external ordering field which results in breaking of symmetry in the Ising model.}
	\label{sign_kurt} 
\end{figure}

One of the widely used theoretical calculations for the search of QCD critical point is with the linear sigma model~\cite{Stephanov:2011pb}. A well-known behavior in critical phenomena is the divergence of response function at criticality due to sudden growth of the correlation length. At the CP, the exponents of correlation length on which various orders of thermodynamic response functions depend on are universal~\cite{hagedron:univer} and hence only rely on degrees of freedom in the theory and its symmetry. In view of the static critical phenomena, the QCD critical point falls into the Ising universality class~\cite{Ising}. The linear sigma model calculations qualitatively predict the universal critical behavior of the kurtosis of the order parameter fluctuations, $i.e.$, of the $\sigma$-field, near the QCD critical point. Introducing coupling of the particles to the $\sigma$-field, allows the connection of fluctuations of the order parameter field, $\sigma$, to the fluctuations of the experimentally observable quantities, like the multiplicity distributions. The kurtosis of the order parameter, the coupling of protons to the $\sigma$-field, as well as the kurtosis of proton multiplicity distribution is discussed in Ref.~\cite{Stephanov:2011pb}. To exploit the universality of critical phenomena, mapping of Ising model coordinates $r$ and $h$ ($r$ is the reduced temperature and $h$ is the external ordering field in Ising) to the $T-\mu_{B}$ plane of QCD phase diagram is needed. This requires the knowledge of QCD equation of state with a critical point. Using information from lattice-QCD at $\mu_{B}$=0, and known features of thermodynamic variables near the QCD critical point, such a mapping is discussed in Ref.~\cite{Stephanov:2011pb,Bzdak:2019pkr}, where certain predictions about the behavior of QCD near the critical point was made. As shown in the Fig.~\ref{sign_kurt}, upon approaching the critical point from the crossover side of phase transition, the kurtosis of the order parameter is universally negative, and hence kurtosis of fluctuation observables like proton multiplicity distributions will be driven to take lower values than purely statistical fluctuations. Receding away from the critical point towards the region of first-order phase transition, the effect which drove the negative kurtosis of the order parameter previously will cause a higher value of kurtosis of proton multiplicity distributions as compared to baseline statistical fluctuations. Hence, non-monotonic energy dependence of the normalized kurtosis of proton multiplicity distributions with respect to baseline fluctuations would suggest the presence of a QCD critical point. The magnitude of the deviation from the baseline fluctuations will depend on the proximity of the freeze-out to the critical point. This model considers the critical fluctuations with the assumption of a static and infinite medium at equilibrium. However, one needs to keep in mind the finite size and time effect of the system created in heavy-ion collisions. Effects like initial geometry fluctuations, diffusion of conserved charge, the time evolution of fluctuations, jets, the role of conservation laws, resonances could also influence the fluctuation measurements in heavy-ion collisions. Nevertheless, the non-monotonic energy dependence of the kurtosis in the presence of a critical point has been confirmed by various other QCD-based model calculations like the Nambu--Jona-Lasinio (NJL)~\cite{Chen:2015dra,NJL_old}, Polyakov loop extended Quark Meson Model (PQM)~\cite{Friman:2011pf,Friman:2014cua,mukherjee2015real}. Functional renormalisation group (FRG) approach to QCD also suggest non-monotonic collision energy dependence of fourth-order baryon number susceptibility($\chi^{B}_{4}/\chi^{B}_{2}$) due to sharpening of chiral crossover at large $\mu_{B}$, which may hint at a change in nature of phase transition at higher $\mu_{B}$~\cite{Fu:2021oaw}.\\

Confirmation of a first-order phase transition would indeed validate the presence of a critical point. This is true, given that lattice-QCD calculations have confirmed the transition to be a crossover at $\mu_{B}$ = 0~\cite{Aoki:2006we}.  If the system formed in heavy-ion collisions is in the proximity of first-order phase transition, the multiplicity distributions could become bi-modal/two-component in nature, reflecting the contributions from two distinct phases. The two-component distribution has a very specific behavior of factorial cumulants, $i.e.$; they increase in magnitude with increasing order and alternate in sign~\cite{volker_case1,volker_case2,volker_case3}. This characteristic behavior is true even if one of the components is very small. This can be verified in the experimental measurements to search for first-order phase transition at finite $\mu_{B}$. Also, there has been no direct experimental evidence of crossover in the small $\mu_{B}$ region so far. LQCD calculations have been extended to non-zero values of $\mu_{B}$ using Taylor series expansion about vanishing $\mu_{B}$. Recent LQCD results for $\mu_{B} <$ 110 MeV, suggests negative sign of ratio of fifth-to-first and sixth-to-second order baryon susceptibility~\cite{Bazavov:2020bjn}. The FRG model calculations also give the negative sign for the two ratios across a large range of $\mu_{B}$ which is accessed in gold nuclei collisions at $\sqrt{s_{NN}}$ = 7.7 -- 200 GeV in the STAR experiment at RHIC~\cite{Fu:2021oaw}. A PQM model study also suggests that for freeze-out close to the chiral phase transition temperature, the ratio of sixth-to-second order baryon susceptibility remains negative~\cite{Friman:2011pf}. The nature of phase transition in these calculations is a crossover. Hence, higher-order cumulant measurements can serve as a litmus test to identify the nature of phase transition.
\section{Experimental search for QCD critical point }
This section deals with various aspects related to  the experimental search for the QCD critical point. First, the observables employed in the search for the QCD critical point are presented. Baselines from various non-critical physics, like the thermal statistical fluctuations, baryon number conservation, and resonances, are also discussed. There are several complexities involved in experimental measurements of the number fluctuations like the volume fluctuations, efficiency correction for the high order cumulants, and centrality resolution. The origin of these experimental challenges and how they are addressed while performing the measurements are dealt with in the subsequent subsections. Finally, the results from the experimental measurements are presented and compared with various model calculations to understand their physics implications.
\subsection{Observables}
\label{rfer_obser_subsectn}
Cumulants quantify the traits of a distribution. The cumulants of a distribution up to the sixth order are defined as follows:
\begin{eqnarray}
C_1 &=& \langle N \rangle\\
C_2 &=& \langle(\delta N)^2\rangle\\
C_3 &=& \langle(\delta N)^3\rangle\\
C_4 &=& \langle(\delta N)^4\rangle -3 \langle(\delta N)^2\rangle^2\\
C_5 &=& \langle(\delta N)^5\rangle -10 \langle(\delta N)^3\rangle \langle(\delta N)^2\rangle\\
C_6 &=& \langle(\delta N)^6\rangle -15 \langle(\delta N)^4\rangle \langle(\delta N)^2\rangle\nonumber\\ &-&10 \langle(\delta N)^3\rangle^2+30 \langle(\delta N)^2\rangle^3.
\label{eqn_cumu}
\end{eqnarray}
where $N$ is the observable whose distribution is being considered, which for our case is the net-particle number in a single event and $\langle N \rangle$ is the average of the $N$ from all the events. The difference of N from $\langle N \rangle $ is given by $\delta N = N - \langle N \rangle $. The first and second order cumulants are the well known mean ($M$)  and variance ($\sigma^2$) of a distribution whereas the third and fourth order cumulants reflect the skewness ($S$) and kurtosis ($\kappa$) of a distribution, respectively. Cumulants can be related with conserved charge number susceptibilities for a system in thermal equilibrium. In LQCD and hadron resonance gas (HRG) model calculations, using the grand canonical ensemble construct, the $n^{th}$ order number susceptibility ($\chi^{q}_{n} $) associated to conserved charges, $q$ ($q = B, Q, S$) can be calculated by taking derivative of dimensionless pressure ($P/T^{4}$) with respect to reduced chemical potential ($\mu_{q}/T$) corresponding to the conserved charge $q$ and is given as,~\cite{Gavai:2010zn,Gupta:2011wh}
\begin{eqnarray}
\chi^{q}_{n} = \frac{\partial^{n} (P/T^{4})}{\partial (\mu_{q}/T)^n}.
\label{sus_eq1}
\end{eqnarray}
Here $P$ and $T$ represents the system pressure and temperature, respectively. Pressure can be written in terms of logarithm of partition function ($Z$) as follows: 
\begin{eqnarray}
P/T^{4} = \frac{1}{VT^{3}}\ln[Z(V,T,\mu_{q})].
\label{sus_eq2}
\end{eqnarray}
One can relate the cumulants of the conserved charge multiplicity distribution to the corresponding number susceptibilities as follows.
\begin{eqnarray}
C^{q}_{n} = VT^{3}\chi^{q}_{n}.
\end{eqnarray}
As can be seen from the above equation, all the cumulant orders directly depend on system volume ($V$). Constructing the ratio of cumulants eliminates the trivial system volume dependence and thus allows for a direct comparison of the experimental measurements to the thermodynamic susceptibilities calculated in lattice-QCD, HRG, and various QCD-based models, although subject to some caveats. The experimental measurements are carried out within specific kinematic phase space, and probe the conserved charge quantities via forming proxies from the detected charged particles. The theoretical calculations like those from the lattice-QCD and QCD-based models instead deal with conserved charge quantities and do not usually involve any phase space cuts. These limitations should be kept in mind when comparing the experimentally measured cumulant ratios to the susceptibility ratios obtained from theoretical calculations presented throughout this review article. In the existing literature, both the ratios of cumulants and the product of moments of the distribution are used. The most widely used products of moments in the literature are the $\sigma^2/M$, $S\sigma$, and $\kappa\sigma^2$.  The exact relation of these moment products to the cumulants is listed below.
\begin{eqnarray}
\sigma^2/M = \frac{C_2}{C_1} \\
S\sigma = \frac{C_3}{C_2}\\
\kappa\sigma^2 = \frac{C_4}{C_2}
\label{eqn_cumu_ratio}
\end{eqnarray}
From cumulants, one can construct the factorial cumulants, also known as the integrated correlation functions. The factorial cumulants ($\kappa_n$) are related to cumulants ($C_n$) as follows:
\begin{eqnarray}
\kappa_1 &=& C_1\\
\kappa_2 &=& -C_1+C_2\\
\kappa_3 &=& 2C_1-3C_2+C_3\\
\kappa_4 &=& -6C_1+11C_2-6C_3+C_4\\
\kappa_5 &=& 24C_1-50C_2+35C_3-10C_4+C_5\\
\kappa_6 &=& -120C_1+274C_2-225C_3\nonumber\\ &+&85C_4-15C_5+C_6.
\label{eqn_kappa}
\end{eqnarray}
In addition, extensive studies on various other fluctuation observables have been done in the existing literature. Dynamical $K/\pi$, $p/\pi$ and $K/p$ fluctuations in terms of an observable which is robust to detector efficiency effects called the $\nu_{dyn}$ have been reported by experiments~\cite{other_observble_1}. Another such robust observable is the balance function, which measures probability of observing net-particle separated in pseudo-rapidity. Narrowing of balance function was observed by the experiments~\cite{other_observble_2}, which is consistent with the concept of delayed hadronization of a deconfined QGP. Measurement of $p_T$ dynamical correlations have also been performed by experiments~\cite{other_observble_3}. Monotonic scaling of the measurements on dynamical correlations has been observed so far.
\subsection{Non-critical point contributions and baselines}
This section discusses various non-critical baseline calculations. Although a non-monotonic collision energy dependence has been suggested as being the signature of the presence of QCD critical point, there could be several other non-critical effects causing appreciable correlations in the experimental measurements. For example, the effect of unavoidable physical effects like the purely statistical fluctuations, baryon stopping at lower energies, baryon number conservation, and  resonances~\cite{Bzdak:2012an,Braun-Munzinger:2016yjz,Luo:2014tga,Behera:2019cds,Zhang:2019lqz}, needs to be studied. In this regard, in addition to statistical baselines, two widely studied models, which do not incorporate any physics of QCD critical point or QCD phase transition, are discussed, namely the HRG~\cite{Karsch:2010ck,Garg:2013ata} and the UrQMD~\cite{Bleicher:1999xi}.
\subsubsection{Statistical baseline}
If the fluctuations of net-particle multiplicity distributions are purely Gaussian in nature, the third and all higher-order cumulants become zero. Hence, the presence of non-zero third, fourth, and all higher-order cumulant is the first sign to take note of while probing critical behavior of cumulant.

One of the widely used statistical baselines is the Skellam baseline. A Skellam distribution is the resulting distribution of difference of two variates that follow independent Poisson distributions. For example, if the protons and anti-protons follow independent Poisson distributions, the distribution of net-protons is Skellam. The Poisson distribution has the unique property of having the same value for all the order of cumulants. Cumulants, in general, follow additive property, $i.e.$, given two independent random variables $X$ and $Y$, cumulants of a new variate $Z$, constructed by addition of $X$ and $Y$ ($Z$ = $X$ + $(-1)^m Y$, where $m$ can be any integer), could be obtained by simply adding the cumulants of the input variates ($C^{Z}_{n} = C^{X}_{n} + (-1)^nC^{Y}_{n} $). Using the previous two statements, it can be easily shown that the ratio of odd-to-odd and even-to-even cumulants calculated from Skellam expectations are unity. Similarly, one can also assume binomial and negative binomial distribution for the constituent variates of the net-particle distributions for statistical baseline. The table~\ref{tab_baseline} tabulates the values of statistical baseline for cumulants of net-particle distributions ($Z$) up to the sixth-order, when the constituent variates of the distributions ($X$ and $Y$: $Z$ = $X$ - $Y$)  follow independent Poisson, Gaussian, Binomial, and Negative Binomial distributions(NBD). In the table, the baseline expectations for Binomial and NBD distributions are expressed in terms of their cumulants. The cumulants of a Binomial distribution up to sixth-order can be obtained in terms of its parameters; the number of trials ($n$), and probability of success ($p$) as follows:
\begin{eqnarray}
C_1 &=& np\\
C_2 &=& np(1-p)\\
C_3 &=& np(1-3p+2p^2)\\
C_4 &=& np(1-7p+12p^2-6p^3)\\
C_5 &=& np(1-15p+50p^2-60p^3+24p^4)\\
C_6 &=& np(1-31p+180p^2-390p^3\nonumber\\ &+&360p^4-120p^5).
\label{eqn_binomial}
\end{eqnarray}
Similarly, cumulants of an NBD distributions up to the sixth order can be expressed in terms of its parameter; the number of failures ($r$), and probability of success ($p$) as follows:
\begin{table}[!htb]
	\caption{Values of statistical baselines up to sixth order cumulant when underlying nature of constituent distributions are Poisson, Gaussian, Binomial and Negative Binomial distributions. A Poisson distribution has only one parameter: the mean ($\lambda$), a Gaussian has two parameters: the mean ($\mu$) and width ($\sigma$), Binomial has two parameters: number of trials ($n$) and probability of success ($p$), and NBD also has two parameters: number of failures ($r$) and probability of success ($p$), respectively. }
	\centering   
	\begin{tabular}{|c|c|c|c|c|}
		\hline	
		Cumulant  & Poisson  & Gaussian &  Binomial & NBD   \\
		\hline 
		$C_1$ & $\lambda_X - \lambda_Y$ & $\mu_{X} - \mu_{Y}$ & $C_{1X}- C_{1Y}$ & $C_{1X} - C_{1Y}$\\
		\hline 
		$C_2$  & $\lambda_X + \lambda_Y$ & $\sigma^2_{X} + \sigma^2_{Y}$  &  $C_{2X} + C_{2Y}$ & $C_{2X} + C_{2Y}$\\
		\hline 
		$C_3$ &  $\lambda_X - \lambda_Y$ & 0   & $C_{3X} - C_{3Y}$ & $C_{3X} - C_{3Y}$ \\
		\hline 
		$C_4$ &  $\lambda_X + \lambda_Y$ & 0  & $C_{4X} + C_{4Y}$& $C_{4X} + C_{4Y}$ \\
		\hline 
		$C_5$ &  $\lambda_X - \lambda_Y$ &  0 &  $C_{5X} - C_{5Y}$& $C_{5X} - C_{5Y}$ \\
		\hline 
		$C_6$ &  $\lambda_X + \lambda_Y$ & 0 & $C_{6X} + C_{6Y}$& $C_{6X} + C_{6Y}$\\
		\hline 
	\end{tabular}
	\label{tab_baseline}
\end{table}

\begin{eqnarray}
C_1 &=& \frac{pr}{(1-p)}\\
C_2 &=& \frac{pr}{(1-p)^2}\\
C_3 &=& \frac{p(1+p)r}{(1-p)^3}\\
C_4 &=& \frac{(p+4p^2+p^3)r}{(1-p)^4}\\
C_5 &=& \frac{(p+11p^2+11p^3+p^4)r}{(1-p)^5}\\
C_6 &=& \frac{(p+26p^2+66p^3+26p^4+p^5)r}{(1-p)^6}.
\label{eqn_neg_binomial}
\end{eqnarray}

With the knowledge of experimental measurement of first and second-order cumulant of the constituent distribution, one can extract the parameters for the underlying statistical distributions considered above for statistical baseline and make predictions for the higher-order cumulants of the net-particle distributions. For Poisson distribution, only the experimental measurement of mean is required as input as it has only one parameter.

\subsubsection{Hadron resonance gas model}
In the ideal or non-interacting hadron resonance gas (HRG) model, the degrees of freedom is comprised of point-like hadrons and resonances which are in thermal equilibrium. The logarithm of the partition function of a hadron resonance gas in the grand canonical ensemble (GCE) is given as 
\begin{eqnarray}
\ln Z^{ideal} = \sum{\ln Z_{i}^{ideal}}
\end{eqnarray}
where the summation is over all hadrons and resonances.  Further, for each particle,
\begin{eqnarray}
\ln Z_{i}^{ideal} = \pm\frac{Vg_{i}}{2\pi^{2}}\int{p^{2}dp \ln[1\pm\exp(\mu_{i} - E_{i})/T]}
\end{eqnarray}
where, $T$ is the temperature, $V$ is the system volume, $E_i$ is the energy, $g_i$ is the degeneracy factor and $\mu_{i}$ = $B_i\mu_{B}+S_i\mu_{S}+Q_i\mu_{Q}$ is the chemical potential of the $i^{th}$ particle. Here in the last expression, $B_i$, $S_i$ and $Q_i$  are the baryon, electric charge, and strangeness number of the $i^{th}$ particle, with corresponding chemical potentials $\mu_B$, $\mu_Q$, and $\mu_S$ , respectively. The $\pm$ sign are for baryons and mesons, respectively. Once the partition function of the hadron resonance gas is known, using the eqns~\ref{sus_eq1} and ~\ref{sus_eq2}, the thermodynamic susceptibilities can be obtained.\\
At higher temperatures and large chemical potential values, the ideal gas assumption may no longer be valid. Van-der-Waals(VDW)-type interactions can be incorporated into the HRG to reflect the qualitative feature of strongly interacting gas of hadrons. In the excluded-volume (EV) approach to HRG, Van-der-Waal type repulsive interactions are introduced by relaxing the argument of the point-like size of hadrons and considering the geometrical size of hadrons. The Van-der-Waals equation of state is given as follows:
\begin{eqnarray}
\big(P + (\frac{N}{V})^2 a\big)\big(V-Nb\big) = NT.
\end{eqnarray}
Here, $P$, $V$, and $N$ are the pressure, volume and number of particles of the system, respectively. The parameters $a$ and $b$ are the Van-der-Waals parameters which describe the attractive and repulsive interactions, respectively.
One can also include both attractive and repulsive Van-der-Waals interactions to the HRG model framework. Susceptibility calculations using varieties of the HRG models in the existing literature are discussed in Ref.~\cite{Karsch:2010ck,Garg:2013ata,Fu:2013gga,Bhattacharyya:2013oya,Bhattacharyya:2015zka,Vovchenko:2017xad,Samanta:2019fef}.

The most commonly used variants of HRG (both ideal and interacting) employs the Grand Canonical Ensemble (GCE), which results in the conservation of quantum number on an average. However, in the low center-of-mass energies for heavy-ion collisions, the canonical ensemble (CE) approach becomes more suitable, which conserves the exact quantum number as opposed to the GCE construct~\cite{Braun-Munzinger:2020jbk}. 
The fluctuation in net-proton is constrained to the net-baryon number in the full phase space, which itself is a conserved quantity. Using the information on the mean number of accepted protons and anti-protons within the kinematic acceptance, fluctuation of net-proton in the selected kinematic acceptance is modelled~\cite{Braun-Munzinger:2020jbk}.

 \subsubsection{UrQMD model}
The Ultra-relativistic Quantum Molecular Dynamics (UrQMD) model is a microscopic transport model~\cite{Bleicher:1999xi}. The model includes transportation of various degrees of freedom ($e.g.$ baryons and mesons) and the production of new particles and their interaction. The production of particles occurs via fragmentation of strings made of valence quarks of the original colliding hadrons, resonance excitation, and decays. There is no quark-hadron phase transition implemented in this model. To study the effect of baryon number conservation, baryon stopping at lower energies, and resonances on the fluctuation measurements, UrQMD calculations serve as a suitable baseline for the collision energies at RHIC.

\subsection{Experimental challenges }
Although obtaining the number of particles in an event may appear simple, several experimental techniques are employed in the measurements of net-particle fluctuations. Event statistics required for proper estimation of cumulants, purity of the charged particles selected, centrality/impact parameter estimates, removing self-correlation between particles used to define centrality and net-particle fluctuation, volume fluctuations and ways to suppress the effect, correction of cumulants for finite detector efficiency, estimation of uncertainties, are some of the topics which are of prime importance in fluctuation measurements. Current subsection discusses these aspects of the measurements. 

\subsubsection{Event statistics and statistical uncertainties}
\begin{figure}[!htb]
	\centering 
	\includegraphics[scale=0.65]{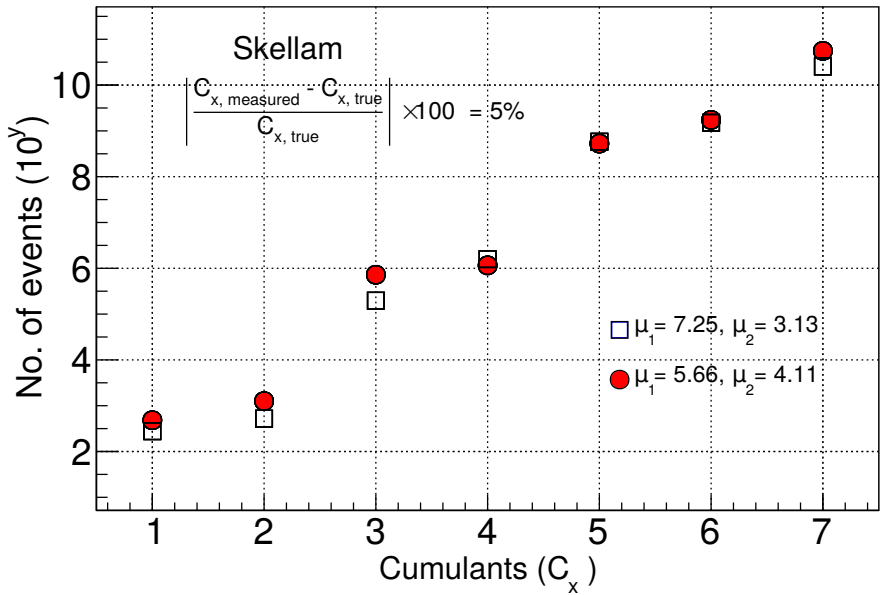}
	\caption{Minimum event statistics required for measurement of net-proton cumulants for various orders. Net-proton distribution is simulated assuming Poisson as the underlying distribution with input parameters $\mu_{1}$ ($\mu_{2}$) taken from the experimental mean of proton (anti-proton) distributions from Au+Au collisions at $\sqrt{s_{NN}}$ = 62.4 GeV (open square) and 200 GeV (filled circle)~\cite{starNP}.}
	\label{stat_limits} 
\end{figure}

As experiments run for a finite amount of time and collect a limited amount of data, estimates should be made a priori on how much event statistics are sufficient enough to measure an observable of interest. The Ref.~\cite{Pandav:2018bdx} performs such a model study and presents the minimum event statistics needed for estimation of cumulants ($C_{n};n<8$) in order to detect a signal (related to phase transition effects or critical point) of 5\% and 10\% above the statistical baseline. The event-by-event net-proton distribution constructed as the difference of the number of protons and anti-protons was simulated using two types of underlying distributions, Poisson and Binomial. The input parameters are taken as the mean and width of (anti-)proton distributions in 0-5\% central Au+Au collisions at $\sqrt{s_{NN}}$ = 62.4 and 200 GeV measured by the STAR detector~\cite{starNP}. Minimum statistics required to estimate the cumulants of net-proton distribution with a precision of 5\% with Poisson being the nature of underlying distributions for protons and anti-protons are shown in Fig.~\ref{stat_limits}. The limits are shown for two sets of input parameters determined from the STAR experiment. Among various orders, we discuss here the statistics needed for $C_4$ and $C_6$ measurements. Estimation of $C_4$ ($C_6$) within a precision of 5\%  requires 1.2 million (1.5 billion) event statistics with input parameters taken from $\sqrt{s_{NN}}$ = 200 GeV while they are found to be 1.6 million (1.5 billion) with input parameters from $\sqrt{s_{NN}}$ = 62.4 GeV. Similar study done with binomial assumption for protons and anti-protons distribution in the simulation fixing the inputs from $\sqrt{s_{NN}}$ = 200 GeV suggests measurement of $C_4$ ($C_6$) at a level of 5\% precision demands 1.3 million (1.8 billion) events~\cite{Pandav:2018bdx}. Taking the inputs from $\sqrt{s_{NN}}$ = 62.4 GeV also gives similar event statistic limits.
One of the important aspects of any measurement is the proper estimation of uncertainties or errors on them. We discuss in this subsection how the statistical uncertainties on the fluctuation measurements are calculated. The limited event statistics available for measurements demands a careful estimation of statistical uncertainties. Statistical uncertainties on the cumulants are usually obtained using the Delta theorem method and Bootstrap method. While the Delta theorem method is an analytical method of standard error propagation, the Bootstrap method is a computer-intensive re-sampling method. \\
Using the Delta theorem, statistical uncertainties on cumulants and moments have been computed~\cite{Luo:2011tp,Luo:2014rea} and have been quite extensively used in the analysis of higher-order cumulants in heavy-ion collision experiments. This error estimation method has at its foundation an important theorem on the convergence of probability known as the central limit theorem (CLT), which goes as follows.\\
\emph{Central Limit Theorem}: Suppose ${X_{1},X_{2}, .., X_{n}}$ is a collection of random variables that are independent and identically distributed (i.i.d.) with E$[X_{i}]=\mu$ and Var$[X_{i}]=\sigma^{2}$, then in the limit $n$ approaches infinity, the random variable $\sqrt{n}(S_{n}-\mu)$ converge to a normal distribution $N(0,\sigma^{2})$, where $S_{n}=(X_{1}+X_{2}+..+X_{n})/n$. In other words, for large value of $n$, $S_{n}$ approximately follows a normal distribution $N(\mu,\sigma^{2}/n)$.
Simply put, the CLT theorem states that several important random variables and estimators are asymptotically normal.\\ 
The Delta theorem method allows approximation of the asymptotic behavior of functions over a random variable if the random variable itself is asymptotically normal~\cite{adasgupta:book}.
The mathematical statement of delta theorem is given as:\\ 
{(Delta theorem)} Let $T_{n}$ be a sequence of statistics such that
\begin{equation}
T_{n}\,\, \xrightarrow{\text{\it{d}}} \,\, N(\theta,\sigma^{2}(\theta)/n), \, \, \sigma(\theta) > 0. \\ 
\end{equation}
Let $g$ be a real function which at least is differentiable at $\theta$ with $g^{'}(\theta)\neq 0$. Then, \\
\begin{equation}
g(T_{n})\, \xrightarrow{\text{\it{d}}} \,\, N(g(\theta),[g^{'}(\theta)]^{2}\sigma^{2}(\theta)/n).
\label{delta_univariate_label}
\end{equation}
Here, the $\xrightarrow{\text{\it{d}}}$ sign represents convergence over the distribution. It is worth mentioning here that the estimation of error using the Delta theorem involves the parameters of the population ($e.g.$ $\sigma$ and $\mu$ in the CLT definition), which is inaccessible, as one only has a fixed sample to work with. Hence, the parameters of the population are estimated using the sample itself.
Delta theorem can be easily extended to multivariate case. Then,
\begin{equation}
\boldsymbol{g(X)} \,\, \xrightarrow{\text{\it{d}}} \,\, N(\boldsymbol{g(\theta)},\boldsymbol{D\Sigma D^{'}}/n)
\label{delta_mulvariate_label}
\end{equation}
where, $\boldsymbol{X}$ = $\{X_{n}\}$ is sequence of random vectors which are normally distributed; $\boldsymbol{X} \,\, \xrightarrow{\text{\it{d}}} \,\, N(\boldsymbol{\theta},\boldsymbol{\Sigma}/n)$, $\boldsymbol{\theta}$ is a constant vector (mean vector), $\boldsymbol{\Sigma}$ is the covariance matrix, and $\boldsymbol{D}$ is the Jacobian of $\boldsymbol{g}$. For estimating error on the cumulants, we will make use of one more theorem employed in convergence of a sequence of sample moments, which actually comes as a consequence of multivariate central limit theorem.\\
If central moments $\mu_{2k} =E[(X - \mu)^{2k}] $ are finite, then the random vector
\begin{equation}
(\hat{\mu_2} - \mu_2,..\hat{\mu_k} - \mu_k) \,\, \xrightarrow{\text{\it{d}}} \,\, N((0,0,0,....,0),\boldsymbol{\Sigma}/n),
\label{convergence_moments_label}
\end{equation}
where, $\boldsymbol{\Sigma}$ is a $(k-1)\times(k-1)$ covariance matrix, with elements,\\
\begin{equation}
\boldsymbol{\Sigma_{i,j}} = \mu_{i+j} - \mu_{i}\mu_{j} -i\mu_{i-1}\mu_{j+1} - j\mu_{i+1}\mu_{j-1}+ij\mu_{i-1}\mu_{j-1}\mu_{2}.
\label{error_formula_label}
\end{equation}
Using the above theorem and putting $i=j=m$ ($m>=2$) in the eqn.~\ref{error_formula_label}, the variances of $m^{th}$ order central moments can be obtained. Using the CLT and the above theorem, error on the cumulants up to the third order are obtain as follows:
\begin{eqnarray}
Var(C_1) &=& \mu_2/n\,\nonumber \\
Var(C_2) &=& (\mu_4-\mu_2^2)/n\, \nonumber \\
Var(C_3) &=& (\mu_6-\mu_3^2+9\mu_2^3-6\mu_2\mu_4)/n.\ 
\label{eqn_err_till3rd}
\end{eqnarray}
To avoid the use of square root symbol repetitively, expressions for variances of the cumulants are presented in eqn.~\ref{eqn_err_till3rd}. The statistical uncertainties/errors on the cumulants will simply be the square roots of the variances. Obtaining error on the fourth order cumulant will require the use of Delta theorem method given in eqn.~\ref{delta_mulvariate_label}. Taking $g(\mu_{2},\mu_{4})=C_4=\mu_{4}-3\mu_{2}^2$, 
\begin{equation}
\boldsymbol{D\Sigma D^{'}} = 
({\begin{array}{cc} \partial g/\partial \mu_{2} &  \partial g/\partial \mu_{4} \end{array}})
(\begin{array}{cc} \Sigma_{11} & \Sigma_{12}\\ \Sigma_{21} & \Sigma_{22} \end{array})
(\begin{array}{c} \partial g/\partial \mu_{2} \\ \partial g/\partial \mu_{4} \end{array}).
\end{equation}
Since the central moments $\mu_{2}$ and $\mu_{4}$ are involved in the transformation, the elements of the covariance matrix are $\Sigma_{11}=Var(\mu_{2})$, $\Sigma_{12}=\Sigma_{21}=Cov(\mu_{2},\mu_{4})$, and $\Sigma_{22}=Var(\mu_{4})$. The variances and covariances of the central moments can be calculated using eqn.~\ref{error_formula_label}. Then,
\begin{equation}
\boldsymbol{D\Sigma D^{'}} = 
\begin{pmatrix} -6 \mu_{2} & 1 \end{pmatrix} 
\begin{pmatrix} \mu_{4}-\mu^2_{2} & \mu_{6}-4\mu^2_{3}-\mu_{4}\mu_{2}\\ \mu_{6}-4\mu^2_{3}-\mu_{4}\mu_{2} & \mu_{8}-\mu^2_{4}-8\mu_{3}\mu_{5}+16\mu_{2}\mu^2_{3} \end{pmatrix}
\begin{pmatrix} -6 \mu_{2} \\ 1 \end{pmatrix}.
\label{main_calc_label}
\end{equation}
Using the above equation, the variance on $C_4$ is as follows:
\begin{equation}
Var(C_4)=\boldsymbol{D\Sigma D^{'}}/n=(\mu_8-12\mu_6\mu_2-8\mu_5\mu_3-\mu_4^2+48\mu_4\mu_2^2+64\mu_3^2\mu_2-36\mu_2^4)/n.
\label{fourth_momen_label}
\end{equation} 
With increase in the order of cumulant, the formula for uncertainties on the cumulant becomes more tedious. Involving the detection efficiency while calculating the statistical uncertainties complexifies the formulae further~\cite{Luo:2014rea}. It is seen that larger the width of a distribution and/or smaller the detection efficiency, the larger is the statistical uncertainties on the cumulants. As an alternative to the complicated error formulae, especially for the higher-order cumulants, it is quite helpful sometimes to use re-sampling methods over the analytical delta theorem method for estimation of statistical uncertainties on cumulants.\\

{(Bootstrap Method)} Bootstrap method is a re-sampling method to estimate statistical error and thus does not involve the complexities of the standard error propagation method. It uses Monte-Carlo algorithm to estimate the statistical error on a statistic by creating many bootstrap samples, where each bootstrap sample comprises of elements randomly drawn from the original sample with replacement. The statistic is calculated for each of these bootstrap samples and the sampling variance of the statistic from these bootstrap samples is the error on it~\cite{bootstrap,bootstrap1,Pandav:2018bdx}.\\ 
\begin{figure}[!htb]
	\centering 
	\includegraphics[scale=0.65]{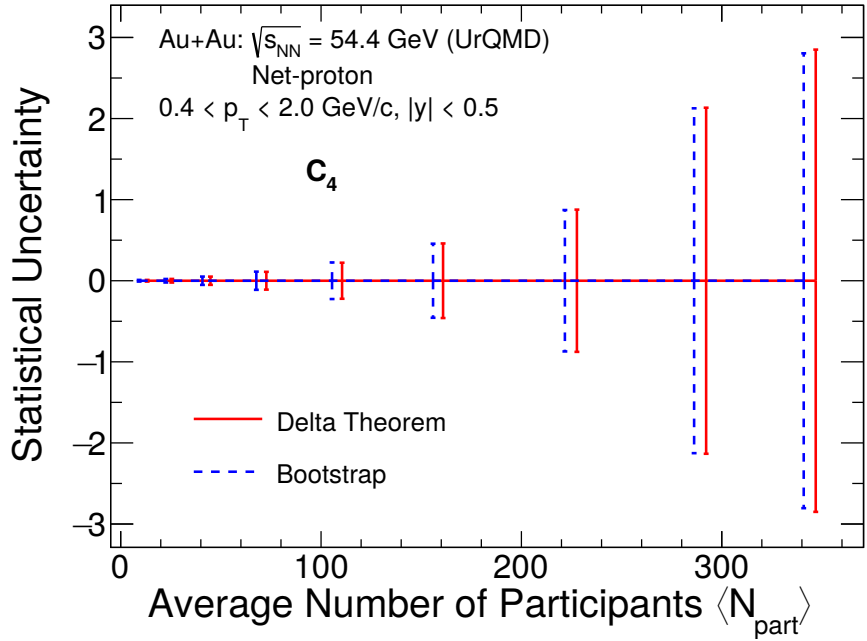}
	\caption{Statistical uncertainty on net-proton $C_{4}$ for Au+Au collisions at $\sqrt{s_{NN}}$ = 54.4 GeV as a function of  $\langle N_{part} \rangle$ (average number of participant nucleons) from the UrQMD model. The solid red and dashed blue lines are uncertainties from Delta theorem method and bootstrap method, respectively.}
	\label{err_cmp} 
\end{figure}

To start with, let $X$ be a random sample with $n$ data points representing the experimental data set randomly drawn from an unknown parent distribution. Let the estimator be denoted by $\hat{e}$, which could be any statistic such as mean or variance, whose standard error we intend to find. The prescription to estimate standard error using the bootstrap method is as follows: \\
1. Construct $B$ number of independent bootstrap samples $X^{*}_{1}$, $X^{*}_{2}$, $X^{*}_{3}$, ..., $X^{*}_{B}$, each consisting of $n$ data points randomly drawn from the random sample X with replacement. \\
2. Evaluate the estimator in each of these bootstrap samples,
\begin{equation}
\hat{e}^{*}_{b}=\hat{e}(X^{*}_{b})      \qquad b=1,2,3, ..., B.
\end{equation}
3. The sampling variance of the estimator is given as follows.
\begin{equation}
Var(\hat{e}) = \frac{1}{B-1}\sum_{b=1}^{B}\Big(\bar{\hat{e}} - \hat{e}^{*}_{b}\Big)^{2},
\end{equation}
where $\bar{\hat{e}}^{*} = \frac{1}{B}\sum_{b=1}^{B}(\hat{e}^{*}_{b})$.\\ 
The sufficient enough value of $B$ for an accurate estimation of error within the bootstrap method varies from case to case depending upon the initial sample size. However, in general, the larger value of $B$ estimates the error better. 
Using a Monte-Carlo procedure, both the methods were subjected to verification of Gaussian nature of the statistical error and found to satisfy the criteria~\cite{Pandav:2018bdx}. Comparison of error estimated by both the methods on fourth order net-proton cumulant $C_{4}$ for Au+Au collisions at $\sqrt{s_{NN}}$ = 54.4 GeV from the UrQMD model as a function of $\langle N_{part} \rangle$ are shown in Fig.~\ref{err_cmp}. The errors estimated from both the methods show good agreement with each other.

\subsubsection{Particle identification}
\begin{figure}[!htb]
	\centering 
	\includegraphics[scale=0.75]{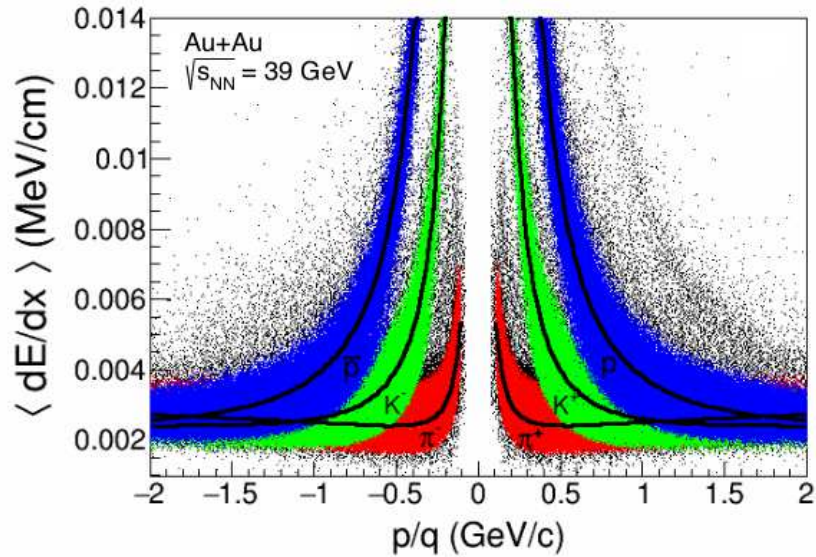}
	\caption{$\langle(dE/dx)\rangle$ of charged tracks measured by TPC as a function of $p/q$ (momentum/charge) in Au+Au collision at $\sqrt{s_{NN}}$ = 39 GeV in STAR~\cite{Adamczyk:2017iwn}. The various bands represents different charged particles, namely the protons ($p$), anti-protons ($\bar p$), kaons ($K^{+}, K^{-}$) and pions ($\pi^+, \pi^-$). The curved lines in black are the Bichsel expectation for various charged particles~\cite{Bichsel:2006cs}.}
	\label{pid_tpc} 
\end{figure}

\begin{figure}[!htb]
	\centering 
	\includegraphics[scale=0.4]{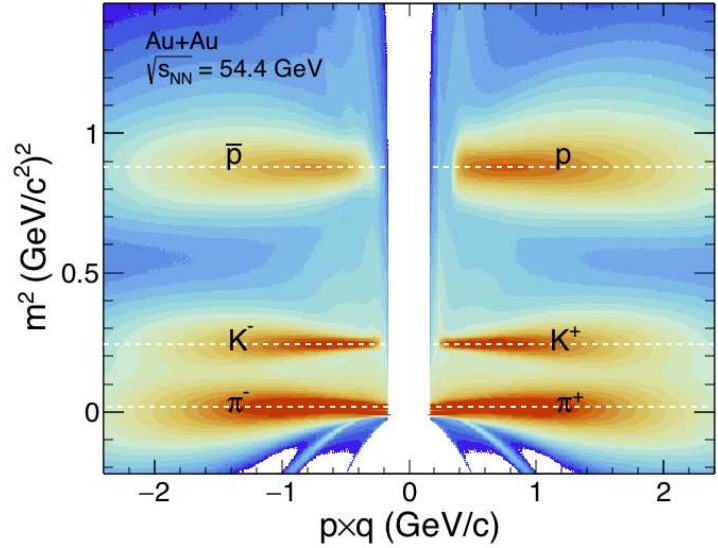}
	\caption{Square of the mass of the charged particles measured using TOF as a function of the product of the momentum ($p$) and the charge ($q$) in Au+Au collisions at $\sqrt{s_{NN}}$ = 54.4 GeV in STAR. The bands corresponds to various charged particles, namely the protons ($p$), anti-protons ($\bar p$), kaons ($K^{+}, K^{-}$) and pions ($\pi^+, \pi^-$).}
	\label{pid_tof} 
\end{figure}

The results presented in this review dominantly cover the measurements from the STAR experiment, hence we discuss the particle identification effects using it as a typical example.
The STAR detector at RHIC using two sub-detectors Time Projection Chamber (TPC) and Time of Flight (TOF) ensures the high purity of the identified particles~\cite{Ackermann:2002ad}. TPC measures the momentum, trajectory and ionization energy loss per unit length ($dE/dx$) of charged particles. Only the particles with transverse momentum greater than 150 MeV/c are recorded. The $dE/dx$ measurement is used for particle identification (PID) by TPC. Figure~\ref{pid_tpc} shows average $dE/dx$ as a function of $p/q$ (momentum/charge) in Au+Au collision at $\sqrt{s_{NN}}$ = 39 GeV at RHIC-STAR~\cite{Adamczyk:2017iwn}. A variable called the $n\sigma_{X}$ is used for selection of identified particles, which can be constructed from the $dE/dx$ information, as shown below,
\begin{eqnarray}
n\sigma_{X} = \frac{\ln[(dE/dx)_{measured}/(dE/dx)_{theory}]}{\sigma_{X}}
\end{eqnarray}
where, $(dE/dx)_{measured}$ is the ionisation energy loss of a charged particle measured by TPC and $(dE/dx)_{theory}$ is the expectation from Bichsel formula~\cite{Bichsel:2006cs} for the charged particle. $\sigma_{X}$ is the $dE/dx$ resolution of the TPC. The particle identification with $|n\sigma_{X}| < 2$ is found to give a good level of purity ($>$ 90\% for kaons and $>$ 98\% for protons~\cite{starNK,starNP}) of the identified charged particle in the momentum range where only TPC is used for fluctuation measurements.
For the lower momentum region, the bands corresponding to various charged particles are well separated while at the larger momentum, the bands can be seen to overlap with each other and thus the purity deteriorates. To identify charged particles in the higher momentum region, the TOF detector is used. The TOF detector measures flight time ($t$) of charged tracks from primary vertex (interaction point where collision between two heavy-ion happens) to the detector. The square of mass of charged particle which requires flight time information from TOF along with information on momentum ($p$) and charge ($q$) from TPC, is used for particle identification. The mass squared ($m^2$) as a function of $p\times q$ for Au+Au collisions recorded by STAR detector at $\sqrt{s_{NN}}$ = 54.4 GeV is shown in Fig.~\ref{pid_tof}. If $L$ is the path length, $v$ is the velocity of the charged particles, then $\beta$ can be written as $\beta$ = $v/c$ = $L/ct$. The $m^2$ can then be written as,
\begin{eqnarray}
m^2c^2 = p^2(\frac{1}{\beta^2} - 1) = p^2(\frac{c^2t^2}{L^2} - 1).
\end{eqnarray} 
The above equation suggests for a given momentum ($p$), charged particles of larger mass ($m$) will lead to larger flight time ($t$) and vice-versa. The $m^2$ measurement by TOF allows for PID with high degree of purity ($>$ 90\% for kaons and $>$ 97\% for protons~\cite{starNK,starNP21}) in the higher momentum region.

\subsubsection{Linking measurements to conserved quantities}
As discussed in the previous section, only charged particles could be measured using the TPC detector. Measuring net-baryon number fluctuations poses a challenge as neutrons, which are also produced in large numbers in experiments and carry a unit baryon number, are not identified. Interestingly, it was found that net-proton number fluctuations could reliably reflect the baryon number susceptibility near the critical point~\cite{Hatta:2003wn}. Theoretical works to link net-proton cumulants to those of net-baryon have been presented in Ref.~\cite{Kitazawa:2011wh,Kitazawa:2012at}. This conversion of cumulants from those of net-proton to net-baryon relies on the fact that isospin of nucleons in the final state are randomized and uncorrelated~\cite{Kitazawa:2011wh}. Correlations between the isospins of nucleons in the final state are found to be negligible over a wide range of collision energy. In central heavy-ion collisions at large energies, the isospin density becomes negligibly small as most of the initial isospin density is absorbed by pions which are abundantly produced. Then the conversion of cumulants of net-proton ($C^{net}_{np}$) to those of net-baryon ($C^{net}_{nB}$) gets simplified and are given as follows~\cite{Kitazawa:2011wh,Kitazawa:2012at}:
\begin{eqnarray}
C^{net}_{1B} &=& 2C^{net}_{1p}\\
C^{net}_{2B} &=& 4C^{net}_{2p} -2C^{tot}_{1p}\\
C^{net}_{3B} &=& 8C^{net}_{3p} -12 C^{net, tot}_{1p,1p}+ 6C^{net}_{1p}\\
C^{net}_{4B} &=& 16C^{net}_{4p} - 48 C^{net, tot}_{2p,1p} +48 C^{net}_{2p}\nonumber\\ &+&12 C^{tot}_{2p} -26 C^{tot}_{1p},
\end{eqnarray}
where $C^{tot}_{np}$ is the $n^{th}$ order cumulant of proton+anti-proton distribution and $C^{net, tot}_{np,mp}$ are the mixed cumulants of net-proton and proton+anti-proton distribution of $(n+m)^{th}$ order. 

\subsubsection{Centrality determination}
\begin{figure}[!htb]
	\centering 
	\includegraphics[scale=0.8]{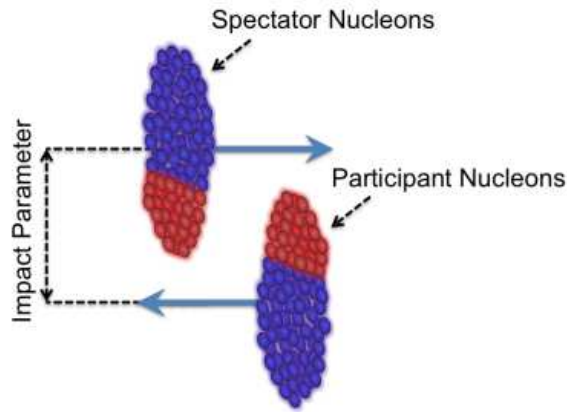}
	\caption{A schematic picture of the geometry of relativistic heavy-ion collision.}
	\label{sketch_coll} 
\end{figure}

In heavy-ion collisions, the impact parameter ($b$) of collisions is defined by the distance between the geometrical centers of the colliding nuclei in the plane transverse to their direction. A schematic diagram of the geometry of heavy-ion collisions is shown in Fig.~\ref{sketch_coll}. As the impact parameter is not measurable in experiments, the charged particle multiplicity of events can be used to define collision centrality as they reflect the initial geometry of the colliding nuclei. To obtain the geometrical quantities like the impact parameter, the number of participant nucleons ($N_{part}$) and the number of binary collisions ($N_{coll}$),  one has to rely on model calculations which involves mapping the charged particle multiplicity from experiment to that of a simulated one~\cite{centra_model1,centra_model2,Miller:2007ri}. The widely used model for determining these quantities is the Glauber Monte Carlo (Glauber MC) model~\cite{Miller:2007ri}. In Glauber model, nucleons in a nuclei are distributed following the Wood-Saxon density distribution. The nuclei are then assigned a random impact parameter and translated to collide. The distance between the centers of nucleons of each nucleus is calculated. If this distance is found to be less than $\sqrt{\sigma^{NN}_{inel}/\pi}$, where $\sigma^{NN}_{inel}$ is the inelastic nucleon-nucleon cross section, the nucleons have undergone a binary collision. Nucleons that have undergone at least one binary collision are said to have participated in the reaction. This procedure is followed to obtain $N_{part}$, $N_{coll}$ event-by-event. 
A class of events belonging to a given centrality (say 0-5\%) is obtained by the collection of events representing the fraction of the total cross-section. Glauber model can only provide the information on $N_{part}$, $N_{coll}$  and impact parameter $b$, but for mapping with the experimental data, one has to use a combination of the Glauber model and a particle production model called the Two-Component Model ~\cite{Kharzeev:2000ph}. Assuming that particle production is governed by contribution from hard component (proportional to $N_{coll}$) and soft component (proportional to $N_{part}$), it simulates the multiplicity distribution according to the eqn~\ref{eqn_twoco}.
\begin{figure}[!htb]
	\centering 
	\includegraphics[scale=0.99]{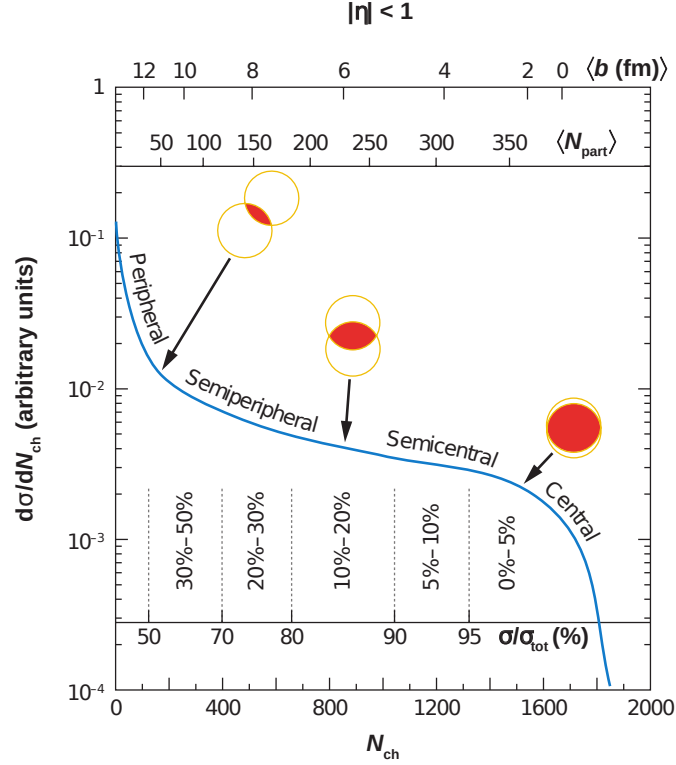}
	\caption{An illustration of the total inclusive charged particles multiplicity $N_{ch}$ and number of participant ($N_{part}$) and impact parameter ($b$) in heavy-ion collisions~\cite{Miller:2007ri}. Figure taken from Ref.~\cite{Miller:2007ri}.}
	\label{glau_coll} 
\end{figure}

\begin{eqnarray}
\frac{dN_{ch}}{d\eta}=n_{pp}\big[xN_{coll}+(1-x)\frac{N_{part}}{2}],
\label{eqn_twoco}
\end{eqnarray}
where the quantity $n_{pp}$ is the multiplicity from a $p+p$ collision of the same center-of-mass energy. $x$  is the contribution of hard component. $n_{pp}$ is drawn from a negative binomial distribution with parameters NBD ($n_{pp}$;$\langle n_{pp} \rangle $,k), where $\langle n_{pp} \rangle $ is the average multiplicity in the $p+p$ collision and $k$ is the width of the distribution. The simulated charged multiplicity distribution is then fitted with the corresponding multiplicity distribution from the experiment. $\langle n_{pp} \rangle$, $k$, and $x$ are kept as free parameters in the fitting to experimentally measured charged particle multiplicity distribution. Initial guess values for $x$ could be set close to those determined in different experiments~\cite{PHOBOS:2004hlv,ALICE:2013hur} for faster convergence of fitting. Figure~\ref{glau_coll} illustrates collision centrality definition in heavy-ion collisions by comparing the charge particle multiplicities with Glauber MC + Two-component model simulation. Centrality classes are classified for the simulated distribution. $\langle N_{part} \rangle $, $\langle N_{coll} \rangle $ in those centrality classes are calculated. 

In addition to using the number of charged particles to define centrality, one can also use detectors different from the ones used for particle identification, such as the Zero Degree Calorimeters (ZDC) and the Event Plane Detector (EPD). These detectors measure the energy deposition in the forward rapidity region and thus could help eliminate any possible correlations between the cumulants measured at mid-rapidity and the centrality.
\subsubsection{Volume fluctuations}
The impact parameter of the collisions could change event-by-event. The fluctuation of the impact parameter and thus the participant nucleons gives rise to fluctuations in the initial system volume created. This effect is called the volume fluctuations (VF), which could result in additional non-dynamical fluctuations in the experimental measurements. Studies on the effect of VF on net-particle cumulants suggest that there is an artificial enhancement of the cumulants due to VF. The methods to correct the cumulants for this effect within the limitation of experimentally defined collision centrality will be discussed in this sub-section.
\paragraph{{Centrality Bin Width Correction (CBWC)}:}
Any given centrality, say $0-5\%$, corresponds to a range of impact parameters. Selection of narrow centrality bins helps to get rid of inherent fluctuations within each centrality bin. For the charged particle multiplicity distribution, the smallest bin-width can be of unity-width. The correction to cumulants of net-particle distributions associated with accounting for the finite centrality bin width is called the centrality bin width (CBW) correction~\cite{Luo:2013bmi}. For a given centrality that spans over a range of charged particle multiplicity, cumulants in each multiplicity bin are evaluated.  A weighted average of cumulants is made with the events in each multiplicity bin as weights to quote the CBW corrected value of the cumulant for that centrality.
The CBW corrected $n^{th}$ order cumulant in a certain centrality is given as\\
\begin{eqnarray}
C_n=\sum_r\omega_r C_{n,r}\\
\omega_r=\frac{n_r}{\sum_rn_r}=\frac{n_r}{N},
\end{eqnarray}
where $\sum_r$, sums over all multiplicity bins in that centrality. $n_r$ is the  number of events in the $r^{th}$ multiplicity bin.
$N$ is the total number of events in the  centrality considered.
\\
The error on the CBW corrected cumulants is evaluated using standard error propagation method to be:
\begin{eqnarray}
\sigma_{C_n}=\sqrt{\sum_r {\omega_r}^2({ \sigma_{C_{n,r}}})^2},
\end{eqnarray}
where $\sigma_{C_{n,r}}$ is the statistical error on $n^{th}$ cumulant in $r^{th}$ multiplicity bin.\\
\paragraph{{Volume Fluctuation Correction (VFC)}:}
\label{sectn_volfluct_label}
Volume fluctuation correction~\cite{Braun-Munzinger:2016yjz,Skokov_volfluc} is another method to suppress volume fluctuations effects on cumulant measurement. The correction procedure assumes each $N_{part}$ value to be an independent source of particle production. Net-particle numbers ($\Delta N$) for a given centrality class are constructed as sum of net-particles ($\Delta n$) from each $N_{part}$ source. As the $N_{part}$ itself fluctuates in a given centrality, the cumulants of net-particle distributions get contribution due to the $N_{part}$ fluctuation. The VFC method requires the estimation of cumulants of $N_{part}$ fluctuation in addition to those of net-particle fluctuations. As the $N_{part}$ is inaccessible in experiments, particle production models which have information on geometrical quantities are used in VFC. The correction prescribed by VFC method for cumulants up to the $4^{th}$ order is as follows~\cite{Braun-Munzinger:2016yjz,Skokov_volfluc}:
\begin{eqnarray}
C_1 (\Delta N) &=& \langle N_{part} \rangle C_1 (\Delta n)\\
C_2 (\Delta N) &=& \langle N_{part} \rangle C_2 (\Delta n) + \langle \Delta n \rangle^2 C_2 (N_{part}) \\
C_3 (\Delta N) &=& \langle N_{part} \rangle C_3 (\Delta n) + 3\langle \Delta n \rangle C_2 (\Delta n) C_2 (N_{part}) \nonumber\\&+& \langle \Delta n \rangle^3 C_3 (N_{part}) \\
C_4 (\Delta N) &=& \langle N_{part} \rangle C_4 (\Delta n) + 4\langle \Delta n \rangle C_3 (\Delta n) C_2 (N_{part}) \nonumber\\&+& 3 C_2^2 (\Delta n)C_2 (N_{part})+ 6\langle \Delta n \rangle^2 C_2 (\Delta n) C_3 (N_{part}) \nonumber\\&+& \langle \Delta n \rangle^4 C_4 (N_{part}).
\label{eqn_VFC}
\end{eqnarray}
The above equations suggests that starting from second order cumulants of net-particle distributions, the $N_{part}$ fluctuation encoded in their higher order cumulants ($C_n (N_p) $, $n\geq2$) also contributes. In the limit $ \langle \Delta n \rangle \xrightarrow {} 0 $, which is realized at very large collision energies, the $N_{part}$ fluctuation contributions up to the third order net-particle cumulants vanish, but there could be sizable contributions for the fourth order cumulant. 
\\\\
\begin{figure}[!htb]
	\centering 
	\includegraphics[scale=0.7]{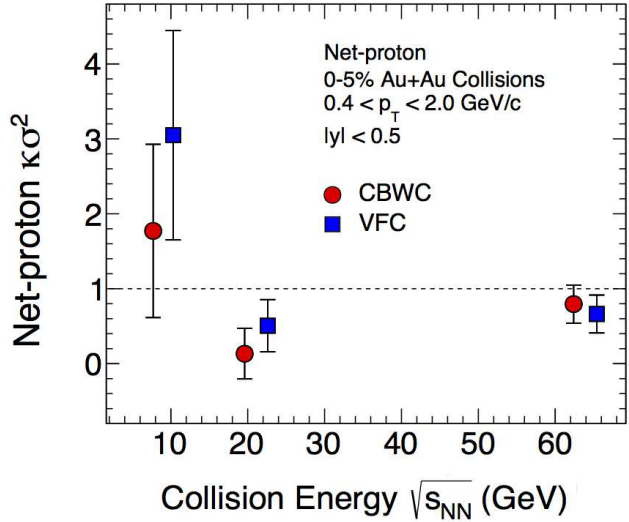}
	\caption{Net-proton $C_{4}/C_{2}$ or $\kappa\sigma^2$ in most central 0-5\% Au+Au collisions at RHIC at $\sqrt{s_{NN}}$ = 7.7, 19.6, 62.4 GeV~\cite{starNP21_long}. Results from CBWC (red circles) and VFC (blue squares) method to correct for volume fluctuations are presented. }
	\label{fig_CBWC_VFC} 
\end{figure}
Comparison of these two methods for the  net-proton $C_{4}/C_{2}$ at three RHIC energies, 7.7, 19.6 and 62.4 GeV for 0-5\% centrality  are shown in Fig~\ref{fig_CBWC_VFC}. The correction factors for the VFC method was obtained from Glauber+Two component model~\cite{Miller:2007ri}. The results of CBWC and VFC are found to be consistent with each other within uncertainties.
While the CBWC  is data-driven correction method, the VFC is model-dependent. While performing CBWC, one needs to note that even in the smallest centrality bin limit of unit multiplicity, participant fluctuation could still persist, although arguably small. The VFC, on the other hand, requires the assumption of independent particle production from sources. Model studies by authors of ~\cite{Sugiura:2019toh} suggest that the VFC method does not work well with the UrQMD model, which is attributed to the breakdown of the assumption of independent particle production in the UrQMD. 

\subsubsection{Centrality resolution and self-correlation effects}
\label{aucorrel_study}
\begin{figure}[!htb]
	\centering 
	\includegraphics[scale=0.8]{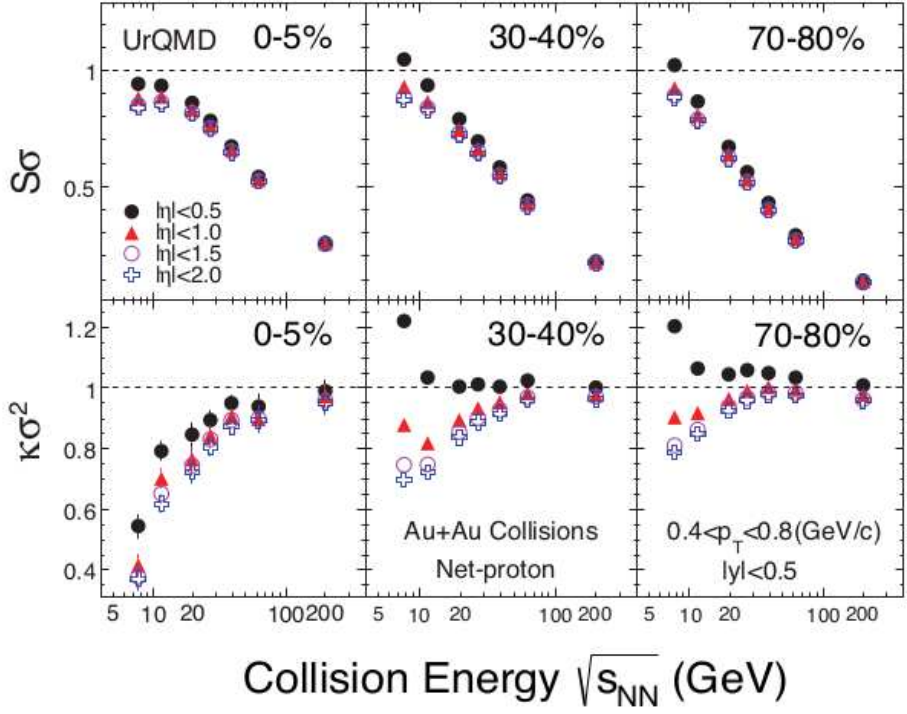}
	\caption{Net-proton $S\sigma$ and  $\kappa\sigma^2$ as a function of collision energy in Au+Au collisions from UrQMD model for 0-5\%, 30-40\% and 70-80\% centralities~\cite{Luo:2013bmi}. Results using CBWC method with different centrality definition are presented.}
	\label{fig_resolu_CBW} 
\end{figure}

Experimentally, while defining centrality using charged-particle multiplicity, one needs to carefully study two important aspects: centrality resolution and self-correlation effect. For the CBWC method to sufficiently suppress the volume fluctuations, the resolution of the centrality estimator should be high. Increasing the charged particle number in centrality definition (for example, by increasing the pseudo-rapidity ($\eta$) acceptance) improves the centrality resolution. Figure ~\ref{fig_resolu_CBW} presents a UrQMD model study on the dependence of net-proton $S\sigma$ and  $\kappa\sigma^2$ on different centrality definitions obtained using varying $\eta$ acceptance~\cite{Luo:2013bmi}. With a smaller $\eta$ acceptance used for centrality definition, an artificial rise is found for the two moment products at all the energies, which is suppressed when larger $\eta$ acceptance are used to define centrality.  However, that could lead to self-correlation effects if the charged particle acceptance used for centrality definition overlaps with the acceptance taken for net-particle selection for fluctuation measurements. Self-correlation effect arises when the charged particles which are selected for fluctuation measurements are included in the centrality definition. To avoid such artificial correlations, either separate  $\eta$ regions with no overlap are considered for centrality definition, and net-particle selection or charged particles of interest are removed from centrality definition in case of any overlap in acceptance. For example, for net-proton cumulant measurement at mid-rapidity ($|y| < 0.5$), all charged particles within pseudo-rapidity coverage $|\eta| < $ 1.0 avoiding protons and anti-protons are selected for centrality determination in the STAR experiment.

\begin{figure}[!htb]
	\centering 
	\includegraphics[scale=0.9]{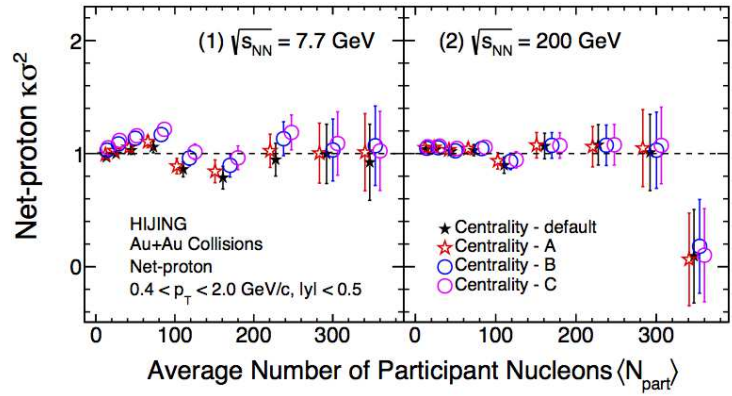}
	\caption{Net-proton $C_{4}/C_{2}$ in Au+Au collisions at $\sqrt{s_{NN}}$ = 7.7 (1) and 200 GeV (2) as a function of average number of participant nucleons from HIJING model. The results from centrality definitions: (a) Centrality - default: centrality definition as used in the STAR net-proton measurement, (b) Centrality - A: same as (a) with pions and kaons from weak decays removed if protons for such decay are selected in net-proton measurement, (c) Centrality - B: same as (a) with pions from $\Delta$ decays are removed, (d) Centrality - C: same as (a) except pions and kaons from both weak and $\Delta$ decays are removed.}
	\label{fig_auco} 
\end{figure} 
Self-correlation effect could also arise due to decay of resonances where one or more of the decay daughters falls into analysis acceptance,  and the other daughters are taken in centrality definition. Weak decays of lambda ($\Lambda$), cascade  ($\Xi$) and omega ($\Omega$) baryons could decay into protons (anti-protons) and charged pions or kaons. 
Self-correlation could arise if protons from such decays are selected for analysis and pions and kaons are included in centrality selection. There are also other resonances $\Delta^{++}$ and $\Delta^{0}$ which could contribute to the effect. Although keeping a tight cut on a track's distance of closest approach (DCA) to the primary vertex suppresses the background from decays and resonances in particle identification to a good extent, but it cannot be fully eliminated. For example, the MC simulation studies done by the STAR experiment with HIJING~\cite{Wang:1991hta} as input indicate that about 81\% of the lambda-decay protons were reconstructed with DCA $<$ 3 cm and 65\% with DCA $<$ 1 cm~\cite{STAR:2001rbj}. As the fluctuation measurements at STAR consider a criterion of DCA $<$ 1 cm for particle identification, the study of self-correlation arising out of such decays becomes important. A HIJING model based study to understand $\Delta$ and weak decays of strange baryon induced self-correlation effects on net-proton cumulant ratio $C_{4}/C_{2}$ is shown in Fig.~\ref{fig_auco}. The $C_{4}/C_{2}$ at $\sqrt{s_{NN}}$ = 7.7 and 200 GeV from various centrality definitions aimed at removing such correlations are found to be consistent with each other,  suggesting that the self-correlation effects from resonance decay are negligible. For the discussed fluctuation measurement results in this review, the effect of pseudo-rapidity range for centrality determination has been understood and optimized using model calculations in order to maximize resolution and suppress self-correlations~\cite{starNP21}.

\subsubsection{Pileup rejection}
The TPC detectors typically have a drift time in the orders of tens of $\mu s$. If this time duration is longer than the bunch crossing rate of the colliding ions in heavy-ion experiments, then there are possibilities that all the charged tracks from an event are not collected, and the subsequent events are triggered. In such scenarios, where successive events are reconstructed as a single event, there is a piling up of charged particles, and the event is called a pileup event. If such events are sufficiently large in number, they can alter the event-by-event net-particle distributions and consequently their cumulants. Detectors that have a faster detection time are used to remove pileup events, for example, the TOF detector at STAR experiment and the START detector at HADES experiment. The correlation between the charged particle multiplicity registered at the TOF and the TPC is examined to remove the background coming from pileup effect. It utilizes the fact that the TOF is a much faster detector than the TPC with a timing resolution of 100ps and thus sees less track fraction from previous events. 

The high interaction rates in the fixed target (FXT) configuration of experiments result in large pileup effects compared to experiments run in collider mode. Removing pileup events then becomes a challenging task. The excellent timing properties of the START detector at HADES help reduce the contamination of identified proton yields by pileup to a probability of $\leq 3 \times 10^{-5}$ in the proton number fluctuation studies in Au+Au collisions collected in the FXT mode at $\sqrt{s_{NN}}$ = 2.4 GeV~\cite{HADES:2020wpc}. The recent measurements of proton cumulants in Au+Au collisions at $\sqrt{s_{NN}}$ = 3 GeV (FXT) by the STAR experiment~\cite{STAR:2021fge}, applied the correction of cumulants for pileup effect suggested in Ref.\cite{Nonaka:2020qcm,Zhang:2021rmu}. The correction deals with two distributions: one is the multiplicity distribution of produced particle ($m$) that is considered for centrality definition,  and the other is the distribution of the particle of interest ($N$) whose fluctuations are to be calculated. The observed particle number distributions $P_m(N)$ at some value of multiplicity $m$ is believed to have a contribution from true or single collision events $P^{t}_m(N)$ and pileup events $P^{pu}_m(N)$. Mathematically, $P_m(N)$ can be written as:
\begin{eqnarray}
P_m(N) = (1-\alpha_m) P^{t}_m(N) + \alpha_m P^{pu}_m(N).
\label{eqn_twoco_pileup}
\end{eqnarray}
where $\alpha_m$ is the contribution of pileup events to the observed particle number at multiplicity $m$. The correction procedure assumes pileup events as the superposition of two independent single-collision events as suggested by the eqn.~\ref{eqn_twoevent_pileup}~\cite{Nonaka:2020qcm}.
\begin{eqnarray}
P^{pu}_m(N) = \sum_{i,j} \delta_{m,i+j} w_{i,j} P^{sub}_{i,j} (N).
\label{eqn_twoevent_pileup}
\end{eqnarray}
Here, $P^{sub}_{i,j}$ is the probability distribution of $N$ in a pileup event which is constituted by two single collision events having multiplicity $i$ and $j$. The combined multiplicity of the two events constituting the pileup, $i.e.$, $i+j$ must be equal to $m$. $w_{i,j}$, here, refers to the probability of observing a pileup event at multiplicity $m$, which is constituted by two single events with multiplicity $i$ and $j$.

The true cumulants in each multiplicity bin (cumulants of $P^{t}_m(N)$), then, can be obtained recursively from lower multiplicity bins applying the correction formulae prescribed in Ref.~\cite{Nonaka:2020qcm}. There are two parameters involved in the correction: $\alpha_m$ and $w_{i,j}$, which need to be determined from simulations. The correction's accuracy depends on how well the simulation mimics the real experimental conditions. Studies by the authors of Ref.~\cite{Zhang:2021rmu} suggest that the parameters involved in the correction formula can be obtained by a model-independent unfolding approach. The whole formalism of the correction can be extended to situations where pileup can also be caused due to more than two single collision events, although the probability of such pileup events will be very small.

\subsubsection{Efficiency correction}
\paragraph{{Binomial efficiency correction}:}
The detection of particles in a detector can be modeled assuming Binomial sampling $B_{\epsilon,N}(n)$ where $\epsilon$ (probability of success) is the efficiency parameter, and $N$ (number of trials) is the total number of charged particles incident on the detector. The assumption of Binomial sampling has been tested in several works from the STAR experiment and is shown to be a good working assumption for event-by-event detection of charged particles~\cite{starNK,starNP21_long,star_C6paper}. The true multiplicity distribution ($P(N)$) gets modified due to detector efficiency. The measured multiplicity distribution is a convolution of true multiplicity distribution and distribution for detector response, which is assumed to be binomial.\\
Individual particles are detected with a probability $\epsilon$, which is assumed to be independent for different particles. Denoting the number of detected particles as $n$ and the distribution of measured multiplicity as $\tilde{P}(n)$, the relation between the measured and true distribution could be given as,
\begin{equation}
\tilde{P}(n) = \sum_{N=n}^{\infty} P(N)B_{\epsilon,N}(n).
\label{eqn_bino}
\end{equation}
As the experiments are interested in fluctuations of net-particle ($n=n_{net}=n_1-n_2$), the bivariate case of the above relation are used to obtain formulae for efficiency correction. Then, using the factorial cumulants generating function, the cumulants of the true net-particle distribution can be obtained using the cumulants of the measured distributions as follows~\cite{Bzdak:2012ab}:
\begin{eqnarray}
C_{1}&=&c_{1}/\epsilon \ ,\\
C_{2}&=&\big[c_{2}-n_{o}(1-\epsilon)\big]/\epsilon^{2}  \ ,\\
C_{3}&=&\big[c_{3}-c_{1}(1-\epsilon^{2})\nonumber\\&-&3(1-\epsilon)(f_{20}-f_{02}-n_{o}c_{1})\big]/\epsilon^{3}  \ ,\\
C_{4}&=&\big[(c_{4}-n_{o}\epsilon^2(1-\epsilon)-3n^2_{o}(1-\epsilon)^2\nonumber\\ &-& 6\epsilon(1-\epsilon)(f_{20}+f_{02})\nonumber\\&+&12c_{1}(1-\epsilon)(f_{20}-f_{02})\nonumber\\
&-&(1-\epsilon^2)(c_{2}-3c_{1}^2)\nonumber\\
&-&6n_{o}(1-\epsilon)(c_{1}^2-c_{2})\nonumber\\
&-&6(1-\epsilon)(f_{03}-f_{12}+f_{02}+f_{20}\nonumber\\
&-&f_{21}+f_{30})\big]/\epsilon^{4}  \ .
\end{eqnarray}
here, $n_{o} = <n_1+n_2>$. $C_{n}$ and $c_{n}$ are the cumulants of true and measured net-particle multiplicity distributions, respectively. $f_{i,j}$ are the factorial cumulants of the measured multiplicity distributions.
This modeling can be extended to multivariate case where there could be different detection efficiency for different particle species and momentum ranges as is the case in most of the experimental measurements. The correction of cumulants for detector efficiency then gets more involved. A detailed discussion on the efficiency correction for many efficiency bins can be found in Ref.~\cite{Luo:2014rea,Nonaka:2017kko,Luo:2018ofd}. 
The complications and computing time involved in efficiency correction with several efficiency bins reduces as compared to the approach in Ref.~\cite{Luo:2014rea}, if cumulants and mixed cumulants are obtained for an observable $q_{(r,s)}$ defined as,
\begin{equation}
q_{(r,s)} = \sum_{i=1}^{M} (a^r_i/\epsilon^s_i)n_i
\label{eqn_qq}
\end{equation}
where, $n_i$ is the number of particles in $i^{th}$ efficiency bin and  $a_i$, $\epsilon_i$ represents the  quantum charge number, efficiency of  particles in $i^{th}$ bin, respectively.  $M$ is the total number of efficiency bins. The efficiency corrected cumulants in terms of the cumulants and mixed cumulant of the above observable are presented in Ref.~\cite{Nonaka:2017kko}. Here, in the eqn.~\ref{eqn_qq},  $\epsilon_i$ is the average efficiency of particle in a bin which requires information of the particle spectrum for that bin. A simplified procedure called the track-by-track efficiency correction eliminates this aspect of correction~\cite{Luo:2018ofd}. If $M\rightarrow \infty$ in eqn.~\ref{eqn_qq}, $i.e.$, there are infinite number of efficiency bins and $q_{(r,s)}$ can be expressed as follows.
\begin{eqnarray}
q_{(r,s)} &=& \sum_{i=1}^{\infty} (a^r_i/\epsilon^s_i)n_i \nonumber\\
&=& \frac{a^r_1}{\epsilon^s_1}n_1 + \frac{a^r_2}{\epsilon^s_2}n_2  +...+ \frac{a^r_i}{\epsilon^s_i}n_i +..
\label{eqn_qq_track}
\end{eqnarray}
With $M\rightarrow \infty$, each efficiency bin could now contain up to one particle. For $n_i$ = 0, there is no contribution to $q_{(r,s)}$ and one can rewrite eqn.~\ref{eqn_qq_track} as,
\begin{equation}
q_{(r,s)} = \sum_{j=1}^{n_{tot}} (a^r_j/\epsilon^s_j)
\label{eqn_qq_track_e2}
\end{equation}
where $n_{tot}$ is the total number of particles in one event. As the efficiency correction using eqn.~\ref{eqn_qq_track_e2} only demands track-by-track efficiency, analytical formula of efficiency with respect to a track variable like $p_T$ could be used. The particle yield information within each efficiency bin is no longer required.
\paragraph{{Unfolding approach}:}
The analytical formula for efficiency correction hinges on the fact that the detector response is binomial. In very high-energy head on collisions, a large number of charged particles are produced. Such dense charged particle environment may result in non-binomial effects in detector response. In the presence of non-binomial effects, the detection probability may not be independent for each particle which could result in unwanted correlation in the measurements. The correction of cumulants in such situations could be done by a method called unfolding~\cite{unfold_1,unfold_2,unfold_3}.

Let the true particle distribution $P(N)$ be linked to the measured particle distribution $\tilde{P}(n)$ by the equation,
\begin{equation}
\tilde{P}(n) = \sum_{N=n}^{\infty} T(N,n) P(N) ,
\label{unfold_eqn}
\end{equation}
which is identical to eqn.~\ref{eqn_bino} except the detector response given by $T(N,n)$ is no longer assumed to be binomial. What unfolding does is to invert the detector response matrix and apply it to the distribution at the measured level to obtain the true net-particle distribution. The equation below depicts the procedure.
\begin{equation}
P(N) = \sum_{n=0}^{\infty} T^{-1}(n,N) \tilde{P}(n) 
\label{unfold_eqn_rev}
\end{equation}
Unfolding method, in general, demands rigorous computation and resources. It has been applied by the STAR collaboration as a cross-check to get back cumulants of true distributions for central Au+Au collisions at $\sqrt{s_{NN}}$ = 200 GeV~\cite{starNP21}. It was found to give values consistent with the cumulants from the binomial efficiency correction procedure within uncertainties. 
\subsubsection{Systematic uncertainties}
Systematic uncertainties are broadly classified under four categories (a) those arising due to the choice of criteria for selection of charged particles, (b) due to specific particle identification method, (c) those associated with the estimation of efficiency of detectors, and (d) due to use of specific event selection criteria. Typically they are estimated by varying the following requirements for charged particle tracks: DCA, track quality reflected by the number of fit points used in track reconstruction algorithm in the TPC, number of fit points associated with the $dE/dx$ measurements, $dE/dx$ and $m^2$ for charged particle identification~\cite{starNC,starNK,starNP,starNP21_long}. A ± 5\% systematic uncertainty associated with determining the efficiency is also considered~\cite{Adamczyk:2017iwn}. All of the different sources of systematic uncertainty are added in quadrature to obtain the final systematic uncertainties on the cumulants and their ratios. \\
The typical value of statistical  and systematic uncertainties on net-proton cumulant ratios $C_{4}/C_{2}$ in most central 0-5\% Au+Au collisions at $\sqrt{s_{NN}}$ = 39 GeV is 20\% and 18\%, respectively~\cite{starNP}.
\subsection{Experimental results }
In this section, the most recent results on the net-particle cumulant measurements from the Au+Au collisions
recorded by the STAR detector~\cite{Ackermann:2002ad} at RHIC during the years 2010 to 2017 are discussed. Au+Au collisions at nine collision energies, $i.e.$, $\sqrt{s_{NN}}$ = 7.7, 11.5, 14.5, 19.6, 27, 39, 54.4, 62.4, 200 GeV were analysed as part of the phase-I of the Beam Energy Scan (BES) program at RHIC~\cite{Adamczyk:2017iwn,Abelev:2009bw}. Table ~\ref{tab1_stats} shows the total event statistics for the nine collision energies along with corresponding $T$ and $\mu_{B}$ at chemical freeze-out.  Time-Projection-Chamber (TPC) and Time-of-Flight (TOF) detectors are used to select charged particles within $p_{T}$ range 0.2 -- 2.0 GeV/c. The measurement are carried out at mid-rapidity ($|y| < $0.5). The collision centrality is determined from the charged particle multiplicity excluding the particles of interest to avoid the self-correlation effect~\cite{Chatterjee:2019fey}. In order to suppress the volume fluctuation effects, centrality bin width correction is applied to the measurement of the cumulants~\cite{Luo:2013bmi}. Cumulants are corrected for acceptance and detection inefficiencies, with the assumption that the distribution of the detector response is binomial~\cite{Luo:2014rea,Nonaka:2017kko}. For estimation of statistical uncertainties of cumulants and their ratios, both Delta theorem method and  Bootstrap are used \cite{Luo:2011tp,Pandav:2018bdx}. Systematic uncertainties of the $C_{n}$'s are estimated varying the tracking efficiency, track selection, and particle identification criteria. In the following subsections, the collision energy dependence of cumulants and cumulant ratios and comparison to various theoretical and model calculations are discussed.

\begin{table}
	\caption{Total event statistics in Au+Au collisions for various $\sqrt{s_{NN}}$ from BES-I program at RHIC and the corresponding $T_{ch}$ and $\mu_{B}$ at chemical freeze-out~\cite{Adamczyk:2017iwn}. }
	\centering   
	\begin{tabular}{|c|c|c|c|}
		\hline	
		$\sqrt{s_{NN}}$ (GeV) &  Events (millions) &$T_{ch}$ (MeV) &  $\mu_{B}$ (MeV)  \\
		\hline 
		7.7 & 3.0 & 144.3 & 398\\
		\hline 
		11.5  & 6.6 & 149.4& 287\\
		\hline 
		14.5 & 20 & 151.6  &264 \\
		\hline 
		19.6 & 15 & 153.9 & 188 \\
		\hline 
		27 & 30 &  155.0 & 144 \\
		\hline 
		39 & 86 & 156.4& 104\\
		\hline 
		54.4 & 550 & 160.0& 83\\
		\hline 
		62.4 & 47 & 160.3 & 70\\
		\hline 
		200 & 238 & 164.3 & 28\\
		\hline 
	\end{tabular}
	\label{tab1_stats}
\end{table}

\subsubsection{ Uncorrected net-proton multiplicity distributions}
\begin{figure}[!htb]
	\centering 
	\includegraphics[scale=0.7]{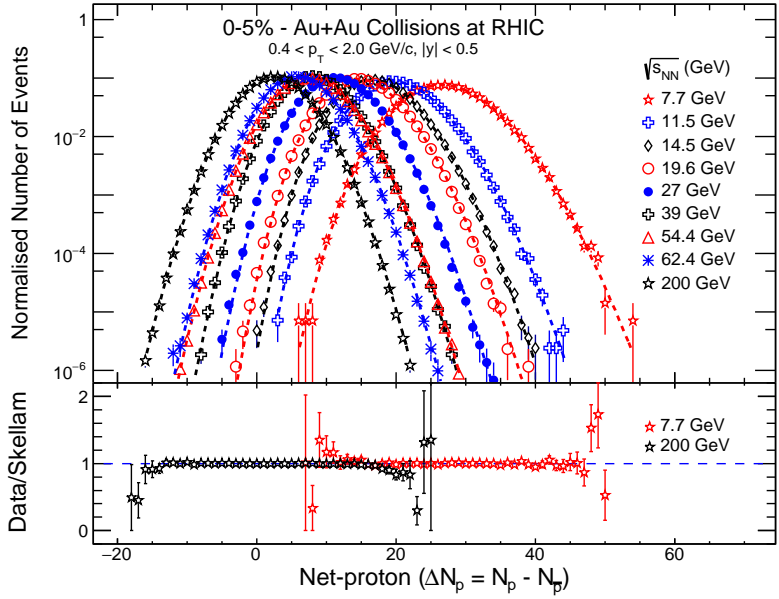}
	\caption{Event-by-event distribution of net-proton multiplicity in top 5\% Au+Au collisions at $\sqrt{s_{NN}}$ = 7.7 - 200 GeV from the BES-I program at RHIC. The distributions are uncorrected for finite centrality width and detector efficiency. The dashed lines are the Skellam fit to the net-proton distributions. Also shown are the ratio of data to Skellam expectations in the lower panel for $\sqrt{s_{NN}}$ = 7.7 and 200 GeV.}
	\label{skelldist_5cen} 
\end{figure}

\begin{figure}[!htb]
	\centering 
	\includegraphics[scale=0.7]{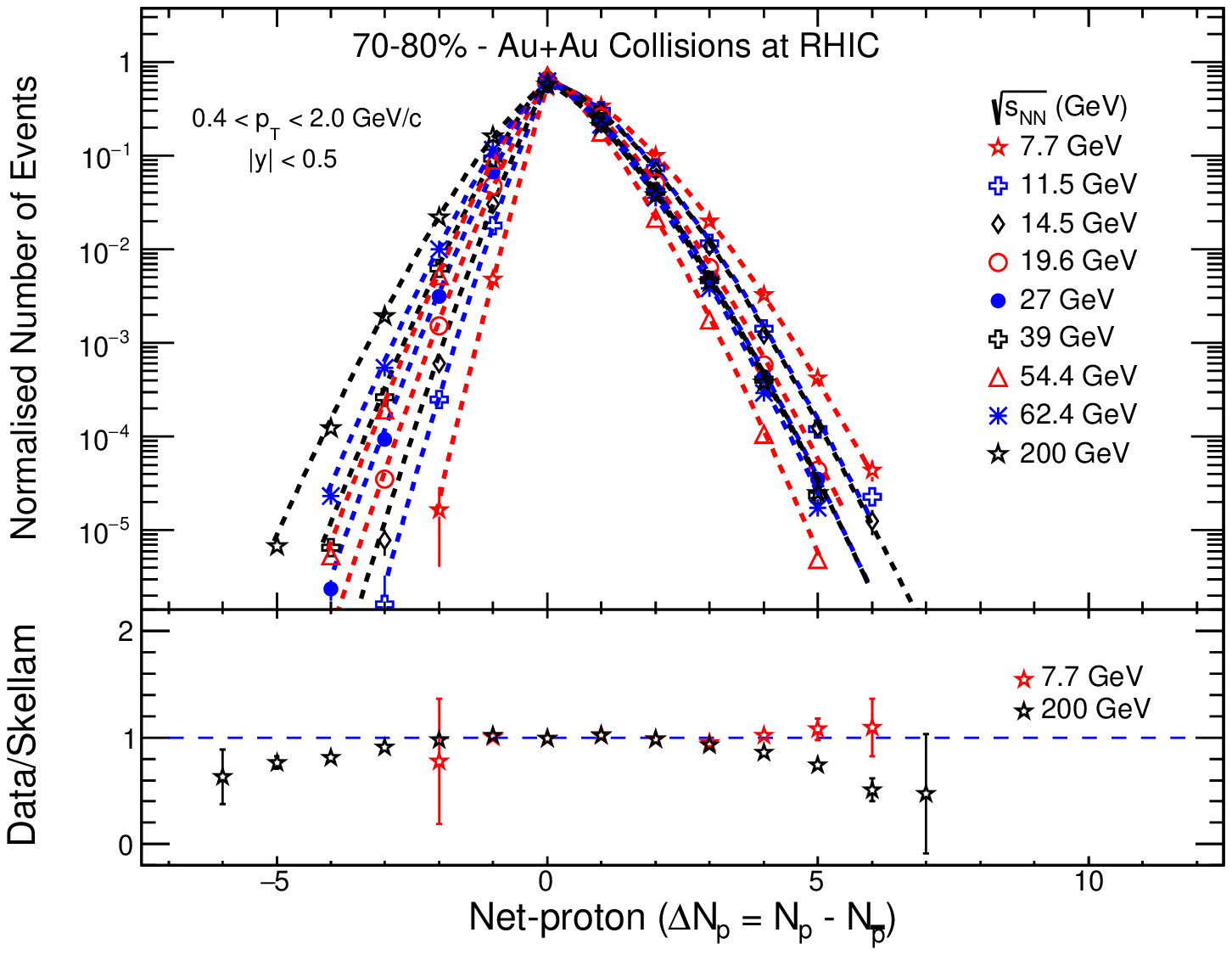}
	\caption{Event-by-event distribution of net-proton multiplicity in peripheral (70-80\%) Au+Au collisions at $\sqrt{s_{NN}}$ = 7.7 - 200 GeV from the BES-I program at RHIC. The distributions are uncorrected for finite centrality width and detector efficiency. The dashed lines are the Skellam fit to the net-proton distributions. Also shown are the ratio of data to Skellam expectations in the lower panel for $\sqrt{s_{NN}}$ = 7.7 and 200 GeV.}
	\label{skelldist_80cen} 
\end{figure}

From the identified protons and anti-protons, event-by-event net-proton distribution are constructed. Figure~\ref{skelldist_5cen} shows raw event-by-event net-proton multiplicity distributions in central (0-5\%) Au+Au collisions for the nine collision energies $\sqrt{s_{NN}}$ = 7.7 - 200 GeV. The distributions are normalized with respect to events in the given centrality and are fitted to Skellam distributions, shown in the dashed lines. They are uncorrected for finite efficiency and acceptance effects of the detector. The mean of the distribution increases with decreasing collision energy. This is understood to be the effect of baryon stopping at mid rapidity, which increases towards lower energy. The width of the net-proton distribution at $\sqrt{s_{NN}}$ = 7.7 GeV is largest as compared to all other energies. The larger width along with low event statistics in $\sqrt{s_{NN}}$ = 7.7 GeV results in larger statistical uncertainties on the cumulants of net-proton distributions as compared to other energies. The Skellam fit to the distributions can well describe the mean and variance of the distributions however if one focuses on the tails of the distributions, then one starts to see deviations of the Skellam fit with respect to the data. The lower panel of the Fig.~\ref{skelldist_5cen} shows the ratio of the data to the Skellam fit for $\sqrt{s_{NN}}$ = 7.7 GeV and 200 GeV where a clear deviation of data from the Skellam expectation can be seen towards the tail of the distribution. Since higher-order cumulants probe the subtleties of distributions, the deviation of the Skellam fit to data would mean the presence of non-Skellam fluctuation in higher-order cumulants. The normalized raw net-proton distributions for the peripheral (70-80\%) Au+Au collisions from the nine collision energies along with the Skellam fit are also shown in Fig.~\ref{skelldist_80cen}. The mean and width of the distribution show a rather small variation vs. collision energy as compared to central collisions. The deviation of the data from Skellam fit shown in terms of their ratios in the bottom panel of the Fig.~\ref{skelldist_80cen} is found to be smaller than that observed for 0-5\% centrality.

\subsubsection{ Collision energy dependence of net-proton cumulants}
\begin{figure}[!htb]
	\centering 
	\includegraphics[scale=0.6]{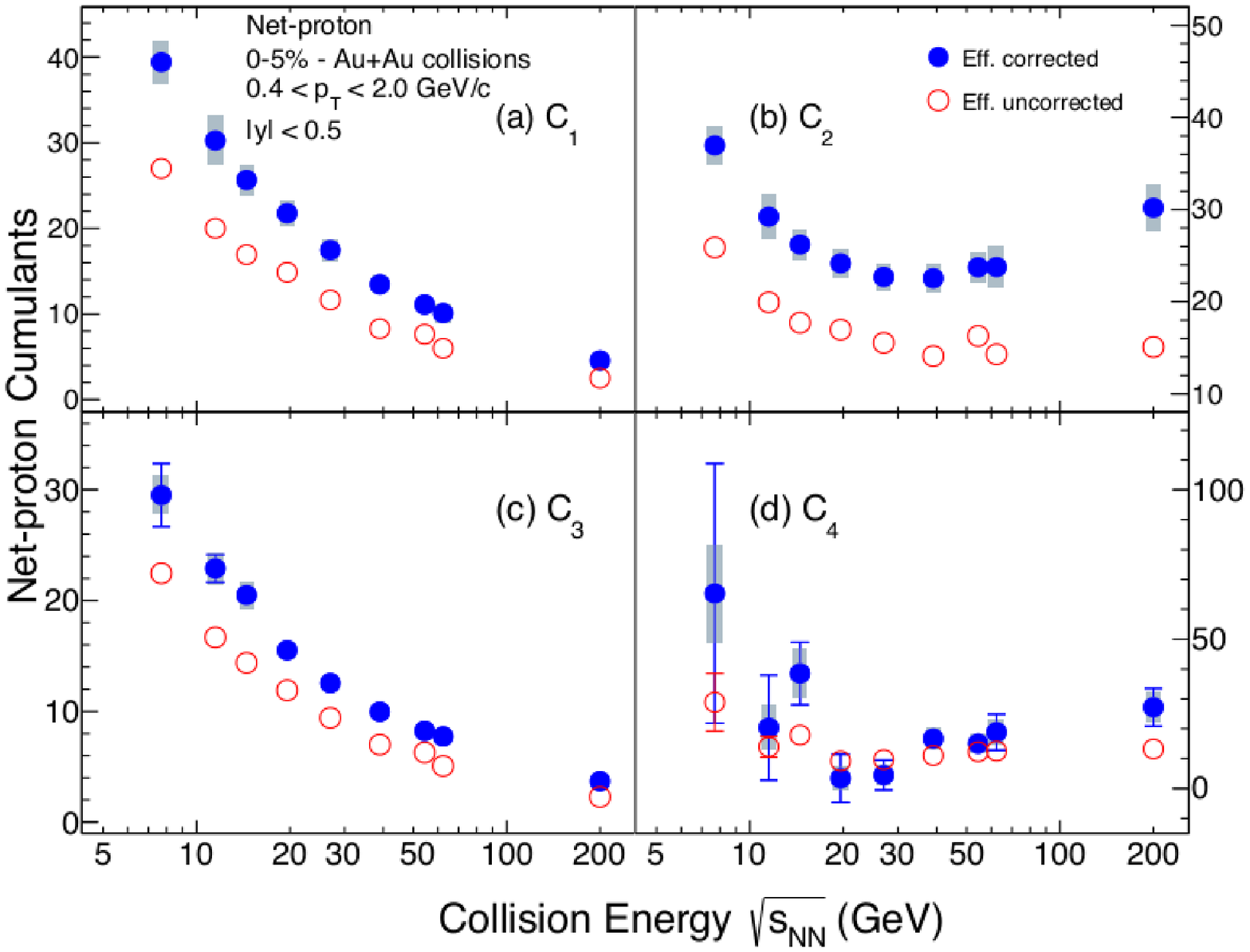}
	\caption{Cumulants of net-proton distributions in most central Au+Au collisions(0-5\%) up to the fourth-order as a function of collision energies from $\sqrt{s_{NN}}$ = 7.7 - 200 GeV from the BES-I program at RHIC. Results are shown for efficiency corrected and uncorrected cumulants as solid blue circles and open red circles, respectively. The bars and shaded band on the data points represent the statistical uncertainties and systematic uncertainties, respectively.}
	\label{cumu_5cen} 
\end{figure}

\begin{figure}[!htb]
	\centering 
	\includegraphics[scale=0.6]{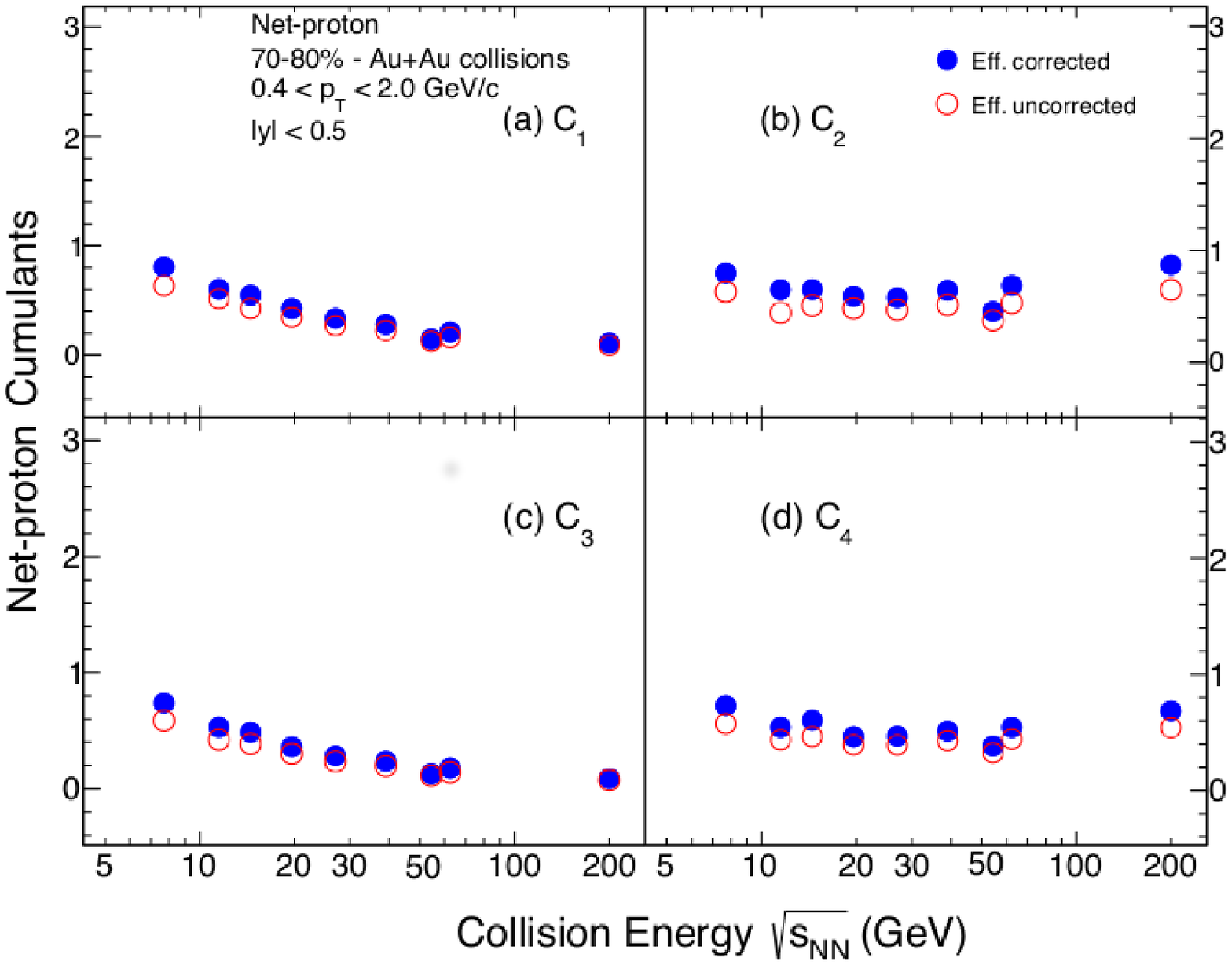}
	\caption{Cumulants of net-proton distributions in peripheral Au+Au collisions(0-80\%) up to the fourth order as a function of collision energies from $\sqrt{s_{NN}}$ = 7.7 - 200 GeV from the BES-I program at RHIC. Results are shown for efficiency corrected and uncorrected cumulants as solid blue circles and open red circles, respectively. The statistical uncertainties and systematic uncertainties are small and within the marker size.}
	\label{cumu_80cen} 
\end{figure}

Cumulants up to the $4^{th}$ order of the event-by-event net-proton multiplicity distributions for most central (0-5\%) Au+Au collisions as a function of collision energy ($\sqrt{s_{NN}}$) are shown in Fig.~\ref{cumu_5cen}.
The data points in blue are the measurements corrected for detector efficiency and acceptance effects while those in red, represent the efficiency uncorrected results. All the presented results are corrected for volume fluctuation effects by employing the CBWC procedure. The efficiency corrected cumulant $C_{1}$ and  $C_{3}$ shows monotonic decrease as a function of collision energy, whereas a non-monotonic collision energy dependence is observed for efficiency corrected $C_{2}$ and $C_{4}$. The trend of collision energy dependence shown by the efficiency corrected cumulants is already captured by the efficiency uncorrected cumulants shown in the open red marker. Upto third order of cumulant ($C_{n}, n\leq3$), the efficiency uncorrected cumulants remain consistently lower than efficiency corrected results at all energies. No such hierarchy is observed for $C_{4}$. The efficiency correction increases the magnitude of statistical uncertainties on the cumulants. Figure~\ref{cumu_80cen}  shows the collision energy dependence of efficiency corrected and uncorrected net-proton cumulants for the peripheral 70-80\% collisions. The efficiency corrected cumulants up to the fourth order is larger than their corresponding efficiency uncorrected counterparts across all the energies. The collision energy dependence trend of cumulants for peripheral 70-80\% collisions is weaker as compared to the 0-5\% central collisions. As seen from the peripheral measurements, the magnitude and collision energy dependence trend of odd order cumulants ($C_{1}$ and  $C_{3}$) are quite similar. The same observation can also be made for $C_{2}$ and  $C_{4}$. This is a characteristic of a Skellam distribution where all the odd order cumulants are the same and are simply the difference of the mean of constituent Poisson distribution, and all the even order cumulants are also the same and given by the sum of the means.

\subsubsection{ Collision energy dependence of net-proton cumulant ratio}
\begin{figure}[!htb]
	\centering 
	\includegraphics[scale=0.65]{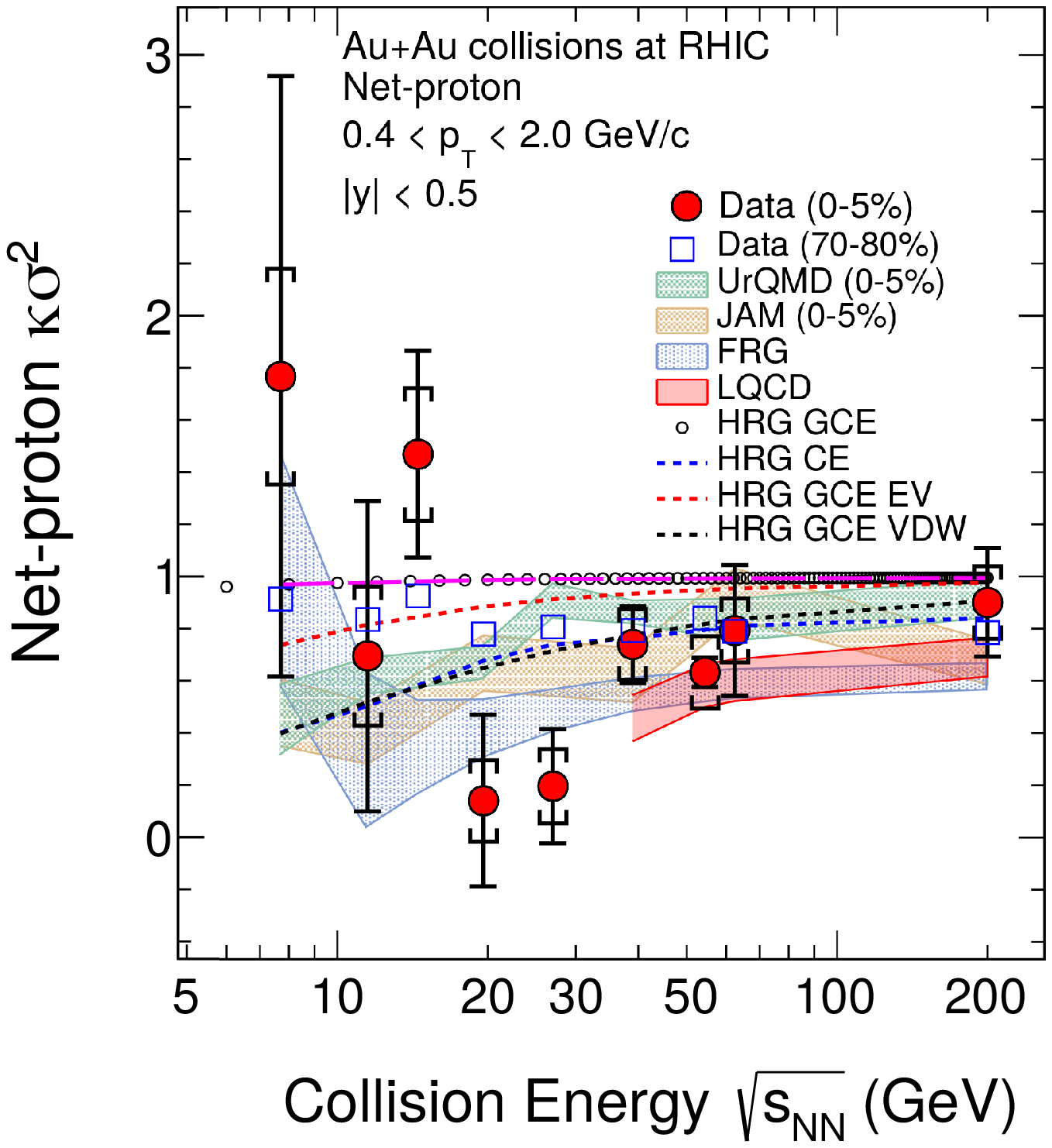}
	\caption{The collision energy dependence of ratio of fourth to second-order cumulant of net-proton distributions($\kappa\sigma^2$) in Au+Au collisions from $\sqrt{s_{NN}}$ = 7.7 - 200 GeV from the BES-I program at RHIC. The measurements are shown for 0-5\% and 0-80\% collision centrality class. The bars and brackets on the data points represent the statistical uncertainties and systematic uncertainties, respectively. UrQMD and JAM model expectation~\cite{Zhang:2019lqz} for 0-5\% central Au+Au collisions are given by the green and brown bands, respectively. Various variants of HRG model calculations are also presented. Lattice-QCD and FRG calculations are shown as red and blue bands, respectively~\cite{Fu:2021oaw,Bazavov:2020bjn}. }
	\label{ener_c42_model} 
\end{figure}

Recent measurements on net-proton cumulant ratios up to fourth -order have shown an intriguing collision energy dependence trend for most central  0-5\% Au+Au collisions. The collision energy dependence of higher-order cumulant ratio $C_{4}$/$C_{2}$, will be discussed in this subsection in particular. As discussed previously, to eliminate the trivial system volume dependence to allow a direct comparison to theoretical calculations, ratios of cumulants are constructed. Figure~\ref{ener_c42_model}, shows the collision energy dependence of cumulant ratio $C_{4}$/$C_{2}$ (equivalently moment product $\kappa\sigma^2$) of net-proton distributions in central 0-5\% and peripheral 70-80\% Au+Au collisions at RHIC. Various variants of HRG and UrQMD model calculations are also shown as baselines to the measurements. The STAR net-proton $\kappa\sigma^2$ measurements for the most central 0-5\% collisions show a non-monotonic collision energy dependence, which is neither reproduced by the transport model UrQMD nor the various variants of HRG model calculations presented. The ideal-HRG (HRG GCE in the figure), remains close to the Skellam baseline value at unity. The HRG calculations in the grand canonical framework taking only repulsive interactions into account are given by the red dotted line. The calculations remain below unity, and the deviation from unity increases with decreasing collision energy. With both repulsive and attractive interactions taken in HRG calculations, the $\kappa\sigma^2$ shows a larger deviation from unity towards lower values as seen from the black dotted line. The UrQMD calculation for the most central 0-5\% collisions suggests a monotonic decrease of $\kappa\sigma^2$ with decreasing collision energy. This suppression of cumulant ratio $C_{4}$/$C_{2}$ in the lower collision energies could be attributed to the effect of baryon number conservation and is also reproduced by HRG canonical ensemble calculations~\cite{Braun-Munzinger:2020jbk}. Towards the higher collision energies, measurements show agreement with lattice-QCD calculations in the energy range $\sqrt{s_{NN}}$ = 39 -- 200 GeV~\cite{Bazavov:2020bjn} subjected to caveats discussed in sub-section~\ref{rfer_obser_subsectn}. Interestingly, FRG calculation~\cite{Fu:2021oaw} shows a decreasing trend from higher to lower energy up to $\sqrt{s_{NN}}$ = 10 GeV, below which it shows an increasing trend, which the authors~\cite{Fu:2021oaw} attribute to the sharpening of crossover. $\kappa\sigma^2$ from peripheral collisions, on the other hand, shows a weak collision energy dependence and always remains smaller than the Skellam baseline at unity. The observation of non-monotonic collision energy dependence of $\kappa\sigma^2$  for most central 0-5\% collisions is qualitatively consistent with expectations from the linear sigma model calculations, which includes a QCD critical point. 

\begin{figure}[!htb]
	\centering 
	\includegraphics[scale=0.65]{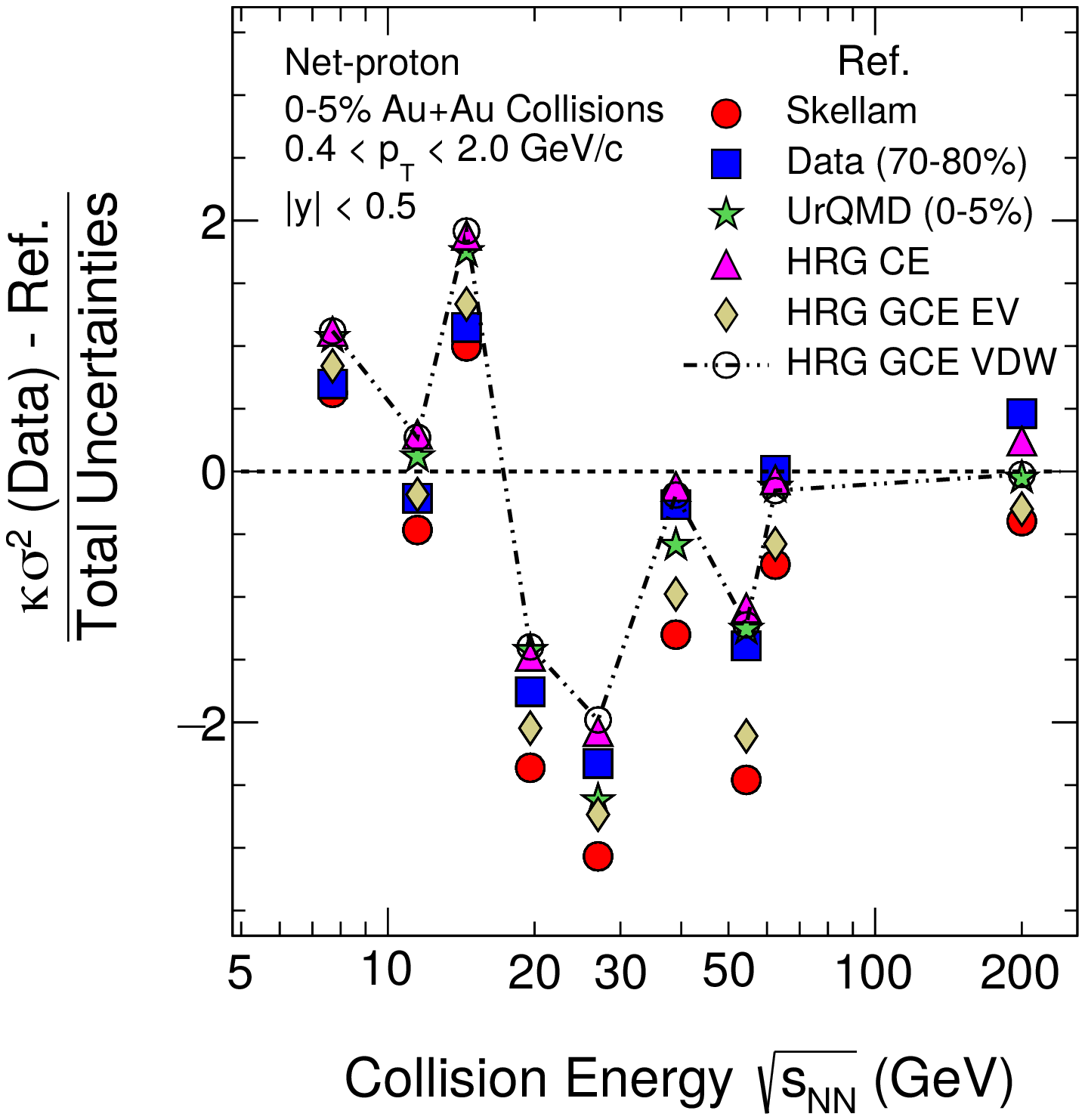}
	\caption{ Difference in $\kappa\sigma^2$ of net-proton distribution for central 0-5\% Au+Au collisions from those obtained from Skellam baseline calculation, UrQMD, HRG and 70-80\% peripheral Au+Au collisions. The difference is divided by the total uncertainties, where total uncertainties are obtained by adding the statistical uncertainties and systematic uncertainties in quadrature. The results are presented as a function of collision energy, in the range $\sqrt{s_{NN}}$ = 7.7 - 200 GeV.}
	\label{fig_deviate} 
\end{figure}

The deviation of the most central  $\kappa\sigma^2$ measurements from various baseline non-critical point calculations quantified in terms of the number of standard deviation ($\sigma$) are shown in Fig.~\ref{fig_deviate} as a function of collision energy. Various baselines, starting from the Skellam baseline at unity to UrQMD and HRG model calculations, are considered. Also shown is the case where the measurements at peripheral 70-80\% collisions serves as a baseline to measurements at central collisions. Irrespective of which baseline one chooses, the most central 0-5\% measurements show both positive and negative deviations from the baseline across collision energies at a level of 2-3$\sigma$. This suggests the robustness of the non-monotonic collision energy dependence with respect to various baseline calculations. This observation is further strengthened by performing a $\chi^2$ test to demonstrate the deviation of the experimental data to various model calculations in the collision energy range $\sqrt{s_{NN}}$ = 7.7 -- 27 GeV~\cite{starNP21,starNP21_long}. The difference between experimental data and model expectations is quantified in terms of $\chi^2$ and then converted to right-tailed $p$-value. The right-tailed $p$-value is obtained as: $p$-value $=1-cdf_{\chi^2,d}(\chi^2_{c})$, where $cdf_{\chi^2,d}$ denotes the cumulative distribution function of the $\chi^2$ distribution with $d$ degrees of freedom and $\chi^2_{c}$ is the calculated $\chi^2$ value. From the table of the probability distribution of $\chi^2$~\cite{Fisher::book}, right-tailed $p$-value can be obtained for a given number of degrees of freedom. A criterion on the right-tailed $p$-value: $p <$ 0.05 is commonly used to ascertain a significant deviation between the data and model expectations. The $p$-value from the $\chi^2$ test between data and various variants of HRG and UrQMD model calculations are tabulated in table~\ref{tab_p_value_calc}. The small $p$-values ($p < $ 0.05) obtained from the test demonstrate that the data significantly differ from model calculations.
\begin{table}[!htb]
	\caption{The right tail $p$ value from the $\chi^2$ test between experimental data on net-proton $C_4/C_2$ for 0-5\% centrality and various variants of HRG: Grand canonical ensemble (GCE), GCE with excluded volume (GCE EV) and canonical ensemble (CE) and UrQMD model calculations in the collision energy range $\sqrt{s_{NN}}$ = 7.7 -- 27 GeV~\cite{starNP21,starNP21_long}.}
	\centering 
	\begin{tabular}{|c|c|c|c|}
		\hline	
		HRG GCE  & HRG EV  & HRG CE &  UrQMD    \\
		\hline 
		0.00553 & 0.0145 & 0.045& 0.0221\\
		\hline 
	\end{tabular}
	\label{tab_p_value_calc}
\end{table}

\begin{figure}[!htb]
	\centering 
	\includegraphics[scale=0.7]{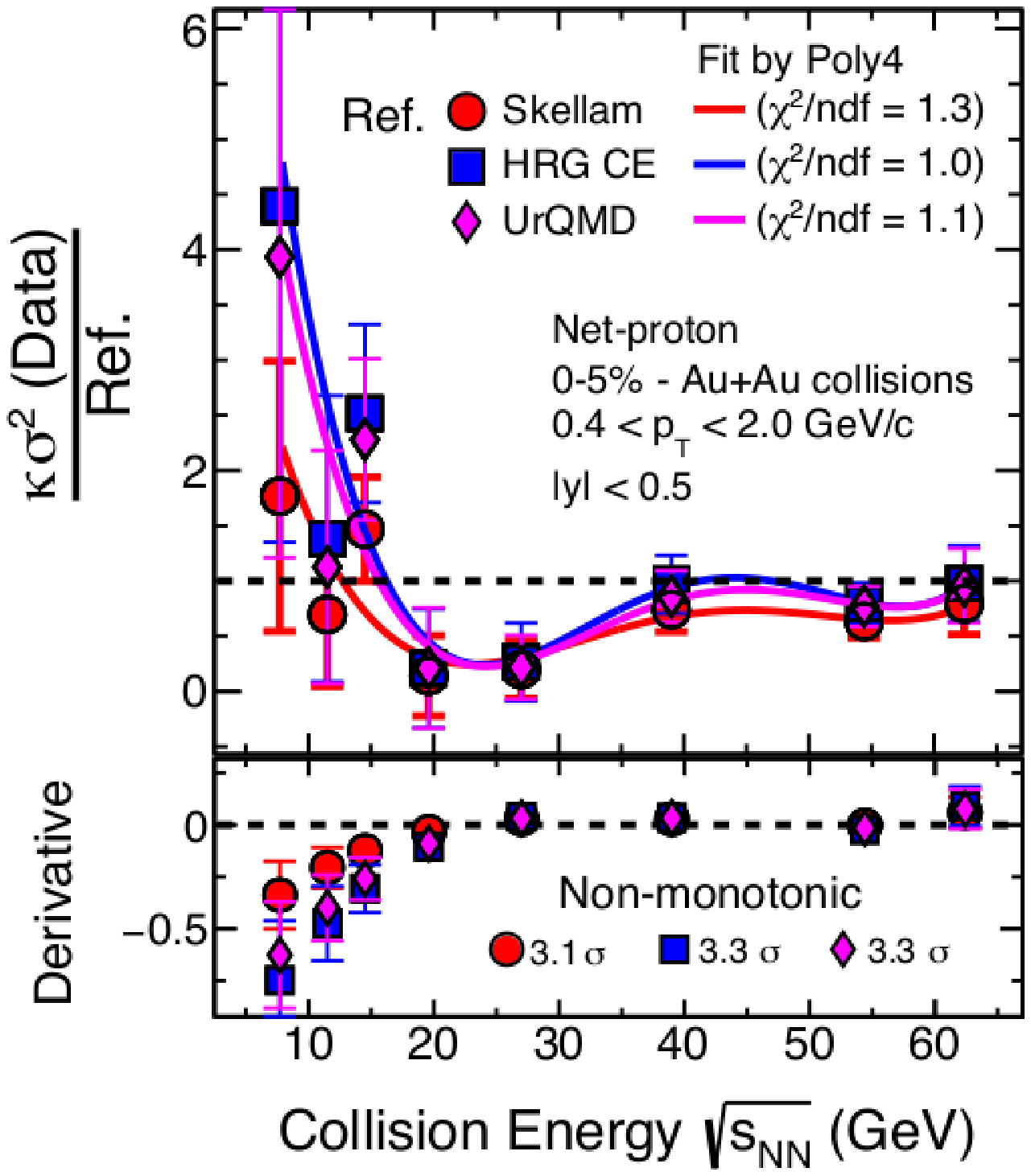}
	\caption{ Upper Panel:  Ratio of $\kappa\sigma^2$ of net-proton distributions for 0-5\% central Au+Au collisions with respect to Skellam baseline , HRG CE and UrQMD expectations from $\sqrt{s_{NN}}$ = 7.7 - 62.4 GeV. The bars on the ratios are statistical and systematic uncertainties added in quadrature. The red, magenta and blue lines are polynomial fit functions to the cumulant ratios. Also mentioned is the chi-square per degree of freedom for the respective fits. The black dashed lines are the Poisson baselines. Lower Panel: Derivative of the fitted polynomials as a function of collision energy for all three cases. The bars represent the total uncertainties on derivatives where total uncertainties are obtained by adding statistical and systematic uncertainties in quadrature.}
	\label{fig_deri_signi} 
\end{figure}

The observed non-monotonicity has been quantified by using a statistical procedure. The most central 0-5\%  measurements on net-proton $\kappa\sigma^2$ as a function of collision energy, in the collision energy range $\sqrt{s_{NN}}$ = 7.7 - 62.4 GeV, is best fitted by a polynomial of degree four with a $\chi^2$ per degree of freedom at 1.3. The baseline here corresponds to the Skellam expectation at unity. The fit along with the derivatives of the fitted polynomial at various collision energies are shown in the upper and lower panels of Fig.~\ref{fig_deri_signi}, respectively. The derivative of the polynomial fit changes sign across the collision energy range $\sqrt{s_{NN}}$ = 7.7 - 62.4 GeV, thus demonstrating the non-monotonic collision energy dependence of measurements. The statistical and systematic uncertainties on derivatives are obtained by randomly varying the data points at each energy within their respective statistical and systematic uncertainties and are added in quadrature when shown in the bottom panel. The significance of the observed non-monotonic dependence of $\kappa\sigma^{2}$ on collision energy, in the energy range $\sqrt{s_{NN}}$ = 7.7 - 62.4 GeV is obtained to be 3.1$\sigma$ based on the fourth order polynomial fitting procedure~\cite{starNP21}. If the baseline is changed to UrQMD and HRG canonical ensemble calculation, the fourth order polynomial fit to data results in $\chi^2$ per degree of freedom $\sim$1.0. The sign change of derivative for the UrQMD and HRG CE baselines are shown in the bottom panels, and using the statistical procedure as used in Ref.~\cite{starNP21}, the significance of non-monotonic dependence of measurements with respect to both these baselines are found to be $3.3\sigma$.

\begin{figure}[!htb]
	\centering 
	\hspace*{-4cm}    
	\includegraphics[scale=0.5]{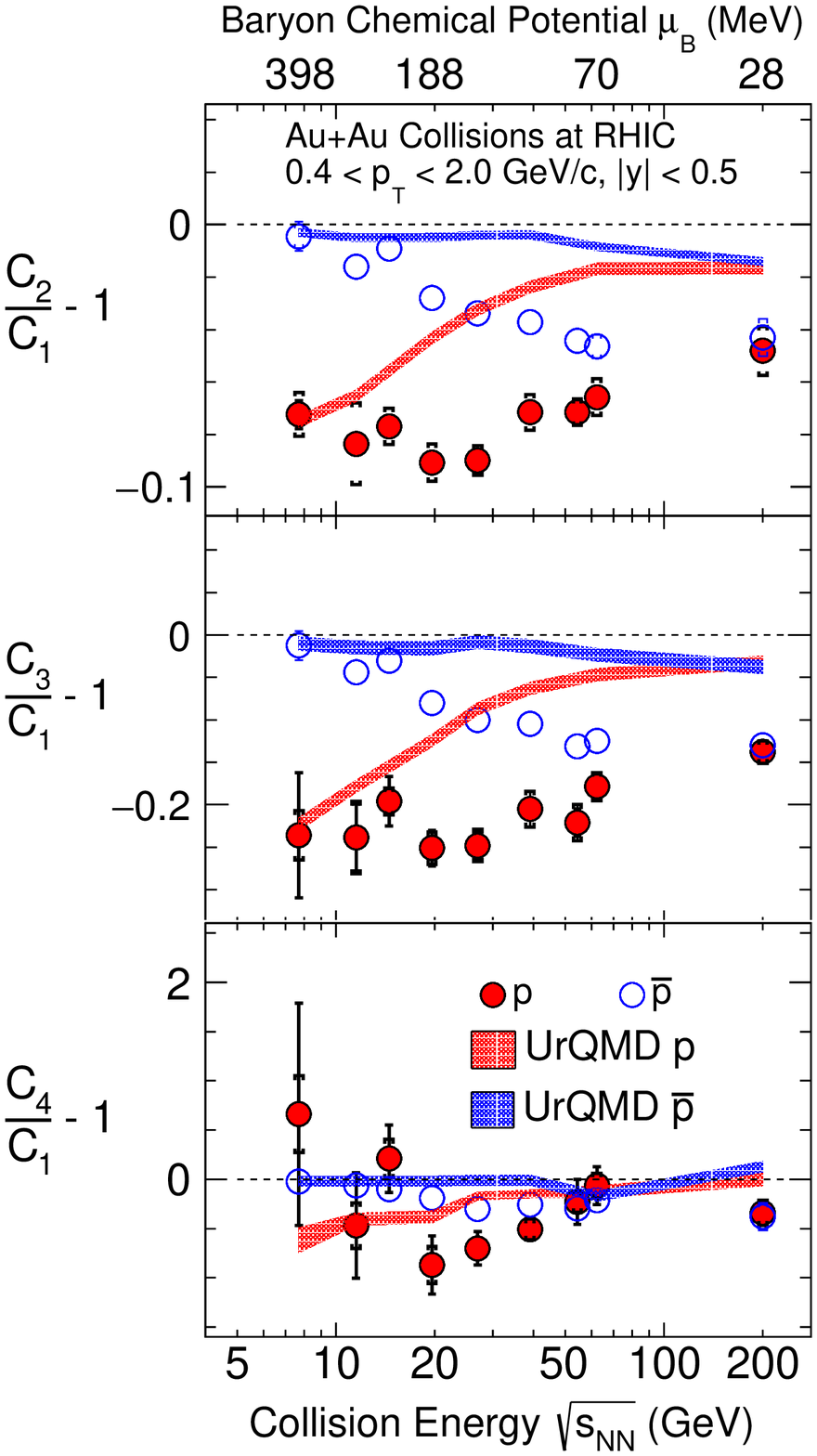}
	\caption{ Normalised cumulant ratios of proton and anti-proton distributions subtracted from unity shown as a function of collision energies in most central (0-5\%) Au+Au collisions from $\sqrt{s_{NN}}$ = 7.7 - 200 GeV. The bars on the  data points represents statistical uncertainties and the brackets represents systematic uncertainties. The UrQMD expectations for the measurements are given by red and blue bands for protons and anti-protons, respectively. }
	\label{pro_apro_norm} 
\end{figure}

Measuring the proton and anti-proton cumulants could reveal what drives the observed non-monotonic collision energy dependence of net-proton cumulants.
The cumulant ratios of proton and anti-proton distributions subtracted from unity as a function of collision energy for most central (0-5\%) and peripheral (70-80\%) Au+Au collisions at RHIC are given in Fig.~\ref{pro_apro_norm}. The second and third-order cumulant ratios minus unity remain negative across all collision energies for both protons and anti-protons. For protons, the deviation of the measurements from Poisson baseline at zero increases with decreasing collision energies, whereas for anti-protons, the deviations from Poisson baseline is maximum at $\sqrt{s_{NN}}$ = 200 GeV. This is understood to be coming from two-particle correlation functions of protons and anti-protons~\cite{starNP21_long}. 
The qualitative trend of both ratios is captured by the UrQMD model for protons and anti-protons. $C_{4}$/$C_{1} -1$ of proton distributions shows a non-monotonic collision energy dependence similar to that observed in net-proton cumulant ratio $C_{4}$/$C_{2}$, whereas $C_{4}$/$C_{1} -1$ of anti-proton distributions show weak dependence on collision energy. The UrQMD calculations show a monotonic collision energy dependence of $C_{4}$/$C_{1} -1$ for protons, whereas for anti-protons, it is mostly consistent with Skellam baseline at zero. It was found in ref~\cite{starNP21_long}, that the non-monotonic collision energy dependence of proton $C_{4}$/$C_{1} -1$ is driven by the presence of four-particle correlations for protons, whereas for anti-protons, only the two-particle correlations drive the collision energy dependence trend.

\subsubsection{ Collision energy dependence of net-charge and net-kaon cumulant ratio $C_{4}$/$C_{2}$ }
\begin{figure}[!htb]
	\centering 
	\includegraphics[scale=0.62]{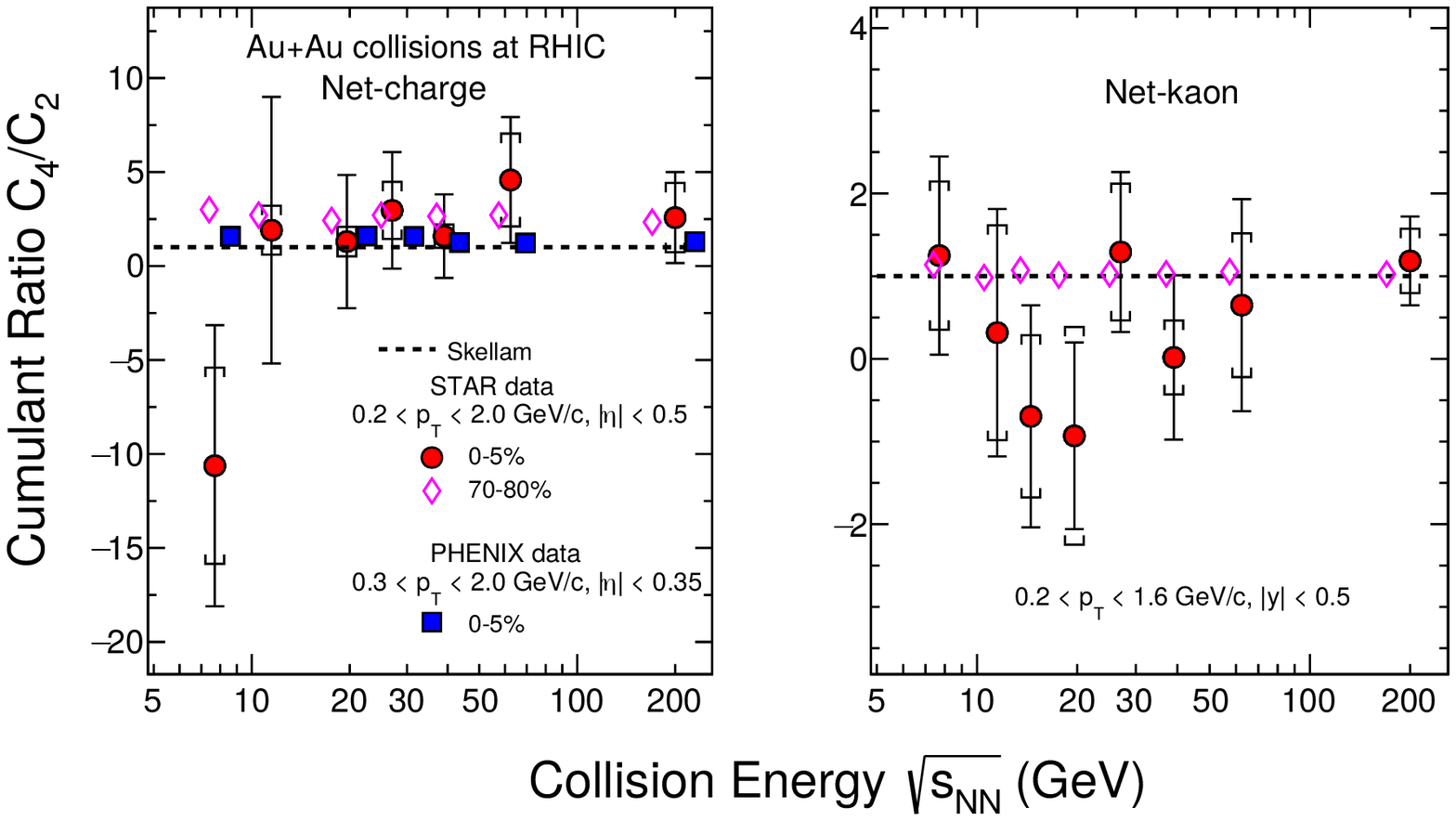}
	\caption{Collision energy dependence of the ratio of fourth to second order cumulant of net-charge and net-kaon distributions($S\sigma$) in Au+Au collisions from $\sqrt{s_{NN}}$ = 7.7 - 200 GeV from the BES-I program at RHIC. The 0-5\% central and 70-80\% results are shown as solid red and open magenta diamond markers, respectively. Results on net-charge $C_4/C_2$ from the PHENIX experiment are also shown as blue squared markers~\cite{PHENIX:2015tkx}. The bars and the caps on the measurements represent the statistical and systematic uncertainties, respectively. The Poisson baseline at unity is shown by the dotted line. }
	\label{cumu_ratio_kao_char} 
\end{figure}
The ratio of fourth to second-order cumulant of net-charge and net-kaon distributions in most central (0-5\%) and peripheral (70-80\%) Au+Au collisions at STAR are shown in Fig.~\ref{cumu_ratio_kao_char}. For net-charge cumulant measurements, all charged particles within $p_{T}$ range 0.2 -- 2.0 GeV/$c$ and $|\eta| < 0.5$ are selected and the centrality is determined uisng charged particle multiplicity in the region $0.5 < |\eta| < 1.0$. Net-kaon cumulant measurements are done selecting charged kaons ($K^{+}$ and $K^{-}$) within $0.2 < p_{T} < 1.6$ GeV/c at mid-rapidity ($|y| < 0.5$). All charged particles in the $|\eta| < 1.0$ coverage excluding kaons ($K^{+}$ and $K^{-}$) are used in the centrality definition. The most central (0-5\%) net-charge and net-kaon cumulant ratio $C_{4}/C_{2}$ have flat collision energy dependence. The measurements, inlcuding those from the PHENIX experiment, are mostly consistent with the Skellam baseline at unity. The statistical uncertainties on the measurements are large, especially at the lower collision energies except for the measurement from the PHENIX experiment, where they are small due to smaller detector acceptance~\cite{PHENIX:2015tkx}. For the net-charge fluctuation measurements, the large statistical uncertainties are attributed to the larger width of net-charge distributions as compared to the other two net-particle measurements, whereas smaller detection efficiency of charged kaons contributes to large statistical uncertainties on net-kaon cumulant ratio $C_{4}/C_{2}$. The larger width of the net-charge distributions can be attributed to the decay of resonances. The net-charge $C_{4}/C_{2}$ in peripheral (70-80\%) collisions remains larger than the Skellam baseline across all the energies, whereas the net-kaon $C_{4}/C_{2}$ for the same collision centrality are either consistent or higher than Skellam baseline at unity, at all the energies. One should also keep in mind that in addition to charged kaons, neutral kaons and multi-strange baryons like the $\Lambda$ are also produced in appreciable numbers in heavy-ion collisions and carry strangeness. Thus, more strange particles should be added to net-kaon fluctuations to serve as a more effective proxy of net-strangeness fluctuations. The lack of statistical precision in these measurements from BES-I forbids us to see any possible critical signal at present. The second phase of the BES program, BES-II, with large event statistics, will shed more light on this. 

\subsubsection{Acceptance  dependence of net-proton cumulant ratios}
The acceptance dependence of number fluctuations are realized by changing the pseudo-rapidity ($\eta$) or rapidity ($y$) coverage considered for fluctuation measurements. Such dependence could reveal important information on the nature of the fluctuations and their origin. The fluctuations could be broadly divided into two categories depending on their source of origin: (a) Thermal fluctuations and (b) Initial state fluctuation. While the thermal fluctuations increase with acceptance and finally saturate after the typical equilibrium correlation length ($\xi \sim 0.5 fm$) is reached, the initial state fluctuations are long-range in nature and grow with acceptance till the effect of conservation laws become significant. In a very small acceptance window, the fluctuations approach the Poisson limit while at broader rapidity coverage, they are driven by conservation laws. Also, it has been suggested in Ref.~\cite{Sakaida:2017rtj}, rapidity dependence of fluctuations could also help understand the diffusion property of the system formed in the heavy-ion collisions. If the critical point lies between two energies in the beam energy scan, then studying the rapidity dependence of cumulants can help pinpoint its location~\cite{Brewer:2018abr}. Rapidity dependence of cumulants measured by the STAR and ALICE experiments with the limitations of finite detector acceptance are discussed in this section.
\begin{figure}[!htb]
	\centering 
		\hspace*{2.5cm}  
	\includegraphics[scale=0.9]{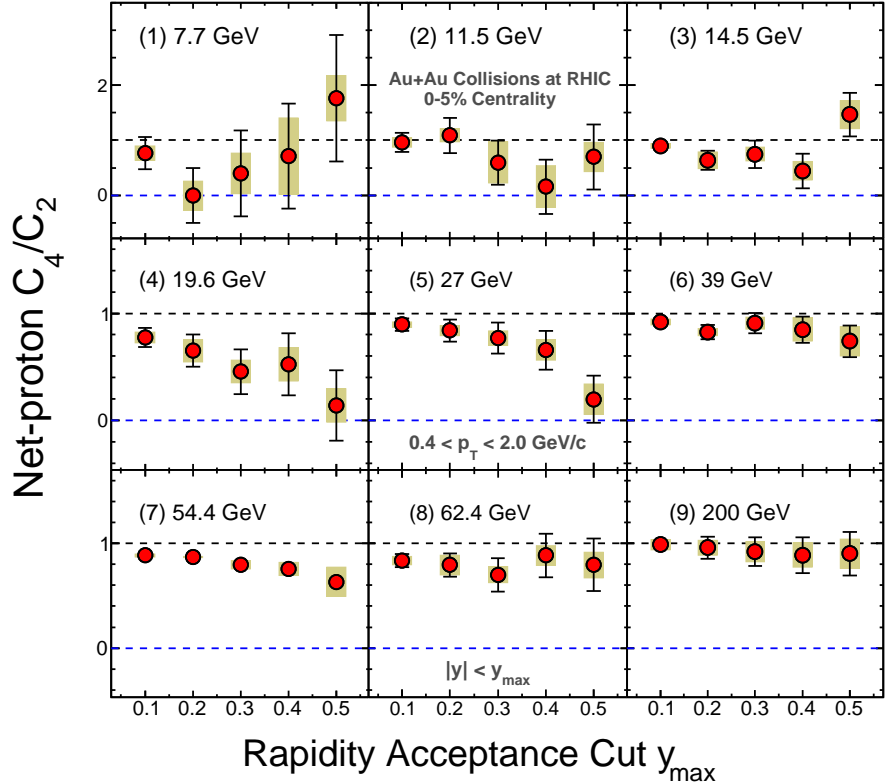}
	\caption{Rapidity dependence of net-proton $C_{4}/C_{2}$ in 0-5\% central Au+Au collisions at $\sqrt{s_{NN}}$ = 7.7, 11.5, 14.5, 19.6, 27, 39, 54.4, 62.4 and 200 GeV. The bars and shaded bands on the data points represents the statistical and systematic uncertainties, respectively. }
	\label{STAR_rapidity} 
\end{figure}
The rapidity dependence of net-proton $C_{4}/C_{2}$ in the collision energy range $\sqrt{s_{NN}}$ = 7.7 -- 200 GeV from the STAR experiment is shown in Fig.~\ref{STAR_rapidity}~\cite{starNP21}. The ratio remains close to the Poissonian limit at unity for the smallest rapidity window accessed at all energies. While decreasing trend of $C_{4}/C_{2}$ with increasing rapidity acceptance is observed for 19.6 and 27 GeV, the ratio at 7.7 GeV first shows a dip followed by an increasing trend, albeit with large uncertainties when acceptance is enlarged. For the higher collision energies $\sqrt{s_{NN}} \geq$ 39 -- 200 GeV, weak dependence is observed on acceptance.  

\begin{figure}[!htb]
	\centering 
	\includegraphics[scale=0.55]{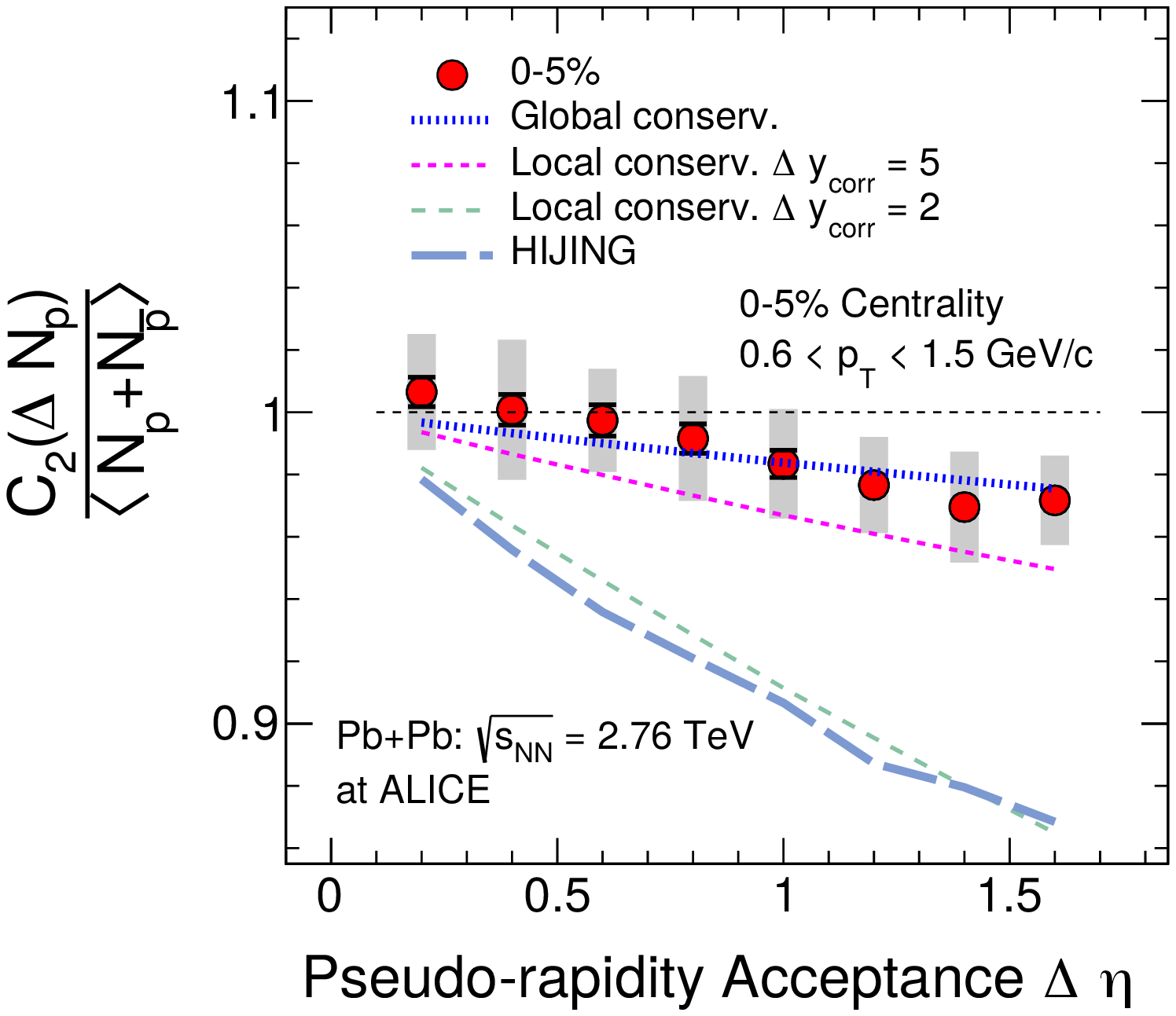}
	\caption{Pseudo-rapidity dependence of net-proton $C_{2}/<N_{p}+N_{\bar{p}}>$ in 0-5\% central Pb+Pb collisions at $\sqrt{s_{NN}}$ = 2.76 TeV at ALICE~\cite{Acharya:2019izy}. The bars and shaded bands on the data points represents the statistical and systematic uncertainties, respectively. The HIJING calculations are shown as dashed pastel blue line. Predictions accounting global baryon number conservation effects are given as blue vertical-dashed line whereas model predictions with local baryon number conservation effects are shown as dashed green and magenta lines. }
	\label{ALICE_rapidity} 
\end{figure}
ALICE also reported its first results on net-proton cumulant measurements up to the second order. The ratio of second order net-proton cumulant to that of mean value of protons+anti-protons are shown as a function of $\eta$ coverage in Pb+Pb collisions at $\sqrt{s_{NN}}$ = 2.76 TeV~\cite{Acharya:2019izy}. Weak acceptance dependence of measurements is observed for second-order ratio, with central values of the measurements differing at a level of 4\% from lowest to largest acceptance. The measurements approach the Poisson limit when acceptance is reduced. The HIJING model calculations fails to describe the measurements. The data seem to be consistent with the global baryon number conservation scenario rather than local baryon number conservation.

\begin{figure}[!htb]
	\centering 
	\includegraphics[scale=0.45]{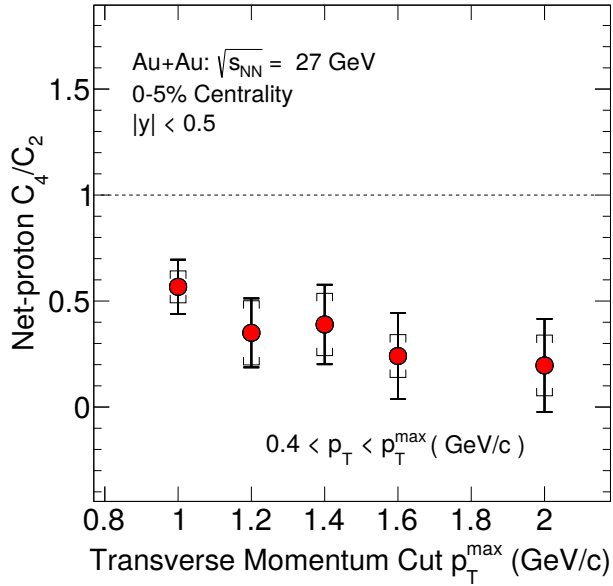}
	\caption{Transverse momentum ($p_T$) dependence of net-proton $C_{4}/C_{2}$ in 0-5\% central Au+Au collisions at $\sqrt{s_{NN}}$ = 27 GeV. The bars and caps on the data points represents the statistical and systematic uncertainties, respectively. }
	\label{STAR_pT} 
\end{figure}

Increasing the $p_T$ range of the charged particle selection in the measurements could also provide interesting results as more charged particles now constitute the system. The recent result on non-monotonic energy dependence of net-proton $C_{4}/C_{2}$~\cite{starNP21} was realised after the $p_{T}$ range for protons was enlarged to $0.4 < p_{T} < 2.0$ GeV/c as compared to the prior measurements with $0.4 < p_{T} < 0.8$ GeV/c~\cite{starNP}. Extending the $p_T$ range was possible due to the addition of the TOF detector. Figure~\ref{STAR_pT} shows the $p_T$ dependence of net-proton $C_4/C_2$ in most central 0-5\% Au+Au collisions at $\sqrt{s_{NN}}$ = 27 GeV. The lower limit of $p_{T}$ is fixed at 0.4 GeV/c so as to avoid contributions from background protons from beam pipe interactions, and upper limit is varied. The data points show a larger deviation from the Skellam baseline at unity with increasing $p_T$ coverage, and the statistical uncertainties on the data points increase as more particles are selected with larger $p_T$ acceptance. The critical phenomenon are of long-wavelength character and thus should be predominant in the lower momenta. Reducing the current lower limit of $p_{T}$ = 0.4 GeV/c is a highly desired and challenging task in the search of the CP. Increasing the phase space coverage by reducing the lower limit of $p_{T}$ would also increase the possibility of $\Delta$ and weak decays of strange baryon induced self correlations, as discussed in subsection~\ref{aucorrel_study}. By performing a similar study as shown in Fig.~\ref{fig_auco}, but systematically reducing the lower limit of $p_{T}$ down to $0.05$ GeV/c, a negligible self-correlation was found due to such decays.

\subsubsection{ A Lattice-QCD inspired fit to experimental measurements}
\begin{figure}[!htb]
	\centering 
	\includegraphics[scale=0.45]{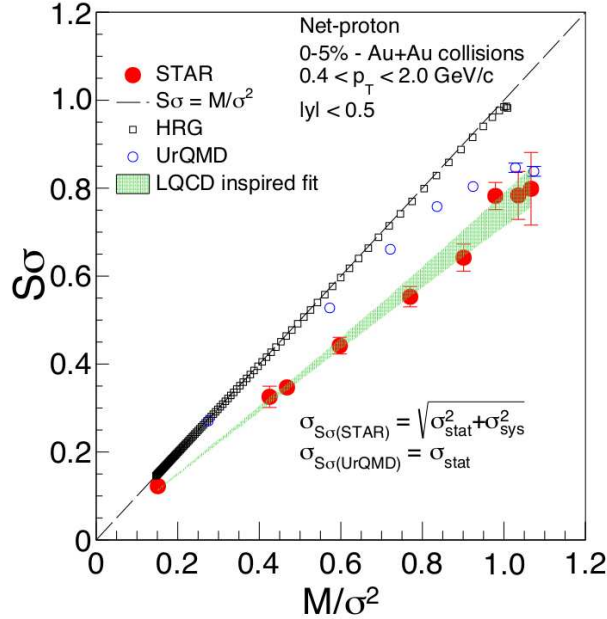}
	\caption{ The of ratio of third to second-order cumulant of net-proton distributions ($S\sigma$) in 0-5\% Au+Au collisions from $\sqrt{s_{NN}}$ = 7.7 - 200 GeV from the BES-I program at RHIC as a function of $M/\sigma^2$. The bars on the measurements represent the total uncertainties obtained by adding statistical uncertainties and systematic uncertainties in quadrature. The calculation from UrQMD and HRG model calculations are also shown as blue and black open markers, respectively. Fit to the data points inspired from lattice calculations is presented by the green band. }
	\label{cumu_latt} 
\end{figure}
In the previous subsection, the cumulant ratios measured at STAR were compared with various model calculation. This section discusses the comparison of net-proton cumulant ratios $C_{3}$/$C_{2}$ to a fit inspired by lattice-QCD calculations for net-baryon number fluctuations.

The net-proton cumulant ratios $S\sigma = C_3/C_2$ from nine collision energies as a function of $M/\sigma^2= C_1/C_2$ is presented in Fig.~\ref{cumu_latt}. As discussed in Ref.~\cite{Bazavov:2017tot}, in lattice-QCD calculations, the baryon number susceptibility ratios $\chi^{B}_{3}/\chi^{B}_{1}$ and $\chi^{B}_{3}/\chi^{B}_{2}$ can be expressed in terms of $\chi^{B}_{1}/\chi^{B}_{2}$ as follows.
\begin{eqnarray}
\chi^{B}_{3}/\chi^{B}_{1} = p_{a1} + p_{b1}(\chi^{B}_{1}/\chi^{B}_{2})^2
\label{eqLQCD_1}
\end{eqnarray}
\begin{eqnarray}
\chi^{B}_{3}/\chi^{B}_{2} = p_{a1} (\chi^{B}_{1}/\chi^{B}_{2}) + p_{b1}(\chi^{B}_{1}/\chi^{B}_{2})^3
\label{eqLQCD_2}
\end{eqnarray}
The quantities $p_{a1}$ and $p_{b1}$ could be fixed by comparing the experimental data on net-proton $C_3/C_1$ as a function of $C_1/C_2$ to eqn.~\ref{eqLQCD_1} for the collision energy range $\sqrt{s_{NN}}$ = 7.7 -- 200 GeV~\cite{starNP21} assuming net-proton to be proxy for net-baryon. The $p_{a1}$ and $p_{b1}$ are then used to calculate $C_{3}/C_{2}$ as a function of $C_{2}/C_{1}$ using the relation mentioned in eqn.~\ref{eqLQCD_2}. The predicted dependence is shown as green band in Fig.~\ref{cumu_latt}. The dashed line at y=x refers to the case where the ratio $S\sigma$ = $M/\sigma^2$. The HRG model calculations is found to be along the dashed line whereas the UrQMD expectations show small deviation at large $M/\sigma^2$  corresponding to low collision energies. The STAR measurements lie on the right side of the dashed line at y=x, indicating that $M/\sigma^2 > S\sigma$ at all energies. This hierarchy is consistent with LQCD calculations reported in Ref.~\cite{Bazavov:2020bjn}. The agreement of the experimental measurements to the LQCD inspired fit within uncertainties suggests the production of strongly interacting and thermalised QCD matter in heavy-ion collisions.

\subsubsection{ Other new measurements}
\begin{figure}[!htb]
	\centering 
	\includegraphics[scale=0.7]{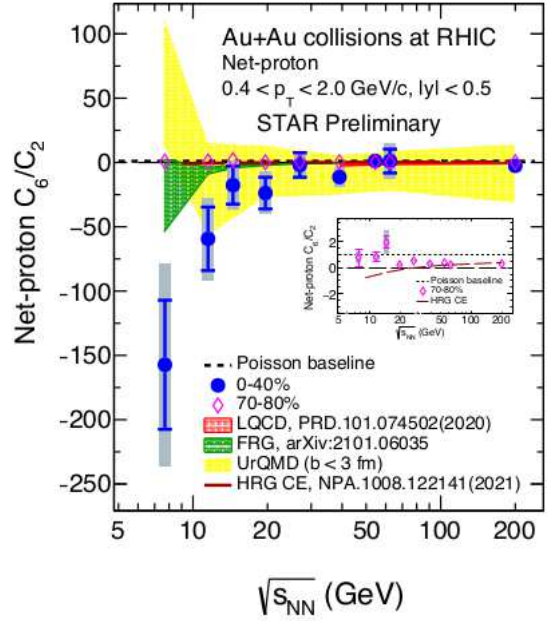}
	\caption{Ratio of sixth to second-order cumulant of net-proton distributions ($C_{6}/C_{2}$) in Au+Au collisions from $\sqrt{s_{NN}}$ = 7.7 - 200 GeV from the BES-I at STAR. The results are shown for 0-40\% (solid blue marker) and 70-80\%(open magenta marker) collision centrality. The bars and the shaded band on the data points represent statistical and systematic uncertainties, respectively. The LQCD and FRG calculations are given by the red and green band, respectively. The expectation from the UrQMD model and HRG canonical ensemble calculations are shown as the yellow band and brown dashed line, respectively. The dashed line at unity represents the Skellam baseline.}
	\label{cumu_C6} 
\end{figure}

\paragraph{{(a) Fixth and sixth order cumulants}:}
\begin{figure}[!htb]
	\centering 
	\includegraphics[scale=0.55]{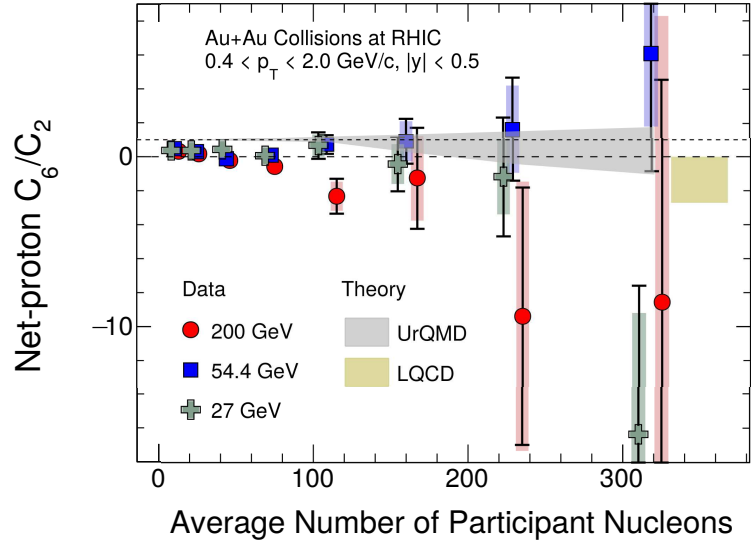}
	\caption{Centrality dependence of sixth to second order cumulant ratio of net-proton distributions ($C_{6}/C_{2}$) in Au+Au collisions at $\sqrt{s_{NN}}$ = 27 (BES-II) (solid cross), 54.4 (solid square) and 200 GeV (solid circle) at STAR-RHIC. Results are shown as a function of average number of participant nucleons for 8 collision centralities, 0-10\%, 10-20\%, 20-30\%, 30-40\%, 40-50\%, 50-60\%, 60-70\%, 70-80\%. The bars and the bands on the data points represent the statistical and systematic uncertainties, respectively. UrQMD calculations for the three collision energies are consistent with each other, thus merged to improve statistics and is shown as grey band. The LQCD calculations from Ref.~\cite{Borsanyi:2018grb,Bazavov:2020bjn} are shown as golden band. }
	\label{cumu_C6_centdep} 
\end{figure}
Experimental measurements of fluctuations of net-proton distributions have been extended to fifth and sixth order. There have been several theoretical works dedicated to fifth and sixth order susceptibility calculations, also known as the hyper-order fluctuations.  Recent results on centrality dependence of $C_6/C_2$ of net-proton distributions in Au+Au collisions at $\sqrt{s_{NN}}$ = 27, 54.4 and 200 GeV are shown in Fig.~\ref{cumu_C6_centdep}~\cite{star_C6paper}. The measurements at $\sqrt{s_{NN}}$ = 200 GeV are increasingly negative with increase in collision centrality while those at $\sqrt{s_{NN}}$ = 27 and 54.4 GeV do not show such trend and remain close to zero. The $C_6/C_2$ values become positive when approaching peripheral collisions. The progressively negative sign of the ratio is qualitatively consistent with the sign predicted by LQCD calculations with $T$ = 160 MeV and $\mu_{B}$ = 0 -- 110 MeV~\cite{Borsanyi:2018grb,Bazavov:2020bjn}. The UrQMD calculations are either positive or consistent with zero.
\begin{figure}[!htb]
	\centering 
	\includegraphics[scale=0.7]{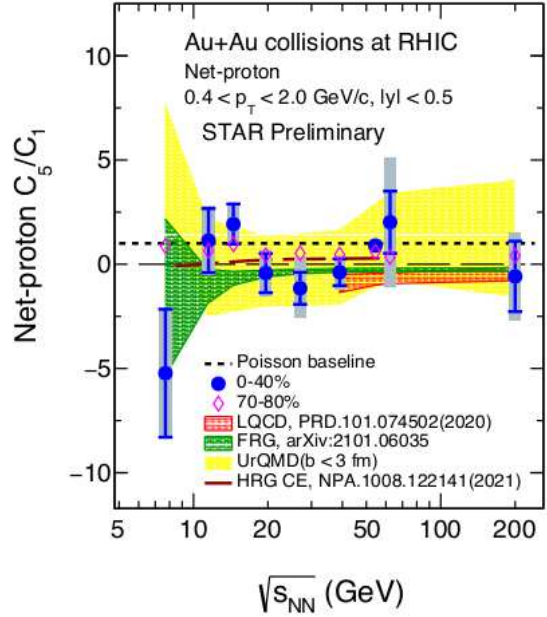}
	\caption{Ratio of fifth to first-order cumulant of net-proton distributions ($C_{5}/C_{1}$) in Au+Au collisions from $\sqrt{s_{NN}}$ = 7.7 - 200 GeV from BES-I at STAR. The results are shown for 0-40\% (solid blue marker) and 70-80\%(open magenta marker) collision centrality. The bars and the shaded band on the data points represent statistical and systematic uncertainties, respectively. The LQCD and FRG calculations are given by the red and green bands, respectively. The expectation from the UrQMD model and HRG canonical ensemble calculations are shown as the yellow band and brown dashed line, respectively. The dashed line at unity represents the Skellam baseline.}
	\label{cumu_C5} 
\end{figure}
Collision energy dependence of fifth and sixth-order fluctuations has also been reported by STAR~\cite{CPOD_pandavA}. Collision energy dependence of net-proton cumulant ratio $C_{5}/C_{1}$ and $C_{6}/C_{2}$  is shown in Fig.~\ref{cumu_C5} and ~\ref{cumu_C6}, respectively~\cite{CPOD_pandavA}. LQCD, FRG calculations along with expectations from the UrQMD model and HRG CE model are also presented. To improve the statistical precision of the measurements, calculations are done over 0-40\% centrality. The results for most peripheral 70-80\% centrality are also presented. The $C_{5}/C_{1}$ shows weak collision energy dependence. Deviations of the measurements from zero at a level of $\leq2\sigma$ significance are observed. The LQCD and the FRG calculations show negative values of the ratio of fifth-to-first order baryon number susceptibilities. The peripheral 70-80\% measurements, on the other hand, remains consistently positive across all collision energies and close to the Skellam baseline at unity. The net-proton $C_{6}/C_{2}$ in central 0-40\% collision shows increasing negative values with decreasing collision energy. This is qualitatively consistent with the sign and trend predicted by LQCD and FRG calculations for the ratio of sixth-to-second order baryon number susceptibilities. The magnitude of deviation of the measurements from zero is of similar significance, as for $C_{5}/C_{1}$. In contrast, the peripheral 70-80\% centrality measurements are positive at all the energies (shown in the inset of Fig.~\ref{cumu_C6}). The HRG CE expectation for both the net-proton cumulant ratios are positive in the higher collision energies and show suppression towards negative values in the lower collision energies due to the effect of baryon number conservation.

\begin{figure}[!htb]
	\centering 
	\includegraphics[scale=0.5]{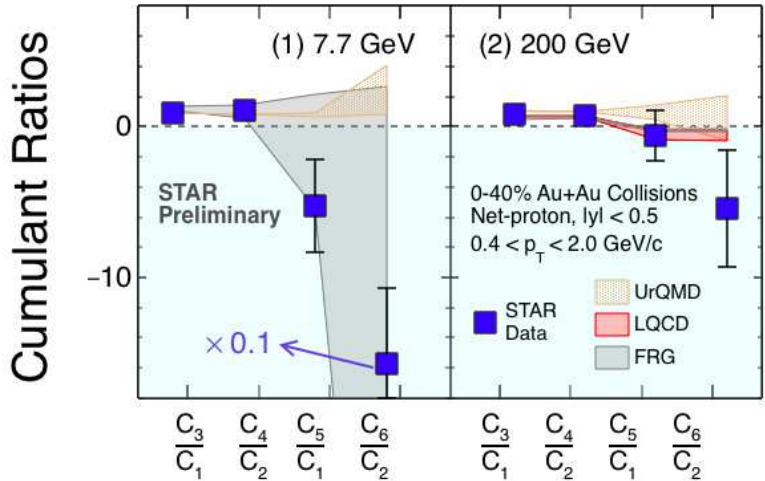}
	\caption{$C_3/C_1$, $C_4/C_2$, $C_5/C_1$ and $C_6/C_2$ of net-proton distributions in central 0-40\% Au+Au collisions for $\sqrt{s_{NN}}$ = 7.7 and 200 GeV (blue solid markers)~\cite{starNP21_long,CPOD_pandavA,star_C6paper}. The results from 0-10\%, 10-20\%, 20-30\%, and 30-40\% centralities from the Ref.~\cite{star_C6paper} are averaged to quote 0-40\% centrality results at 200 GeV. Only statistical uncertainties are shown. Calculations from LQCD (200 GeV)~\cite{Bazavov:2020bjn}, FRG~\cite{Fu:2021oaw} and UrQMD calculations (0-40\% centrality) are shown as red, grey and brown bands, respectively.}
	\label{cumuratio_ordering} 
\end{figure}

Furthermore, a particular ordering of net-proton cumulant ratios: $C_3/C_1 > C_4/C_2 > C_5/C_1 > C_6/C_2$ as predicted by lattice-QCD was also observed in data for Au+Au collisions at $\sqrt{s_{NN}}$ = 7.7 GeV and 200 GeV. The ordering of the four ratios $C_3/C_1$, $C_4/C_2$, $C_5/C_1$ and $C_6/C_2$ for net-proton distributions in 0-40\% collision centrality are shown in the Fig.~\ref{cumuratio_ordering}. While the FRG calculations also follow the predicted hierarchy, the UrQMD does not show such ordering of ratios and gives similar positive values for the four ratios. Here, one should keep in mind the caveats when comparing experimental data to LQCD and FRG model calculations. As opposed to the LQCD and FRG, which calculates net-baryon fluctuations, the experimental measurements are made for net-proton fluctuations within a specific kinematic phase space allowed by the experiment. Also, the former does not take into account the dynamics associated with nuclear collisions, for example, the collective expansion. These limitations should be accounted for when making a quantitative comparison between data and theory.

\begin{figure}[!htb]
	\centering 
	\includegraphics[scale=0.8]{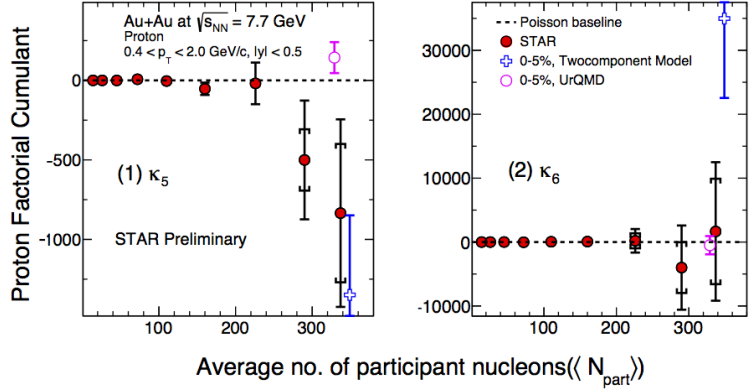}
	\caption{Collision centrality dependence of fifth and sixth order factorial cumulants ($\kappa_5$ and $\kappa_6$, respectively) of proton multiplicity distributions in Au+Au collisions at $\sqrt{s_{NN}}$ = 7.7 GeV from the BES-I at STAR. The bars and the brackets on the data points represent statistical and systematic uncertainties, respectively. The open magenta circles are the expectation from the UrQMD model for the most central 0-5\% collisions. The expectations from a two-component model assumed for protons multiplicity distribution are also shown for 0-5\% collision centrality. The dashed line at unity represents the Poisson baseline.}
	\label{kappa_56} 
\end{figure}

The fifth and sixth order factorial cumulants ($\kappa_5$ and $\kappa_6$, respectively) of proton multiplicity distributions as a function of collision centrality in Au+Au collisions at $\sqrt{s_{NN}}$ = 7.7 GeV are presented in Fig.~\ref{kappa_56}~\cite{CPOD_pandavA}. The $\kappa_5$ shows a decreasing trend with increasing collision centrality, whereas a weak collision-centrality dependence is found for $\kappa_6$. The UrQMD model calculations for both the measurements are close to the Poisson baseline at zero. The $\kappa_5$ measurement for 0-5\% centrality is consistent with the expectation from a two-component proton multiplicity distribution model, which takes proton cumulants up to fourth-order as input in its construction. The most central (0-5\%) $\kappa_6$, however, is 1.8$\sigma$ away from expectation from such a model and consistent with the Poisson baseline at zero, albeit with large uncertainties.

These higher-order cumulant measurements are expected to be measured with higher precision in the second phase of the beam energy scan program.

\paragraph{{(b) Multiplicity dependence of net-proton cumulants in pp collisions }:}
\begin{figure*}[!htb]
	\centering 
	\includegraphics[scale=0.87]{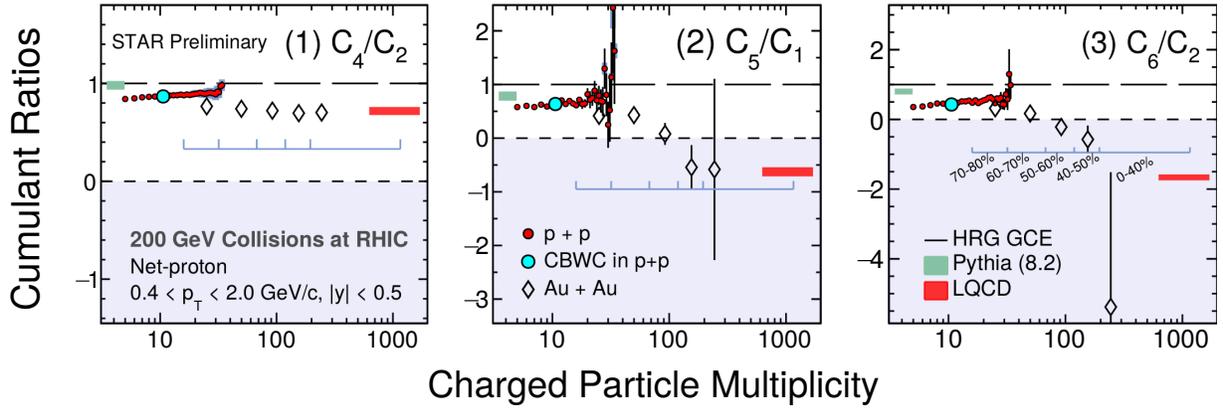}
	\caption{Charged particle multiplicity dependence of net-proton cumulant ratios (shown in red markers): (1)  $C_4/C_2$, (2) $C_5/C_1$ and (3)  $C_6/C_2$ in p+p collisions in $\sqrt{s}$ = 200 GeV at RHIC~\cite{CPOD_nishitaniR}. CBWC average of the ratios are shown in light blue marker. Bars and the blue band on the data points represent the statistical and systematic uncertainties, respectively. Results from Au+Au collisions at  $\sqrt{s_{NN}}$ = 200 GeV for 0-40\%, 40-50\%, 50-60\%, 60-70\%, 70-80\% collision centrality are shown in open diamonds~\cite{starNP21_long,CPOD_pandavA,star_C6paper}. Only statistical uncertainties are shown for Au+Au results. The dashed line at unity represents the HRG GCE calculations. The lattice-QCD and CBWC-averaged Pythia calculations are shown as red and dark green bands, respectively.   }
	\label{pp_cumuratios} 
\end{figure*}
Cumulant measurements have also been recently explored in small systems by STAR experiment~\cite{CPOD_nishitaniR}. Multiplicity dependence of net-proton cumulant ratios $C_4/C_2$, $C_5/C_1$ and $C_6/C_2$ in p+p collisions at $\sqrt{s}$ = 200 GeV are shown in Fig.~\ref{pp_cumuratios}.  The three ratios approach to their Skellam and HRG GCE baseline value of unity with increasing charged particle multiplicity while the centrality dependence measurements of these ratios in Au+Au collisions at $\sqrt{s_{NN}}$ = 200 GeV show opposite trend. CBWC average of p+p collision results are consistent with the multiplicity dependence trend. The $C_5/C_1$ and $C_6/C_2$ in p+p collisions are positive across the entire charged particle multiplicity range which is in contrast to the negative sign predicted by LQCD calculations for net-baryon $\chi^{B}_{5}/\chi^{B}_{1}$, $\chi^{B}_{6}/\chi^{B}_{2}$ for QCD matter. On the other hand, the $C_5/C_1$ and $C_6/C_2$ for Au+Au collisions at $\sqrt{s_{NN}}$ = 200 GeV show increasingly negative value towards central collisions. The Pythia calculations are found to overestimate the values of the ratios.

The higher-order cumulant measurements in a small system provide a baseline for comparison with results from heavy-ion collisions. They also provide important data to test thermalization and freeze-out conditions.

\begin{figure}[!htb]
	\centering 
	\includegraphics[scale=0.65]{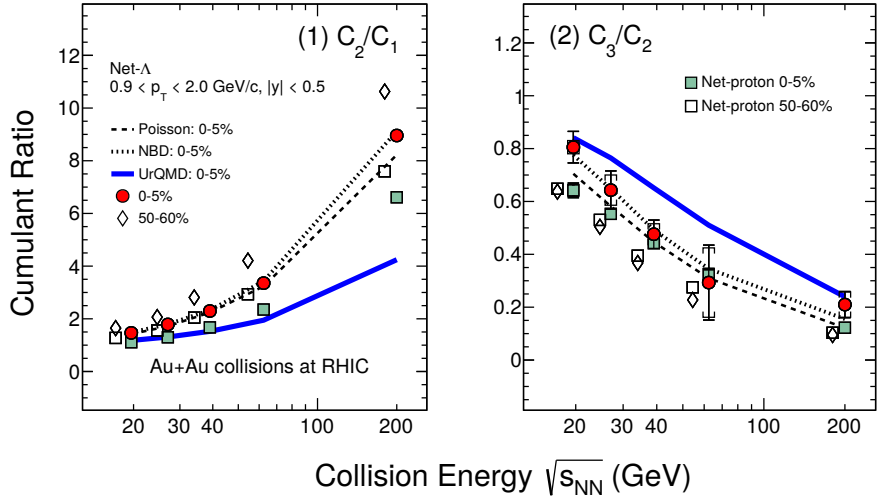}
	\caption{Cumulant ratio (1) $C_{2}/C_{1}$ and (2) $C_{3}/C_{2}$ of net-$\Lambda$ distributions in Au+Au collisions from $\sqrt{s_{NN}}$ = 19.6 -- 200 GeV at RHIC. Results are shown for 0-5\% and 50-60\% centralities. The bars and the caps on the data points represent the statistical and systematic uncertainties, respectively. Poisson and NBD baseline for 0-5\% centrality are shown by dashed lines and the UrQMD calculations are shown in blue lines. Also shown for comparison are the STAR net-proton results for the two centralities~\cite{starNP21_long}.}
	\label{net_lambda} 
\end{figure}

\paragraph{{(c) Net-$\Lambda$ fluctuations}:}
Net-$\Lambda$ fluctuations have been studied at RHIC in the energy region $\sqrt{s_{NN}}$ = 19.6 -- 200 GeV~\cite{star_lambda}. The collision energy dependence of net-$\Lambda$ fluctuations is shown in Fig.~\ref{net_lambda}. Monotonic collision energy dependence is observed for the cumulant ratios $C_{2}/C_1$ and $C_{3}/C_2$ in both central (0-5\%) and peripheral (50-60\%) collisions. NBD baseline seems to show better agreement with the data for central collisions than compared to the Poisson baseline. The UrQMD model calculations could only predict the trend of the measurements, while quantitative differences exist. In the reported range of collision energies, the data shows no features of critical fluctuations. Study of net-$\Lambda$ fluctuations is important as $\Lambda$ has both baryon and strangeness quantum numbers. Adding net-$\Lambda$ to net-kaon fluctuations will provide a more complete measurement of net-strangeness fluctuations in the future. Comparison of net-proton measurements with those of net-$\Lambda$ suggests that the net-$\Lambda$ results follow the qualitative energy dependence as shown by net-proton.
Experimentally $\Lambda$ particles are detected via invariant mass reconstruction of their decay daughters, protons, and pions. To treat the net-$\Lambda$ and net-proton fluctuations as independent, experimental measurements should ensure the decay contributions from $\Lambda$ do not contaminate the net-proton sample. If this can be achieved either by applying kinematic cuts for particle selection or feed-down corrections, then these two net-particle fluctuations could be added to construct a more reliable proxy of net-baryon fluctuations. Since, $\Lambda$ has both baryon and strangeness quantum numbers, their inclusion is essential in the baryon-strangeness correlation studies. Studies on the non-diagonal baryon-strangeness correlator by authors of Ref.~\cite{Bellwied:2019pxh}, indicate that considering only net-protons and net-kaons in such correlation studies are not sufficient. In fact, proton-kaon correlation has negligible contribution in the baryon-strangeness correlations, and a significant contribution to the latter comes from the variance of net-lambda distribution.

In future, it will be very interesting to see if the $C_{4}/C_2$ for net-$\Lambda$ as a function of collision energy shows a similar non-monotonic variation as observed for net-protons. 

\paragraph{{(d) Deuteron and nuclei fluctuations }:}
Since deuteron is the second lightest stable baryon detected in the heavy-ion collision experiments, measurement of their fluctuations might have interesting prospects in the search for the QCD critical point. Especially, in the high baryonic density region (low collision energies), an abundance of the light nuclei is enhanced, and their contribution to the dynamics of baryonic charge fluctuations in the system becomes important. Higher-order moments of event-by-event number distribution of deuterons are recently suggested also as a probe for their production mechanism~\cite{Feckova:2016kjx}. Synthesis mechanism of light nuclei detected in high-energy nuclear collisions is primarily discussed in two scenarios: (a) statistical thermal model and (b) coalescence model. Though both scenarios, to a good extent are able to explain the experimental data on yields and their ratios, a complete picture of production mechanism of light nuclei is still missing. In this context, measurement of light nuclei fluctuations is worthwhile to probe their production mechanism and explore signals for CP~\cite{Shuryak:2020yrs}.
\paragraph{{(e) Probing magnetic field via fluctuations }:}
Presence of a strong magnetic field in heavy-ion collisions can bring interesting effects on QCD phase structure. The chiral-crossover transition temperature $T_{pc}$ is expected to be reduced due to the phenomenon of inverse magnetic catalysis~\cite{Bali:2014kia,Ilgenfritz:2013ara,Bornyakov:2013eya,Tomiya:2019nym}. Lattice-QCD calculation studies done by authors of Ref.~\cite{Ding:2021cwv} suggest using fluctuations and correlation of conserved charge quantities to probe the existence of magnetic field in the heavy-ion collisions. It has been known from previous studies, in the absence of a magnetic field, the second-order fluctuations of $B$, $Q$, and $S$ increase with increasing temperature~\cite{HotQCD:2012fhj,Bazavov:2020bjn}. Recent studies indicate that they seem to develop a peak structure due to the presence of a strong magnetic field, with the location of the peak decreasing towards lower temperatures with increasing magnetic field. In particular, observables constructed from baryon-strangeness, charge-strangeness and baryon-charge second order correlations and fluctuations: (2$\chi_{11}^{QS}$-$\chi_{11}^{BS}$)/$\chi_{2}^{S}$, (2$\chi_{11}^{BQ}$-$\chi_{11}^{BS}$)/$\chi_{2}^{B}$, $\chi_{2}^{B}/\chi_{2}^{S}$, $\chi_{2}^{B}/\chi_{11}^{QS}$ and $-3\chi_{11}^{BS}/\chi_{2}^{S}$ have been proposed to detect presence of magnetic field in early stages of heavy-ion collision and breaking of isospin symmetry in non-zero magnetic field~\cite{Ding:2021cwv}. The proposed magnetic-field-dependence of the observables could be realized by performing measurements on the centrality dependence of these quantities in experiments as the magnetic field is expected to vary as a function of collision-centrality.

\section{Future directions}
Discovering the QCD critical point will be a landmark in the QCD phase diagram. Progress in both theory and experimental measurements is needed to ascertain the existence and location of the QCD critical point. We discuss in this section the outlook from the theoretical side on the study of critical fluctuations along with the future opportunities with upcoming new experimental facilities in the hunt for the QCD critical point.
\begin{figure}[!htb]
	\centering 
	\includegraphics[scale=0.5]{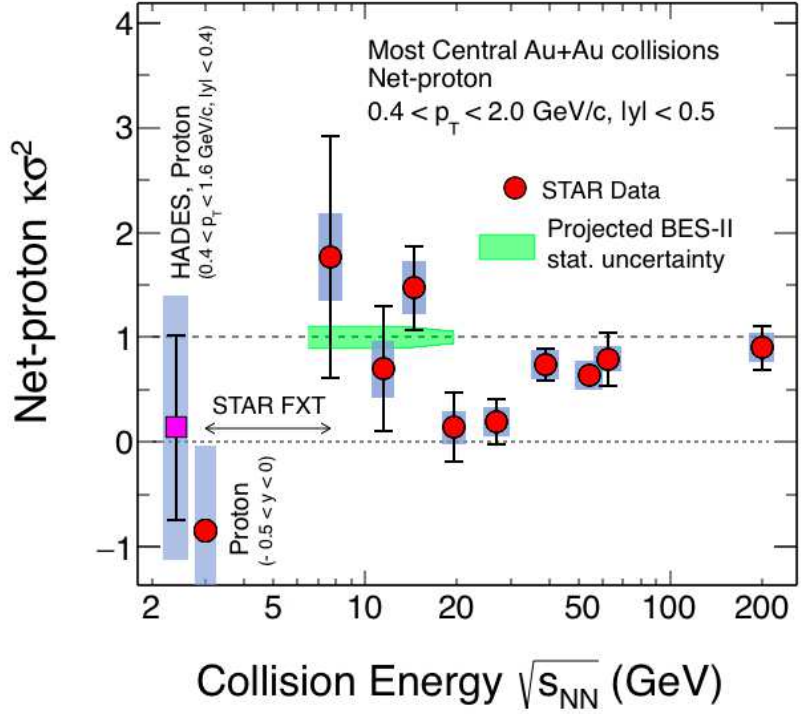}
	\caption{ Ratio of fourth to second order cumulant ($C_4/C_2$) of net-proton/proton distributions($\kappa\sigma^2$) in top 0-5\% Au+Au collisions from $\sqrt{s_{NN}}$ = 3.0 to 200 GeV from the STAR experiment along with measurement on proton $C_4/C_2$ from the HADES experiment~\cite{HADES:2020wpc} for 0-10\% central Au+Au collisions at $\sqrt{s_{NN}}$ = 2.4 GeV with $|y| < $ 0.4. The bars and blue bands on the data points represent the statistical and systematic uncertainties, respectively. The projection of uncertainties on the cumulant ratio $\kappa\sigma^2$ from the ongoing phase II of the BES program at RHIC are shown in green band around unity. The double-headed arrow shows the collision energy range ($\sqrt{s_{NN}}$ = 3 - 7.7 GeV) covered by the STAR's fixed target program (STAR FXT).}
	\label{cumu_exp} 
\end{figure}
\subsection{Theory }
A concrete answer to the existence and location of the QCD critical point will require a complete understanding of the dynamics of the system created in the heavy-ion collisions. A quantitative theoretical framework that concerns the dynamical evolution of the system formed in heavy-ion collisions till it reaches its final state will be crucial to the search of QCD critical point. Studies of fluctuations and their coupling with critical modes in stochastic-hydrodynamics as well as hydro-kinetics have been ongoing. Stochastic fluid dynamics generally involves event-by-event simulation of the viscous relativistic fluid dynamical evolution with stochastic conservation law, where energy-momentum tensor and conserved current 4-vector include Ideal + Viscous + Noise parts. On the other hand, the hydro-kinetics method is used to calculate the time dependence of the two-point correlation functions of the fluid dynamical fields using a set of linearized equations of motion. Hydrodynamical fluctuations and their time evolution considering the effect of critical slowing down are being studied~\cite{dyna_case1,dyna_case2,dyna_case3,dyna_case4,dyna_case5,dyna_case6,dyna_case7,dyna_case8}. 

Another important aspect of stochastic fluid dynamics and hydro-kinetics study of fluctuations is the {\it particlization}. Experimental observables measure correlations of produced particles. The conversion from fluid degrees of freedom to the particle degrees of freedom includes the effects like finite particle statistics, limited acceptance and conservation of energy and global charges at the freeze-out hypersurface.

Further, the thermodynamics of the system formed in these collisions can be studied via comparison of the fluctuations measured in the experiments to the various thermal model and lattice-QCD calculations~\cite{Gupta:2011wh,Alba:2014eba,Borsanyi:2013hza}. Since the net-particle distribution could be fully characterized by its all-order moments, consistent thermal description of mean yields as well higher moments becomes important to test the thermal nature of the whole system~\cite{Gupta:2021dmo}. Deviation from thermal equilibrium might indicate interesting physics, such as the presence of a QCD critical point.

\subsection{Experiment }
The observed non-monotonic collision energy dependence of net-proton $C_{4}/C_{2}$ has been the most interesting result so far in the experimental search of the QCD critical point. Recently STAR collaboration reported the proton $C_{4}/C_{2}$ in Au+Au collisions at $\sqrt{s_{NN}}$ = 3 GeV from STAR's FXT program~\cite{STAR_3GeV}. At such low energies, the number of event-by-event anti-protons are negligibly. As seen from the Fig.~\ref{cumu_exp}, a suppression in $C_4/C_2$ was observed compared to higher collision energies. This is consistent with expectations from baryon number conservation and suggests that matter is dominantly hadronic at such low collision energies. The non-monotonic collision energy dependence of the $C_4/C_2$ measurements from the BES-I program together with the observed suppression at $\sqrt{s_{NN}}$ = 3 GeV (FXT) indicates that the possible critical point could only exist at collision energies higher than 3 GeV. Nonetheless, the measurements at collision energies  $\sqrt{s_{NN}}$ = 7.7 -- 27 GeV suffer from large statistical uncertainties which are the key energies that drive the non-monotonic collision energy dependence trend. The measurement on proton $C_{4}/C_{2}$ by HADES~\cite{HADES:2020wpc}, for 0-10\% central Au+Au collisions at $\sqrt{s_{NN}}$ = 2.4 also has large uncertainties and require more event statistics to serve as effective experimental baseline in the large $\mu_{B}$ region. 
Here, in this section, we discuss future prospects of fluctuation measurements in experiments that are currently active and upcoming in the near future.

\subsubsection{STAR BES-II and FXT }
In order to make precise measurements in the collision energy range  $7.7$ GeV $\leq \sqrt{s_{NN}} \leq 19.6$ GeV, phase II of the beam energy scan (BES-II) has been ongoing using the STAR detector at RHIC~\cite{BES2_whitepaper}. As compared to the presented results from the BES-I program, additional collision energy at $\sqrt{s_{NN}}$ = 9.2 GeV from the BES-II program will be available for fluctuation measurements with large statistics. Table~\ref{tab1_stats_besii} contains the event statistics for each collision energy being collected in the ongoing BES-II program and the corresponding $\mu_{B}$ at chemical freeze-out.
\begin{table}
	\caption{Event statistics to be collected in Au+Au collisions for $7.7$ GeV $\leq \sqrt{s_{NN}} \leq 19.6$ GeV from the ongoing BES-II program at RHIC and the corresponding $\mu_{B}$ at chemical freeze-out. }
	\centering   
	\begin{tabular}{|c|c|c|}
		\hline	
		$\sqrt{s_{NN}}$ (GeV) &  Events (millions) &  $\mu_{B}$ (MeV)  \\
		\hline 
		7.7 & 100 &  398\\
		\hline 
		9.2 & 160 &  355\\
		\hline 
		11.5  & 230 & 287\\
		\hline 
		14.5 & 300 & 264 \\
		\hline 
		19.6 & 400 &  188 \\
		\hline 
	\end{tabular}
	\label{tab1_stats_besii}
\end{table}
The statistical uncertainties of the measurements in BES-II will be significantly reduced as compared to current measurements. The magnitude of projected statistical uncertainties from BES-II measurements are shown in Fig.~\ref{cumu_exp}. In addition to precision measurements of cumulants, the goal is to also improve the quality of the potential critical signal. Several detector upgrades have been made in the STAR detector at RHIC. Upgrades such as that of the inner chambers of the TPC, the event plane detector, the end-cap time of flight detector, bring new scopes to the fluctuation measurements, like enlarged rapidity coverage (from $|\eta| < 1$ to $|\eta| < 1.5$), improved particle identification and centrality definition at forward rapidity. The QCD critical point induced power law behaviour of the fluctuation measurements can be tested using the extended rapidity coverage~\cite{Ling:2015yau}. STAR fixed target (FXT) program will carry Au+Au collisions from $\sqrt{s_{NN}}$ = 3.0 -7.7 GeV (shown in the Fig.~\ref{cumu_exp}) allowing to scan the QCD phase diagram up to $\mu_B$ = 720 MeV. The collision energies, event statistics and the corresponding $\mu_{B}$ values for the STAR FXT program is listed in table~\ref{tab1_stats_besii_fxt}.
\begin{table}
	\caption{Event statistics to be collected in Au+Au collisions from the ongoing STAR FXT program at RHIC and the corresponding $\mu_{B}$ at chemical freeze-out. }
	\centering   
	\begin{tabular}{|c|c|c|}
		\hline	
		FXT Energy (GeV) &  Events (millions) &  $\mu_{B}$ (MeV)  \\
		\hline 
		7.7 & 160 &  398\\
		\hline 
		6.2 & 120 &  487\\
		\hline 
		5.2  & 100 & 541\\
		\hline 
		4.5 & 100 & 589 \\
		\hline 
		3.9 & 120 &  633 \\
		\hline 
		3.5 & 120 &  666 \\
		\hline 
		3.2 & 200 &  699 \\
		\hline 
		3.0 & 260 &  720 \\
		\hline 
	\end{tabular}
	\label{tab1_stats_besii_fxt}
\end{table}
At beam energies above $\sqrt{s_{NN}} \geq$  19.6 GeV, the reaction rates of STAR are limited to a thousand Hertz (Hz) by the TPC read-out, and drop down to a few Hz for energies below $\sqrt{s_{NN}}$ = 8 GeV in collider mode due to low luminosity. Future new experiments, which are designed for operation at high rates, large acceptance, and the state-of-the-art particle identification, at the energy region where baryon density is high, will be needed in the search for the critical point. The new facilities for studying high baryon density matter includes (a) Nuclotron-based Ion Collider fAcility (NICA) at the Joint Institute for Nuclear Research (JINR), Dubna, Russia ~\cite{Geraksiev:2019fon}, (b) Compressed Baryonic Matter (CBM) at Facility for Anti-proton and Ion Research (FAIR), Darmstadt, Germany~\cite{Ablyazimov:2017guv}, (c) CSR External-target Experiment (CEE) at High Intensity heavy-ion Accelerator Facility (HIAF), Huizhou, China~\cite{Ruan:2018fpo}, and (d) Heavy-Ion program at Japan Proton Accelerator Research Complex (JPARC-HI), Tokai, Japan~\cite{JPARC_HI}. 
\begin{figure}[!htb]
	\centering 
	\includegraphics[scale=0.7]{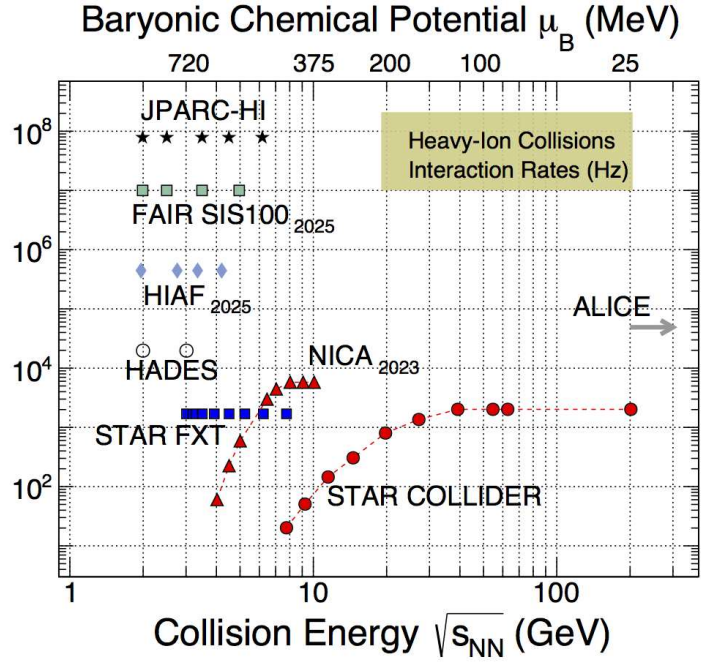}
	\caption{Interaction rates from various heavy-ion collision facilities~\cite{Galatyuk:2019lcf,Bzdak:2019pkr}: BES-II at RHIC (7.7 $< \sqrt{s_{NN}} < $ 19.6 GeV, solid red circles), STAR FXT (solid blue squares), NICA (red triangles), CBM at FAIR (solid green squares), HADES (open circles), HIAF (solid diamonds), and JPARC-HI (solid stars)~\cite{JPARC_HI,Hachiya:2020bjg}. }
	\label{interac} 
\end{figure}
\subsubsection{NICA }
Two dedicated experiments are designed at NICA for the study of QCD phase diagram: a fixed target experiment, Baryonic Matter at Nuclotron (BM$@$N)~\cite{NICA_A}, which will cover the lower energy region, and the Multi-Purpose Detector (MPD)~\cite{NICA_B} which will run in collider mode. The MPD experiment includes critical point search using fluctuation measurements among its physics priorities~\cite{Kekelidze:2016dgy}. The MPD is a $4\pi$ spectrometer which can detect charged hadrons, electrons and photons in heavy-ion collisions in the energy range of the NICA collider. The detector setup includes the Central Detector (CD) covering $\pm 2$ units in pseudo-rapidity ($\eta$),  Time-Projection Chamber (TPC), barrel Time-Of-Flight system (TOF), Zero-Degree Calorimeter (ZDC) and Fast Forward Detector (FFD)~\cite{Golovatyuk:2016fut}. The TPC is the main tracking detector of the MPD and will provide tracking and PID with resolution better than 8\% in the  range $|\eta| < 1.2$, and momentum resolution for charge particles better than 3\% in  $0.1 < p_T < 1$ GeV/c. The cylindrical part of TOF also has full azimuthal and $|\eta| < 1.2$ coverage and will allow PID in the $p_T$ range 0.1 -- 2 GeV/c for charged hadrons. Combining the measurements from TPC and TOF will provide an efficient $\pi/K$ PID separation up to 1.5 GeV/c and $\pi/p$ separation up to 3 GeV/c~\cite{Geraksiev:2018okj}.

Initial run of NICA will be done with Bi+Bi collisions at $\sqrt{s_{NN}}$ = 9.2 AGeV~\cite{SQM_kiselA}. Au+Au collisions are planned with beam energies
spanning from $\sqrt{s_{NN}}$ = 12 - 27 GeV with luminosity $L \geq 10^{30} cm^{-2}s^{-1}$ and $\sqrt{s_{NN}}$ = 4 - 11 GeV with average luminosity
$L \geq 10^{27} cm^{-2}s^{-1}$. The interaction rate of 6 kHz for minimum bias is planned for the higher collision energies but decreases
to about 10 Hz because of low luminosity at  $\sqrt{s_{NN}}$= 4 GeV. A technique to sustain the beam luminosity in low energies for prolonged period
of time, called the electron cooling, which is also being used in BES-II at STAR,  will be employed. 
\subsubsection{Compressed Baryonic Matter (CBM) experiment}
The CBM experiment is a fixed-target detector capable of identifying hadrons, leptons ($e$, $\mu$) and photons.
The interaction rate will be higher for the CBM experiment at FAIR. The experiment will operate over the energy range, $\sqrt{s_{NN}}$= 2.7 - 4.9 GeV.
The goal of the CBM experiment at SIS100  is to discover fundamental properties of QCD matter at high baryon density using rare probes. The experiment plans to perform high-precision study of higher-order fluctuations at various beam energies in baryon density region of 500 MeV $ < \mu_{B} < $ 800 MeV. In order to achieve high precision, the measurements will be performed at reaction rates up to 10 MHz. The CBM detector acceptance of polar emission angles between 2.5 and 25 degrees allows it to cover mid-rapidity and the forward rapidity for symmetric collision systems over the FAIR energy range. The lab pseudo-rapidity range coverage of the CBM detector system is $1.5 < \eta < 3.8$.  The detector setup includes Micro Vertex Detector (MVD), Silicon Tracking System (STS), Ring Imaging Cherenkov Detector (RICH), Transition Radiation Detector (TRD), and Time-of-Flight Detector (TOF)~\cite{Senger:2017oqn,Senger:2020fvj}. The collision vertex of the event will be provided by the MVD. The STS inside the dipole field records the momentum information of the charged particle tracks. Particle identification will be done by the TOF. Measurement of centrality of collision will be possible with Project Spectator Detector (PSD). Feasibility study of performing fluctuation measurements at CBM are ongoing and show promising prospects~\cite{Samanta:2020fho}.\\\\
The new experiments will collect enormous event statistics due to high interaction rates. Large event statistics will be beneficial for higher-order fluctuation measurements. Handling such rates would demand fast and state-of-the-art detectors, which are currently being envisaged. The interaction rates from various upcoming experiments are summarised in the Fig.~\ref{interac}~\cite{Galatyuk:2019lcf,Bzdak:2019pkr}. In addition, understanding the systematics of fluctuation measurements is crucial. For example, determining the centrality from different detectors (PSD at CBM and EPD at STAR) than those used for PID, enlarging the kinematic acceptance for PID (iTPC at STAR) will help us to study the measurements better. Due to low multiplicity of produced particles in low energy collisions, centrality definition using charge particles or energy deposition measurements suffer from poor resolution. This could result in a artificial correlation in fluctuation measurements as discussed in subsection~\ref{sectn_volfluct_label}. So, proper investigation of this effect should be studied in the fluctuation measurements. As nearly half of the net-baryon is carried out by (anti-)neutrons, fluctuations of neutron number, in principle, should also reflect criticality. Adding neutrons to proton fluctuations would serve as a better proxy of net-baryon number.
Other important factors will be the detector efficiencies and acceptance for various hadrons at low energies, especially in the fixed target configuration. Uniform acceptance of protons at mid-rapidity coverage would be crucial for studies on the critical point via proton number fluctuations.

\subsubsection{Acceptance: collider vs fixed target}
\begin{figure}[!htb]
	\centering 	
	\includegraphics[scale=0.5]{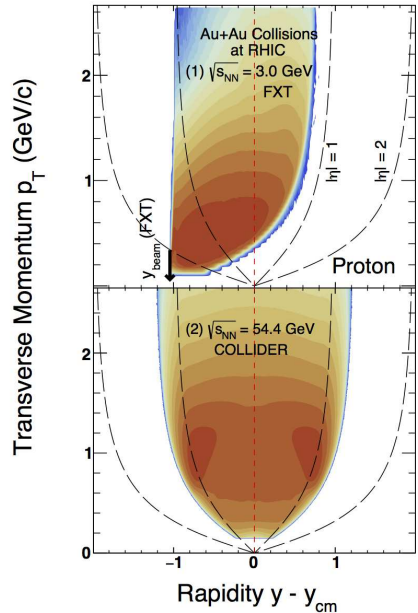}
	\caption{y-$p_{T}$ acceptance figure of protons measured in Au+Au collisions in the STAR experiment. The upper panel is from the collision in the fixed-target mode at $\sqrt{s_{NN}}$ = 3.0 GeV. The bottom panel represents the collision in collider mode at $\sqrt{s_{NN}}$ = 54.4 GeV.}
	\label{fxtacpt} 
\end{figure}
In nuclear collision experiments, for a given detector setup, the mode of collisions determines the phase space acceptance for the charged particles. For example, as shown in Fig.~\ref{fxtacpt}, the acceptance of protons is asymmetric in fixed target mode (upper panel) as compared to the collider mode (lower panel), where it is symmetric. 

In the collider mode, the mid-rapidity is always at zero, and this allows STAR TPC to perform mid-rapidity measurements at all collision energies. However, in the fixed target mode, the location of mid-rapidity depends on collision energy. The acceptance of the STAR detector allows coverage of mid-rapidity distributions of protons in fixed target mode at $\sqrt{s_{NN}}$ = 3 GeV, while at higher fixed target energies, additional detectors like iTPC (inner - TPC) and eTOF (extended - TOF) are needed for the phase-space coverage around the mid-rapidity.

\section{Summary}
In the absence of any first-principle theoretical calculations on the location of the QCD critical point, one has to rely on predictions from QCD-based theories to search for critical signals in heavy-ion collisions. One such calculation is done with the linear sigma model, which predicts a non-monotonic dependence of kurtosis of number fluctuations when approaching the critical point. 
Experimental measurement on the net-proton cumulant ratio $C_{4}/C_{2}$ in 0-5\% Au+Au collisions, shows an evidence of a non-monotonic collision energy dependence of the measurements in the range $\sqrt{s_{NN}}$ = 7.7 - 200 GeV. 

The normalised fourth-order cumulant ratio of proton shows collision energy dependence similar to net-proton measurements, which is attributed to the presence of four-proton correlations, the anti-proton, on the other hand, have a weak dependence on collision energy. None of the model calculations are able to reproduce the observed energy dependence of the net-proton cumulant $C_{4}/C_{2}$. Comparison with a wide pool of baseline calculations suggests that the 0-5\% net-proton $C_{4}/C_{2}$ measurements deviate below and above the baseline calculations at a level of 2-3 $\sigma$ at various collision energies. The exhibited non-monotonicity of net-proton cumulants with respect to Skellam baseline, UrQMD baseline, and HRG canonical ensemble calculations are found to be of 3.1-3.3 $\sigma$ significance. The recent measurement of proton $C_4/C_2$ in Au+Au collisions at $\sqrt{s_{NN}}$ = 3 GeV show suppression for 0-5\% centrality and is consistent with the expectation from the baryon number conservation effect suggesting the dominance of hadronic matter in such low energies. The suppression of proton cumulant ratio $C_4/C_2$ at 3 GeV, together with the reported non-monotonic collision energy dependence of net-proton $C_4/C_2$ at higher energies, indicates that the possible critical point could exist at collision energies higher than 3 GeV.

The experimental results on net-charge and net-kaon cumulant ratio $C_{4}/C_{2}$ in most central 0-5\% collisions have large uncertainties associated with the measurements and show a flat collision energy dependence with most of the data points showing agreement with the Skellam baseline within uncertainties. The current net-$\Lambda$ fluctuation results are only up to the $3^{rd}$ order and do not shown any signs of critical fluctuations. Addition of net-$\Lambda$'s to net-kaon fluctuation could serve as a better proxy for net-strangeness fluctuations as most of the strangeness is carried by kaons and $\Lambda$'s in the system.

Negative values were observed from measurements of the fifth and sixth-order net-proton cumulant ratios, $C_{5}/C_{1}$ and $C_{6}/C_{2}$ for 0-40\% Au+Au collisions at most of the collision energies, although with large uncertainties. $C_{6}/C_{2}$ becomes increasingly negative with decreasing collision energy. At $\sqrt{s_{NN}}$ = 200 GeV, the $C_{6}/C_{2}$ measurements progressively become negative from peripheral to central collisions. These observations, albeit with large uncertainties, are qualitatively consistent with the trend and sign predicted by LQCD calculations for the ratios of fifth-to-first and sixth-to-second order baryon number susceptibilities for QCD matter in the range $\mu_{B} <$ 110 MeV. In contrast, positive $C_{5}/C_{1}$ and $C_{6}/C_{2}$ were reported in peripheral (70-80\%) collisions at all energies. Multiplicity dependence studies of the two ratios in p+p collisions at $\sqrt{s_{NN}}$ = 200 GeV also yeild positive sign. Current measurements on higher order factorial cumulants at $\sqrt{s_{NN}}$ = 7.7 GeV indicate that more event statistics is needed for the search of first-order phase transition in the high $\mu_{B}$ ($\sim$ 400 MeV) region. 

The system formed in the heavy-ion collision is of femto-scale level and evolves dynamically with time. Hence, it is quite challenging both theoretically and experimentally to probe into the physics of QCD critical point. Theoretical calculations considering the effects of non-equilibrium and critical slowing down near the CP will guide the experimental search for the CP. As a first step, one tries to establish the possibility of the existence of the QCD critical region, which could span about 100 MeV range in baryon chemical potential. The current result on net-proton kurtosis is an important step in that direction as these are the first fluctuation measurements exhibiting non-monotonic variation with collision energy. To reduce the statistical uncertainties on the current net-particle cumulant ratios measurements at lower collision energies and improve the quality of the observed signal, the BES-II program has been carried out at RHIC. The STAR FXT program, along with several future experiments like, NICA and CBM, will collect large event statistics and explore the baryon-rich energy region. The overlapping $\mu_{B}$ region covered by these experiments will allow for independent confirmation of possible critical signals. With the state-of-the-art detector technology at the helm, the upcoming experiments carry huge potentials to discover the elusive critical point and map the QCD phase diagram with utmost precision.

{\bf{Acknowledgements:}} The authors thank the colleagues from STAR and ALICE collaborations for several discussions. B.M. was supported in part by the J C Bose Fellowship from Department of Science of Technology and Department of Atomic Energy, Government of India.

\end{document}